\documentclass[imslayout,preprint]{imsart} 
\def\journal@name{} 
\usepackage{amssymb,amsthm,mathtools}
\usepackage{mathtools}
\usepackage{graphicx}
\usepackage{geometry}
\usepackage[colorlinks,citecolor=blue,urlcolor=black,linkcolor=blue,pdfborder={0 0 0}]{hyperref}
\usepackage{comment}
\usepackage{graphicx}
\usepackage{enumitem}
\usepackage{framed}
\usepackage{dcolumn}
\usepackage{subfig}
\usepackage[sort]{natbib}
\usepackage[capitalize]{cleveref}  
\usepackage{datetime}
\usepackage{algorithm}
\usepackage{algpseudocode}
\usepackage[normalem]{ulem}
\usepackage{booktabs}
\usepackage{url} %
\RequirePackage[usenames,dvipsnames]{color}
\usepackage{jhh-misc2}
\usepackage{xr}
\newlist{scenarios}{enumerate}{1}
\setlist[scenarios,1]{label=\textit{Scenario~\arabic*},wide=0pt, leftmargin=*}

\newlist{subassump}{enumerate}{1}
\setlist[subassump]{label=(\alph*)}
\crefname{subassumpi}{Assumption}{Assumptions}

\newlist{assump}{enumerate}{1}
\setlist[assump]{label=(A\arabic*)}
\crefname{assumpi}{Assumption}{Assumptions}

\usepackage{mathrsfs}
\usepackage{autonum}

\newcommand{\indicatorfn}{\mathds{1}}

\newcommand{\littleo}{o}
\newcommand{\littleoP}{\littleo_{P}}

\newcommand{\bigo}{O}

\newcommand{\statistic}[1]{T_{#1}}
\newcommand{\p}{p}

\newcommand{\truesym}{\circ}

\newcommand{\obsspace}{\mathbb X}
\newcommand{\numobs}{N}
\newcommand{\data}{\mathbf{x}}
\newcommand{\datarv}{\mathbf{X}}

\newcommand{\obs}[1]{x_{#1}}
\newcommand{\obsrv}[1]{X_{#1}}

\newcommand{\postdist}[1]{\Pi_{#1}}
\newcommand{\postdistfull}[2]{\Pi(#1 \given #2)}

\newcommand{\postdensity}[1]{\pi_{#1}}
\newcommand{\postdensityfull}[2]{\pi(#1 \given #2)}

\newcommand{\priordist}{\postdist{0}}

\newcommand{\likdist}[1]{P_{#1}}
\newcommand{\likfun}[1]{p_{#1}}
\newcommand{\lik}[2]{\likfun{#2}(#1)}

\newcommand{\priormarginal}{m_\textrm{prior}}
\newcommand{\postmarginal}{m_\textrm{post}}

\newcommand{\optsym}{\star}
\newcommand{\param}{\theta}
\newcommand{\paramspace}{\Theta}
\newcommand{\trueparam}{\param_{\truesym}}
\newcommand{\optparam}{\param_{\optsym}}

\newcommand{\mle}[1]{\hat\param}

\newcommand{\Ehessloglik}[1]{J_{#1}}

\newcommand{\obsdist}{P_{\truesym}}

\newcommand{\pcreprv}{X_{\text{pred}}}  

\newcommand{\spc}{\text{SPC}}  
\newcommand{\spcpvalue}[1]{ \p_{\spc}(#1)}
\newcommand{\spcprop}{q}
\newcommand{\spcobsrv}{\datarv_{\text{obs}}}    
\newcommand{\spcobsobs}[1]{X_{\text{obs},#1}}
\newcommand{\spcnewrv}{\datarv_{\text{ho}}}    
\newcommand{\spcnewobs}[1]{X_{\text{ho}, #1}}
\newcommand{\spcreprv}{\datarv_{\text{pred}}}  
\newcommand{\spcrepobs}[1]{X_{\text{pred},#1}}
\newcommand{\spcobssize}{\lceil \spcprop\numobs \rceil}
\newcommand{\spcnewsize}{ \numobs - \spcNobs}
\newcommand{\spcNobs}{\numobs_{\text{obs}}}
\newcommand{\spcNnew}{\numobs_{\text{ho}}}
\newcommand{\spcpostmarginal}{\postmarginal^{\spc}}
\newcommand{\intBVMapprox}{\widehat{\int}}
\newcommand{\dspcK}{k}
\newcommand{\dspc}{\text{dSPC}}
\newcommand{\dspcdata}[1]{\datarv^{(#1)}}
\newcommand{\dspcpval}[1]{\p_{\dspc}(#1)}
\newcommand{\dspcobspval}[1]{u_{#1}}

\newcommand{\dspcnumobs}{N_\dspcK}

\newcommand{\spcrepobsobs}[1]{\datarv^{\spc}_{\text{obs}}}

\newcommand{\pvalbvmdist}[1]{ \p'_{\spc}(#1)}
\newcommand{\pvalbvmdistball}[1]{ \p''_{\spc}(#1)}

\newcommand{\pvaltoK}[1]{ \p'_{\spc}(#1)}
\newcommand{\spcpvalueapprox}[1]{ \p''_{\spc}(#1)}

\newcommand{\asympmean}[1]{\nu_{#1}}
\newcommand{\trueasympmean}{\nu_{\truesym}}
\newcommand{\diffasympmean}[1]{\dot{\nu}_{#1}}
\newcommand{\asympsd}{\sigma}
\newcommand{\trueasympsd}{\asympsd_{\truesym}}
\newcommand{\diffasympvar}[1]{\dot{\asympsd}(#1)}

\newcommand{\mlevar}{\Sigma}
\newcommand{\truemlevar}{\Sigma_{\optsym}}
\newcommand{\distBVM}[2]{\widehat{\Pi}(#1\given#2)}
\newcommand{\densityBVM}[2]{\widehat{\pi}(#1\given#2)}
\newcommand{\paramstar}{\param'}

\newcommand{\paramseq}[1]{\tilde{\param}_{#1}}
\newcommand{\distKS}[1]{F_\mathrm{KS}(#1)}
\newcommand{\qdistKS}[1]{K_{#1}}

\newcommand{\truesigma}{\sigma_{\truesym}}
\newcommand{\liksd}{\sigma}
\newcommand{\priorsd}{\tau}
\newcommand{\priormean}{\mu}
\newcommand{\sdratio}{\kappa}
\newcommand{\ess}{\numobs_0}
\newcommand{\unadjess}{r}
\newcommand{\adjess}{r_{\text{adj}}}
\newcommand{\samplemean}[1]{\overline{#1}}

\newcommand{\pvaldist}{\mathbb{P}_{k}}
\newcommand{\pvalempdist}{\hat{\mathbb{P}}_{k}}

\newcommand{\fnclass}{\mathscr{F}}

\newcommand{\VCnum}[1]{\nu(#1)}
\newcommand{\BB}[1]{\mathbb{G}^{(#1)}}
\newcommand{\empGn}[2]{\mathbb{G}^P_{#1}#2}
\newcommand{\empPn}[1]{\hat{\mathbb{P}}_{#1}}

\newcommand{\indclass}{\mathscr{F}_{\text{ind}}}

\newcommand{\BL}{BL}

\newcommand{\BLnorm}[1]{\Vert #1 \Vert_{\BL}}

\newcommand{\Int}[1]{\operatorname{Int}({#1})}

\def\norm#1{\left\|{#1}\right\|} 

\graphicspath{{figures/}}

\allowdisplaybreaks[2]

\begin{document}

\begin{frontmatter}

\title{Calibrated Model Criticism Using \\ Split Predictive Checks}
\runtitle{~Split Predictive Checks}
\runauthor{J.\ Li and J.\ H.\ Huggins~}

\begin{aug}
\author[MS]{\fnms{Jiawei} \snm{Li}\ead[label=jl]{jwli@bu.edu}}
\and
\author[MS,CDS]{\fnms{Jonathan H.} \snm{Huggins}\ead[label=jh]{huggins@bu.edu}}
\address[MS]{Department of Mathematics \& Statistics, Boston University, \printead{jl}}
\address[CDS]{Faculty of Computing \& Data Sciences, Boston University, \printead{jh}}
\end{aug}

\begin{abstract}
Checking how well a fitted model explains the data is one of the most fundamental parts of a Bayesian data analysis.
However, existing model checking methods suffer from trade-offs between being well-calibrated, automated, and computationally efficient. 
To overcome these limitations, we propose \emph{split predictive checks} (SPCs), which combine the ease-of-use and speed of posterior predictive checks with the good calibration properties of 
predictive checks that rely on model-specific derivations or inference schemes. 
We develop an asymptotic theory for two types of SPCs: \emph{single SPCs} and the \emph{divided SPCs}.
Our results demonstrate that they offer complementary strengths.
Single SPCs work well with smaller datasets and provide excellent power when there is substantial misspecification,
such as when the estimate uncertainty in the test statistic is significantly underestimated. 
When the sample size is large, divided SPCs can provide better power and are able to detect more subtle form of misspecification. 
We validate the finite-sample utility of SPCs through extensive simulation experiments in exponential family and hierarchical models, and 
provide three real-data examples where SPCs offer novel insights and additional flexibility beyond what is available when using posterior predictive checks. 
\end{abstract}


\end{frontmatter}

\section{Introduction}
\label{sec:intro}

Assessing how well a fitted model explains the data is one of the most fundamental parts of Bayesian data analysis \citep{Box:1980,Gelman:2013,Blei:2014}. 
A well-calibrated method for model checking must correctly reject a model that does not capture aspects of the data viewed as important by the modeler
and fail to reject a model fit to well-specified data.
In addition to being well-calibrated, a model check should be computationally efficient and easy-to-use.
These latter requirements have evolved in the last 10--15 years due to the now-widespread use of \emph{probabilistic programming} systems 
such as BUGS~\citep{Lunn:2009:BUGS}, Stan~\citep{Carpenter:2017}, 
PyMC3~\citep{Salvatier:2016}, NIMBLE~\citep{deValpine:2015} and Pyro~\citep{Bingham:2019:Pyro},
which allow the data analyst to carry out estimation using Bayesian methods in essentially any model they would like.
It has thus become invaluable to have general-purpose (``black-box'') model checks that can use the outputs of such systems
without requiring any additional effort from the data analyst beyond (1) determining what aspects of data are important (2) writing a few lines of code to call
the model check. 

However, current approaches to Bayesian model checking face a trade-off between being (i) computationally efficient, (ii) general-purpose and (iii) well-calibrated.
Many well-calibrated methods are either computationally burdensome \citep{Gelfand:1992,Marshall:2003,Moran:2024,Robins:2000,Hjort:2006}
or require model-specific derivations \citep{Johnson:2004:Bayesian-chi-squared,Johnson:2007:pivotal,Yuan:2011:gof-diagnostics,Bayarri:2000,Dahl:2007:conflict-measure,Bayarri:2007}.
On the other hand, computationally efficient and general-purpose approaches %
often fail to be well-calibrated  \citep{Bayarri:2000,Robins:2000,Bayarri:2007}. 
In practice, the  general-purpose and easy-to-use posterior predictive check \citep[PPC;][]{Guttman:1967,Rubin:1984,Gelman:1996} is probably the most widely used. 
PPCs allow the user to specify a data summary statistic of interest, allowing them to focus the check on the aspects 
of the data they care most about. 
However, PPCs cannot be expected to be well-calibrated as a consequence of using the observed data twice. 
Predictive checks that use some form of cross-validation aim to avoid double use of the data \citep{Gelfand:1992,Vehtari:2002}. 
However, these methods can be computationally prohibitive \citep{Marshall:2003, Gronau:2019} and are not necessarily well-calibrated \citep{Piironen:2017,Ambroise:2002,Cawley:2010}.

To overcome the limitations of existing model checking approaches, we propose a simple new approach to constructing general-purpose predictive checks based on splitting the data into ``train''
and ``test'' sets. 
Hence, we call them \emph{split predictive checks} (SPCs). 
We develop an asymptotic theory for two types of SPCs, which we call \emph{single SPCs} and \emph{divided SPCs}. 
Our results show that both SPCs provide asymptotically well-calibrated $p$-values and the divided SPC is guaranteed to have power approaching one asymptotically.
Single SPCs, on the other hand, tend to fail only when there is subtle model misspecification, which can be 
an advantage when slight misspecification is not of concern. 
We also show how SPCs offer substantial flexibility to match the model's use case (e.g., prediction, description, extrapolation). 
We provide practical guidance on when and how to use SPCs in hierarchical, time-series, and other complex models. 
We demonstrate the benefits of SPCs in large- and small-sample regimes through simulation studies 
in some conjugate exponential family models and a Gaussian hierarchical model.
Finally, we verify the practical utility of SPCs in three real-data examples, using a range of models from simple exponential 
families to Gaussian processes.  

\paragraph*{Notation.}
For a sequence $a_{1}, a_{2}, \dots, a_{N}$ (where $a$ may be replaced with any other letter-like symbol), 
we define $\boldsymbol{a} \defined \boldsymbol{a}_{N} = (a_{1}, a_{2}, \dots, a_{N})$. 
For a vector $v \in \reals^{d}$, let $\norm{v}_{2}$ denote the usual Euclidean norm. 
For a set $A$, denote the interior of the set as $\Int{A}$.
We write $f(\numobs) = O(g(\numobs))$ if $\limsup_{\numobs \to \infty} |f(\numobs)/g(\numobs)| < \infty$. 
We write $z_{\alpha}$ to denote the $\alpha$th quantile of standard normal distribution.
We write $\convD$ to denote convergence in distribution and $\convP$ to denote convergence in probability. 

\section{Background}

\subsection{Approaches to Bayesian model assessment and model checking}

The philosophy between Bayesian model assessment and Bayesian model checking has been extensively explored in the past decades \citep{Gelman:2013,Presnell:2004,Sprenger:2013,Gustafson:2001,Stern:2005}. 
Bayesian model assessment concerns testing for
\textit{correctness} of a model, while Bayesian model checking aims to understand how much of a discrepancy there is between the assumed model and the observed data.
One approach to Bayesian model assessment is to extend classical goodness-of-fit methods (e.g., $\chi^{2}$ tests) 
\citep{Johnson:2004:Bayesian-chi-squared,Johnson:2007:pivotal,Yuan:2011:gof-diagnostics}.
These approaches produce \emph{(asymptotically) frequentist $\p$-values} when the model is correct.
That is, (in the large-data limit) the $\p$-value has a uniform distribution $\distUnif(0,1)$ when the model is correctly specified. 
However, these methods often require model-specific derivations and do not allow the data analyst to interrogate whether the model correctly captures the aspects of the data the analyst believes are most important. 

Bayesian model checking methods, on the other hand, can be formulated as assessing how well the model fits the data \emph{in 
	particular directions} (via some test statistics) rather than contrasting it with alternative models or attempting to match it with the true model \citep{Gelman:1996, Steiger:2007, Wang:2023, Ibrahim:2001, Czado:2009,Berkhof:2000}.
A common Bayesian model checking method is called \emph{predictive check} \citep{Guttman:1967,Box:1980,Rubin:1984,Gelfand:1992,Marshall:2003,Bayarri:2000,Robins:2000,Hjort:2006,Dahl:2007:conflict-measure,Bayarri:2007,Gasemyr:2019:ppp}.
Consider a model $\theset{\likdist{\param} \st \param \in \paramspace}$ and denote the density of $\likdist{\param}$ with respect to some reference measure $\lambda$
by $\likfun{\param}$.  
Assume random variables $\obsrv{1},\dots,\obsrv{\numobs} \overset{\iid}{\dist} \obsdist$, where $\obsrv{n} \in \obsspace$ and $\obs{i}$ is a realization of $\obsrv{i}$. 
Define $\datarv \defined (\obsrv{1},\dots,\obsrv{\numobs})$, $\data \defined (\obs{1},\ldots, \obs{\numobs})$ and, with a slight abuse of notation, 
$\likfun{\param}(\data) \defined \prod_{n=1}^{\numobs}\likfun{\param}(\obs{n})$.
Let $\optparam \defined \sup_{\param \in \paramspace} \EE\{\log \likfun{\param}(\obsrv{1})\}$ denote the KL-optimal parameter, which we assume throughout
exists and is unique. 
Predictive checks rely on a family of statistics $(\statistic{\numobs}: \obsspace^{\numobs} \rightarrow \mathbb{R})_{\numobs \in \nats}$ indexed by the dataset size.
The analyst can determine which aspects of the data the predictive check will validate via the choice of $\statistic{} = \statistic{\numobs}$.
The statistic of the observed data is compared to the same statistic under a ``null distribution'' with density $m$. Let $\priordist$ denote the prior distribution on $\paramspace$.
Standard choices for $m$ include the prior predictive density $\priormarginal(\mathbf{y}) \defined \int\lik{\mathbf{y}}{\param} \priordist(\dee\param)$ \citep{Box:1980} 
and the posterior predictive density $\postmarginal(\mathbf{y}) \defined \int \lik{\mathbf{y}}{\param}\postdistfull{\dee\param}{\data}$ \citep{Guttman:1967,Rubin:1984},
where %
$\postdistfull{\dee\param}{\data} \defined\lik{\data}{\param} \priordist(\dee\param) /\priormarginal(\data)$ is the posterior distribution. 
The $p$-value for the predictive check is given by
$\p(\data) \defined \Pr_{\mathbf{Y}\dist m}\{ \statistic{}(\mathbf{Y})\ge \statistic{}(\data) \}$, where the superscript $m$ denotes the distribution of 
$\mathbf{Y} = (Y_{1},\dots,Y_{\numobs})$.
Given any $p$-value $\p(\data)$, define the two-sided version by $\p_{2}(\data) = 2 \min\{\p(\data), 1 - \p(\data)\}$.

Some predictive checks require elaborations on the basic definition. 
For example, the statistic can also be allowed to depend on the model parameter, in which case it is of the form $\statistic{}(\data, \param)$ and the null distribution $m$ is a joint distribution
on $\obsspace \times \paramspace$ \citep{Mimno:2011,Gelman:1996}.
In the posterior predictive case, the null density would be $\postmarginal(\mathbf{y}, \param) \defined  \lik{\mathbf{y}}{\param} \postdensityfull{\param}{\data}$,
where $\postdensity{}$ is the density of $\postdist{}$ with respect to $\lambda$. 
A second example is the cross-validated predictive check, where one $p$-value $\p_{n}$ is computed for each observation: 
$\p_{n}(\data) \defined \Pr_{Y_n \dist \postmarginal^{-n}}\{ \statistic{}(Y_{n})\ge\statistic{}(\obs{n}) \}$,
where $\postmarginal^{-n}(y_{n})  \defined \int \lik{y_{n}}{\param}\postdistfull{\dee\param}{\data^{-n}}$ is the posterior predictive
density of the $n$th observation conditional on $\datarv^{-n} \defined (\obsrv{1},\dots,\obsrv{n-1}, \obsrv{n+1},\dots,\obsrv{\numobs})$,
the data without the $n$th observation \citep{Marshall:2003}.
A final example is the first version of the population predictive check (POP-PC) described in Supplementary Materials B of \citep{Moran:2024}.
POP-PC avoids the double use of data by comparing the observed data with replicated data generated from the true distribution. 
The POP-PC $\p$-value is defined as
$
\p_{\text{POP-PC}}(\data) \defined \EE\left[\Pr_{\mathbf{Y}\dist\postmarginal}\left\{\statistic{}(\mathbf{Y}) \ge \statistic{}(\datarv_{\text{new}})\right\}\right],
$
where the expectation is with respect to $\obsrv{\text{new},1}, \dots, \obsrv{\text{new},\numobs} \distiid \obsdist$. 
However, as empirically verified by \citep{Moran:2024}, POP-PC fails to provide well-calibrated $\p$-values.
Finally, rather than computing $p$-values, predictive checks can be also performed graphically \citep{Gelman:1996}.

\subsection{Limitations of existing predictive checks}
The posterior predictive check (PPC) is particularly widespread use because it only requires samples from the posterior 
predictive distribution.
Hence, it is easy to implement and has minimal computational overhead. 
However, posterior predictive $\p$-values are not, in general, (asymptotically) frequentist, %
which makes their interpretation challenging.
Even if the analyst chooses to carry out, say, a graphical predictive check, the non-uniformity of the posterior predictive $p$-values 
makes the interpretation of the graphical check suspect.  
PPCs tend to produce conservative $\p$-values (that is, that concentrate away from $0$ and $1$) and thus fail to detect model misspecification. 
The conservatism of posterior predictive checks results from ``double use'' of the observed data, which assesses the model by comparing the observed data with some replicated data drawn from a distribution conditional on the same observed data \citep{Robins:2000}. 

More generally, previous approaches to constructing predictive $p$-values suffer from a trilemma: they fail to simultaneously produce (asymptotically) frequentist $p$-values, 
be computationally efficient, and be general-purpose. 
While it is possible to calibrate an arbitrary $p$-value, the calibration process is, in general, computationally prohibitive \citep{Robins:2000,Hjort:2006}. 
Other proposals that produce uniform $p$-values do not seem to have been adopted in practice because they require model-specific derivations 
\citep{Bayarri:2000,Dahl:2007:conflict-measure,Bayarri:2007}.
Predictive checks -- such the cross-validated predictive check  \citep{Gelfand:1992,Marshall:2003} --
which try to avoid double use of the data can be computationally prohibitive and do not necessarily produce uniform $p$-values.

\section{Split Predictive Checks}
\label{sec:singleSPC}

We introduce split predictive checks (SPCs) as a way to solve the trilemma discussed in the previous section. 
The idea of SPCs is to compute predictive $\p$-values by splitting the original dataset into observed ``training'' data and held-out ``test'' data,
thereby avoiding double use of the data while retaining computational efficiency. 

\subsection{Single SPCs} \label{sec:single-SPC}

We start with the simplest possible SPC, which we call the \emph{single SPC}. 
For a given proportion $\spcprop \in (0,1)$, we split the data into two disjoint subsets: the observed data $\spcobsrv$ of size $\spcNobs \defined \spcobssize$  and the held-out data  $\spcnewrv$ of size  $\spcNnew \defined\spcnewsize$. 
The \emph{single $\spcprop$-SPC $p$-value} is defined as
\begin{equation}
	\spcpvalue{\datarv} \defined \Pr_{\spcreprv\dist\spcpostmarginal}\{ \statistic{\spcNnew}(\spcreprv)\ge \statistic{\spcNnew}(\spcnewrv)\},
\end{equation}
where $\spcpostmarginal \defined \postmarginal(\spcreprv \mid \spcobsrv) \defined \int \lik{\spcreprv}{\param}\postdistfull{\dee\param}{\spcobsrv}$ denotes the distribution of $\spcreprv$. Here we consider $\spcpvalue{\datarv} $ as a random variable with $\spcpvalue{\data}$ as its realization.
Single SPCs inherit the ``black-box'' nature of general predictive checks as well as the flexibility provided to the analyst via the choice of statistic. 
The additional computation required to obtain the single SPC $p$-value is relatively small because it only requires the estimation of posterior of dataset smaller than than original. 

We first show that single $\spcprop$-SPC $p$-values are asymptotically uniform under the well-specified model and characterize their asymptotic power. 
The assumptions required for our result are relatively mild. 
First, we require some regularity conditions on the test statistic under the assumed model. 
\begin{assumption} Assume that 
	\label{assump:regularity}
	\begin{subassump}[label = (\alph*),ref=(\theassumption-\alph*)]
		\item The asymptotic mean $\asympmean{}(\param) \defined \lim_{\numobs\to\infty} \int\statistic{\numobs}(\data)\likdist{\param}(\dee\data)$ and
		asymptotic variance $\asympsd^2(\param) \defined \lim_{\numobs\to\infty} \int\statistic{\numobs}^2(\data)\likdist{\param}(\dee\data) -\asympmean{}^2(\param)$
		exist; 
		\item \label{assump2:asymp_mean_diff} the asymptotic mean is twice-differentiable and finite at $\optparam$, so
		$\diffasympmean{}(\optparam) \defined \lim\limits_{\numobs\to\infty}\frac{\partial \asympmean{}(\param)}{\partial \param}|_{\param = \optparam}$
		and $\ddot{\asympmean{}}(\optparam)\defined \lim\limits_{\numobs\to\infty}\frac{\partial \diffasympmean{}(\param)}{\partial \param}|_{\param = \optparam}$ are well-defined and
		$\Vert \ddot{\asympmean{}}(\optparam)\Vert_2 < \infty$; and
		\item 	\label{assump4:bdd_diff_sigma} there exists an open neighborhood $U_{\optsym,1} \ni \optparam$ and constant $M> 0$ such that for all $\param \in U_{\optsym,1}$, $\diffasympvar{\param} \defined \frac{\partial \asympsd(\param)}{\partial\param}$ exists and $\Vert\diffasympvar{\param}\Vert_2\leq M$.
	\end{subassump}
\end{assumption}

Our second assumption requires the scaled test statistic to satisfy a central limit theorem, which will typically be the case (e.g., when $\statistic{}$ takes the form of an average or is a $U$-statistic).
\begin{assumption} \label{assump:asymp_norm_stat} 
	There exists an open neighborhood $U_{\optsym,2} \ni \optparam$ such that for $\param \in U_{\optsym,2}$ 
	and $Y_1, \dots, Y_{\numobs} \distiid \likdist{\param}$, 
	\begin{equation}
		\frac{\numobs^{1/2}\{\statistic{\numobs}(\mathbf{Y}_{\numobs}) - \asympmean{}(\param)\}}{\asympsd(\param)} \convD \distNorm(0,1). 
	\end{equation}
\end{assumption}

Our third assumption requires asymptotic normality of the maximum likelihood estimator $\mle{}(\datarv)\defined \argmax_{\param \in \paramspace}\lik{\datarv}{\param}$ and the posterior. 
Denote the usual information matrices by $\Ehessloglik{\optsym}\defined -\EE_{\obsdist}[\nabla^2\log \likfun{\optparam}(\obsrv{1})]$ and $\mlevar_{\optsym}\defined \cov_{\obsdist}[\nabla\log \likfun{\optparam}(\obsrv{1})]$,
and denote the total variation distance between distributions $\mu$ and $\omega$  by $d_{TV}(\mu, \omega) \defined 2\sup\{|\mu(B) - \omega(B)| \,:\,\text{measurable } B \subseteq \paramspace \}$. 
\begin{assumption}
	\label{assump:bvm_and_mle} 
	Assume that
	\begin{subassump}[label = (\alph*),ref=(\theassumption-\alph*)]
		\item  \label{assump:MLE_conv}	
		$\numobs^{1/2}\{\mle{}(\mathbf{X}_{\numobs}) - \optparam\} \convD \distNorm(0, \Ehessloglik{\optsym}^{-1}\mlevar_{\optsym}\Ehessloglik{\optsym}^{-1})$
		and
		\item \label{assump:bvm}
		$d_{TV}(\postdist{}(\cdot\mid\mathbf{X}_{\numobs}), \distNorm(\hat{\param}(\mathbf{X}_{\numobs}),\Ehessloglik{\optsym}^{-1}/ \numobs))  \convP 0$.		
	\end{subassump}
\end{assumption}
The conditions for \cref{assump:MLE_conv} to hold are standard while \citet{Kleijn:2012} gives general conditions under which \cref{assump:bvm} holds. 
Denote the asymptotic means and standard deviations under the data distribution by, respectively,
$\trueasympmean \defined \lim_{\numobs\to\infty} \int\statistic{\numobs}(\data)\obsdist(\dee\data)$
and $\trueasympsd^{2} \defined \lim_{\numobs\to\infty} \int\statistic{\numobs}^2(\data)\obsdist(\dee\data) - \trueasympmean^{2}$.

\bthm \label{THM:SINGLESPC}
Under \cref{assump:regularity,assump:asymp_norm_stat,assump:bvm_and_mle}, the following hold  for any $\alpha \in (0,1)$: 
\begin{enumerate}
	\item If $\trueasympmean = \asympmean{}(\optparam)$, then
	\[
	\Pr\left[\spcpvalue{\datarv}< \alpha\right] \convP 1-\Phi\left(\frac{z_{1-\alpha} }{\rho}\right) 
	\]
	where
	\[
	\rho^2 \defined \frac{\spcprop\trueasympsd^2 + (1 - \spcprop )\diffasympmean{}(\optparam)^\top\Ehessloglik{\optsym}^{-1}\mlevar_{\optsym}\Ehessloglik{\optsym}^{-1}\diffasympmean{}(\optparam)}{\spcprop\asympsd(\optparam)^2+ (1 - \spcprop ) \diffasympmean{}(\optparam)^\top\Ehessloglik{\optsym}^{-1}\diffasympmean{}(\optparam)}.
	\]
	In particular, if the model is correctly specified, then $\Pr[\spcpvalue{\datarv}< \alpha] \convP  \alpha$. 
	\item If $\trueasympmean < \asympmean{}(\optparam)$,  then
	$
	\Pr\left[\spcpvalue{\datarv}> 1 - \alpha\right] \convP 1. 
	$
	\item If $\trueasympmean > \asympmean{}(\optparam)$, then
	$
	\Pr\left[\spcpvalue{\datarv}< \alpha\right] \convP 1. 
	$
\end{enumerate}

\ethm

The proof of \cref{THM:SINGLESPC} is in Supplementary Materials \ref{appx:proof-of-singleSPCthm}. 
The asymptotic behavior of single SPC $p$-values is impacted by three factors.
The first factor is whether the model can match the mean of the test statistic under the data distribution; that is, 
whether $\trueasympmean = \asympmean{}(\optparam)$. 
If $\trueasympmean \ne \asympmean{}(\optparam)$, then the model is misspecified, and Parts 2 and 3 
state that the single SPC $p$-value will have asymptotic power 1. 
We refer the case where the model cannot match the mean of the statistic as indicating ``major'' misspecification.

However, when $\trueasympmean = \asympmean{}(\optparam)$, the behavior of the single SPC $p$-value depends 
on the two other factors: the quality of the uncertainty quantification of the test statistic and of the model parameters,
which is summarized by the quantity $\rho^{2}$. 
The numerator of $\rho^{2}$ is a mixture of the asymptotic variance of the test statistic under the data distribution, $\trueasympsd^{2}$, and a more complicated term,
$\diffasympmean{}(\optparam)^\top\Ehessloglik{\optsym}^{-1}\mlevar_{\optsym}\Ehessloglik{\optsym}^{-1}\diffasympmean{}(\optparam)$.
This latter term can be interpreted as the variance of the MLE, $\Ehessloglik{\optsym}^{-1}\mlevar_{\optsym}\Ehessloglik{\optsym}^{-1}$, 
in the direction $\asympmean{}(\optparam)$ is changing most quickly (i.e., $\diffasympmean{}(\optparam)$).
The denominator has a similar form, but now $\trueasympsd^{2}$ is replaced by the test statistic variance under the 
model and the second term now quantifies the posterior variance $\Ehessloglik{\optsym}^{-1}$ in the direction of $\diffasympmean{}(\optparam)$.

\begin{figure}[tp]
	\centering
	\subfloat{\includegraphics[width=75mm]{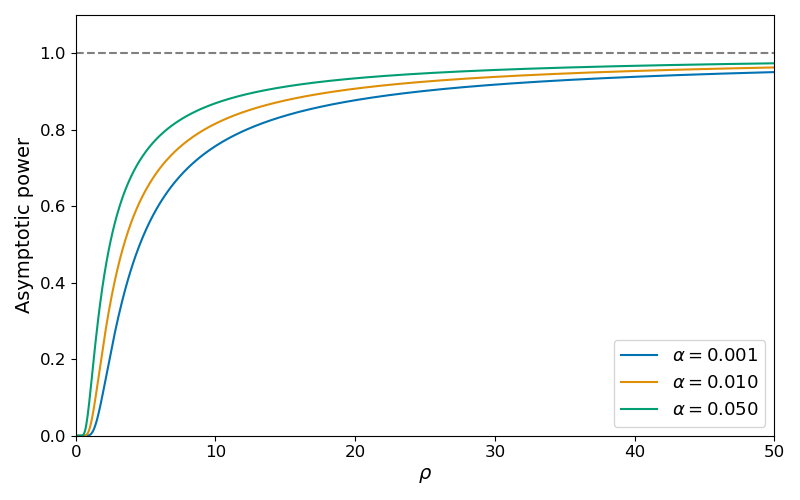} }
	\subfloat{\includegraphics[width=75mm]{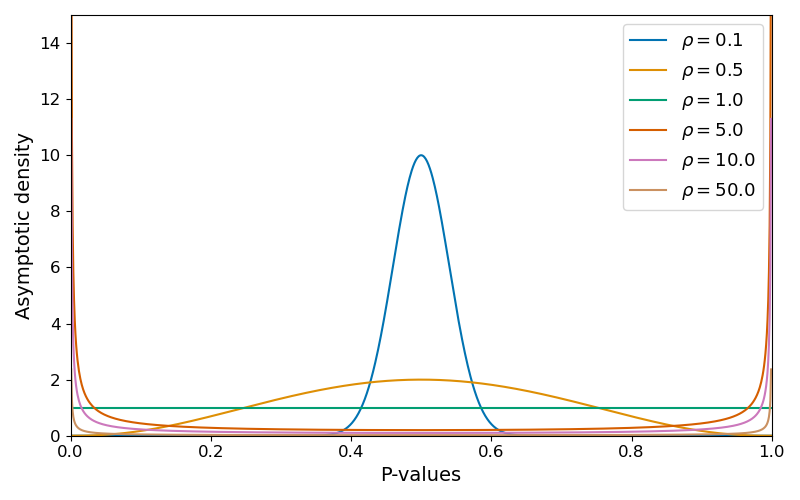} }
	\caption{\textbf{Left: }The asymptotic power of single SPCs against $\rho$ using a two-sided $\p$-value with fixed $\alpha \in \{0.001, 0.01, 0.05\}$. The horizontal dashed line indicates power 1. \textbf{Right:} The asymptotic density of the single SPC p-values for different values of $\rho$.}
	\label{fig:gaussian_rho_pw}	
\end{figure}

When the model \emph{under}estimates the two forms of uncertainty, $\rho^{2} > 1$.
In this case the power of a test using the single SPC $p$-value is greater than $\alpha$, 
and increases to 1 as $\rho^{2} \to \infty$.
Thus, the greater the underestimation of uncertainty, the greater the probability of obtaining a $p$-value close to 0 or 1.
When the model correctly quantifies both forms of uncertainty, which includes the well-specified case,
$\rho^{2} = 1$ and the $p$-value is uniform. 
Hence, single SPCs are asymptotically frequentist. 
When the model \emph{over}estimates the two forms of uncertainty, $\rho^{2} < 1$.
In this case the $p$-values concentrate close away from 0 and 1. 
In this scenario the single SPC $p$-value is not useful for detecting model misspecification.
How the asymptotic power and $p$-value distribution changes as a function of $\rho$ are illustrated in \cref{fig:gaussian_rho_pw}.

We can think of the case where the model can match the mean of the statistic as indicating at most ``mild-to-moderate'' misspecification, with the exact degree quantified by the magnitude $\rho$.
So, as a slogan, we can summarize \cref{THM:SINGLESPC} as showing that single SPCs will be effective at 
detecting ``moderate-to-major misspecification.''
It will depend on the use-case whether it is desirable that single SPCs fail to detect ``mild'' misspecification --
i.e., when (1) the model can match the statistic and (2) the model provides either fairly accurate or conservative uncertainty quantification.

\begin{example}
	\label{exa:gaussian-loc-model}
	Consider a Gaussian location model with known variance $\liksd^2$ and the parameter of interest is the mean $\param$:
	\begin{equation}
		\begin{aligned}
			\obsrv{i} &\distiid \mathcal{N}(\param, \liksd^2), \quad &
			\param &\dist \mathcal{N}(\priormean, \priorsd^2).
			\label{GaussianMod}
		\end{aligned}
	\end{equation}
	Assume the data-generating distribution is $\obsdist=\distNorm(\trueparam,\truesigma^2)$, so the model is misspecified if $\liksd \ne \truesigma$. 
	If we use the (sufficient) mean statistic $\statistic{\numobs}(\datarv_{\numobs}) = \samplemean{\obsrv{}} \defined \frac{1}{\numobs}\sum_{i = 1}^{\numobs} \obsrv{i}$,
	then $\trueasympmean = \asympmean{}(\optparam) = \optparam$.
	It follows from Part 1 of \cref{THM:SINGLESPC} that the asymptotic power of the two-sided single $\spcprop$-SPC  
	converges to $2\Phi(z_{\frac{\alpha}{2}}/\rho)$ with $\rho= \truesigma/\liksd$. %
	Thus, the single $\spcprop$-SPC will be effective at detecting misspecification when true variance is substantially 
	underestimated (i.e., $\truesigma^{2}/\liksd^{2} \gg 1$).
	However, when $\truesigma^{2}/\liksd^{2}$ is close to or less than 1, the single $\spcprop$-SPC test will fail to reject the model. 
\end{example}

\brmk[Extension to realized discrepancies] \label{RMK:GAUSSIAN_MSE}
While our focus is on the case of statistics depending only on the data, our approach and results can be extended to 
the case of realized discrepancies, where the statistic depends on the data and the parameter. 
For example, in Supplementary Materials \ref{appx:MSE_gaussianmod} we show that, for the Gaussian location model and mean-square error statistic $\statistic{}(\datarv,\param) = \frac{1}{\numobs}\sum_{i = 1}^{\numobs}(\obsrv{i} - \param)^2$,
$
\Pr[\spcpvalue{\datarv} <  \alpha] \convP 1-\Phi(z_{1-\alpha}/\rho),
$
where $\rho = \truesigma^2/\sigma^2$. 
Hence, we reach similar conclusions about the $\spcprop$-SPC $p$-values for realized discrepancies being asymptotically well-calibrated but possibly having asymptotic power less than one. 
\ermk

\subsection{Divided SPCs}
\label{sec:dividedSPC}

\begin{figure}[tp]
	\centering
	\includegraphics[width=95mm]{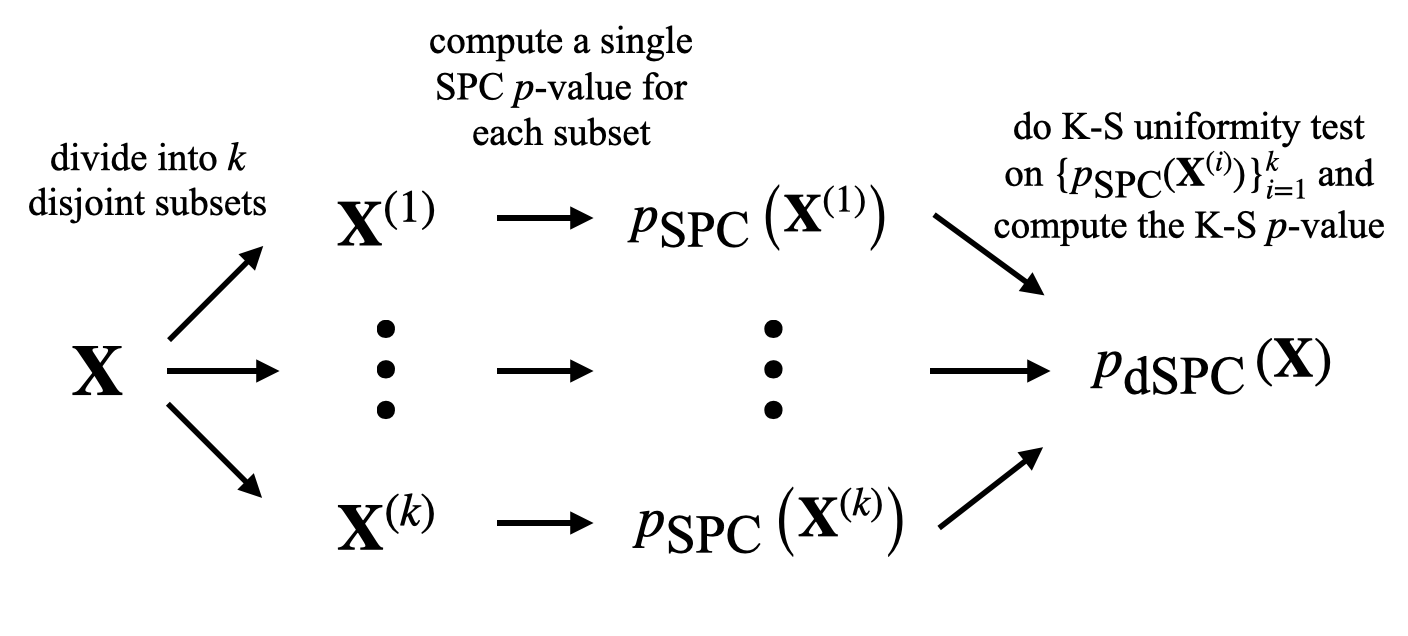}
	\caption{A schematic diagram of divided SPC. Divided SPC splits the original dataset into $\dspcK$ disjoint subsets $\datarv^{(i)}, i = 1, \ldots, \dspcK$ and computes a single SPC $\p$-value for each subset. The divided SPC $\p$-value is the $\p$-value from the Kolmogorov-Smirnov test on all single SPC $\p$-values.}
	\label{figur:dspc_diagram}
\end{figure}

In light of the possibly poor power of the single SPC for the ``mild'' misspecification case, we aim to develop an alternative predictive check that is guaranteed to
have asymptotic power of one when the model is misspecified. 
We build upon the single SPC by observing that, when the model is misspecified, we still expect the asymptotic 
distribution of the single SPC $p$-values to be non-uniform. 
Hence, if we have many such $p$-values, we can instead test for their uniformity. %
More precisely, given data $\datarv$, divide the original data into $\dspcK$ equal folds of size $\dspcnumobs \approx \numobs/\dspcK$.
Denoting these folds by $\dspcdata{1}, \ldots, \dspcdata{\dspcK}$, the next step is to compute the single $\spcprop$-SPC $\p$-value $\dspcobspval{\dspcK j} \defined \spcpvalue{\dspcdata{j}}$ for each fold
$j = 1, \ldots, \dspcK$. 
To assimilate all the $\dspcK$ single SPC $\p$-values, we use the Kolmogorov--Smirnov (KS) test for uniformity \citep{Massey:1951}. 
Specifically, we define the \emph{divided SPC $\p$-value} as
\[
\dspcpval{\spcobsrv} \defined \p_{\text{KS}}(\dspcobspval{\dspcK1}, \ldots, \dspcobspval{\dspcK\dspcK} ),
\]
where $\p_{\text{KS}}(\cdot)$ denotes the $\p$-value of the KS uniformity test. 
A schematic diagram of how to calculate the divided SPC $\p$-value is presented in \cref{figur:dspc_diagram}. 

The choice of the KS test is reasonable in light of our next result, which establishes that, when the model is correctly specified,
we can bound the Kolmogorov distance between the distribution of the single SPC $\p$-values and the uniform distribution. 
We recall that for distributions $\mu$ and $\omega$ on $\reals$, the Kolmogorov distance is given by 
$
d_{K}(\mu, \omega) \defined \sup_{u\in \reals}  |\mu((-\infty, u]) - \omega((-\infty, u])|.
$ 
We sometimes slightly abuse notation and write the distance $d_{K}(U,V)$ instead of $d_{K}(\mu,\omega)$ for random variables $U$ and $V$ having distributions $\mu$ and $\omega$. 
In addition to \cref{assump:regularity}, the assumptions required are essentially quantitative versions of \cref{assump:asymp_norm_stat,assump:bvm_and_mle}. 
For example, we require high-probability convergence of the maximum likelihood estimator $\mle{}(\datarv) \defined \argmax_{\param \in \paramspace}\lik{\datarv}{\param}$
and asymptotic normality of the posterior.  
For a set $S \subseteq \reals^{d}$,  
define the constrained total variation distance %
$
d_{TV,S}(\mu,\omega) \defined \sup_{B \subseteq \mathcal{B}(S)}  |\mu (B)- \omega(B)|,
$
where $\mathcal{B}(S)$ is the collection of all Borel subsets of $S$.
\begin{assumption}
	\label{assump:mle_and_bvm_rate} 
	There exists a constant $s \geq 2$ such that, 
	for any compact set $\mathcal{K}$ satisfying $\optparam \in  \Int{\mathcal{K}}$, 
	there exist constants $c_{\mathcal{K}}, c'_{\mathcal{K}} > 0$ such that 
	\begin{subassump}[label = (\alph*),ref=(\theassumption-\alph*)]
		\item  \label{assump:a:mle_convto_true_rate}$\Pr\left[\Vert \hat{\param}(\mathbf{X}_{\numobs}) - \optparam \Vert > c_{\mathcal{K}}\numobs^{-1/2}\right] = \bigo(\numobs^{-s/2})$ and 
		\item\label{assump:b:BVMrate} $\Pr\left[d_{TV,\mathcal{K}}(\postdist{}(\cdot\mid\mathbf{X}_{\numobs}), \distNorm(\hat{\param}(\mathbf{X}_{\numobs}),\mlevar_{\optsym}/
		\numobs)) > c'_{\mathcal{K}}\numobs^{-1/2}\right] = \bigo(\numobs^{-s/2})$.
	\end{subassump}
\end{assumption}

\cref{assump:a:mle_convto_true_rate,assump:b:BVMrate} hold under the regularity conditions listed in \citet[Section 4]{Hipp:1976}. 
These conditions essentially impose some restrictions on the smoothness of the model density and the existence of lower-order moments of the log-likelihood, 
as characterized by the regularity parameter $s\ge 2$. 
We postpone the statement and discussion of the other conditions (\cref{assump:asymp_norm_stat_rate,assump:MLE_conv_rate}) to Supplementary Materials \ref{appx:proof-of-singleSPCthm}. 

\bthm \label{THM:SINGLESPC_RATE}
Suppose $\obsdist = \likdist{\optparam}$. 
If \cref{assump:regularity,assump:asymp_norm_stat_rate,assump:mle_and_bvm_rate,assump:MLE_conv_rate} hold, then there exists an absolute constant $C > 0$
such that 
\[
d_{K}\big(\spcpvalue{\datarv}, U\big)  < C \log^{\frac{s}{1+s}}\numobs/\numobs^{\frac{s}{2(1+s)}}. \label{eq:K-spc-bound}
\]
\ethm

The proof of \cref{THM:SINGLESPC_RATE} is in Supplementary Materials \ref{appx:proof-of-singleSPCthm}. 
The guarantee holds for any choice of split proportion $\spcprop$, although the constant $C$ may depend on $\spcprop$. 
Hence, in the well-specified case, the KS $\p$-value for these SPC $\p$-values (i.e., divided SPC $\p$-values) should be asymptotically uniform as long as $\dspcK \to \infty$ as $\numobs \to \infty$. 
On the other hand, if the model is misspecified, the single SPC $\p$-values of the subsets are not uniform asymptotically,
so the KS test (and hence the divided SPC test) will have asymptotic power 1.\footnote{It follows from the proof of \cref{THM:SINGLESPC_RATE} that we
	could also use a Wasserstein distance-based uniformity test. However, in preliminary experiments we found the KS approach to be superior.}

\subsection{Calibration and power of divided SPCs}

While the arguments in favor of the divided SPC are suggestive, we need to be careful with the choice of the number of divided splits $\dspcK$ to ensure it provided well-powered and asymptotically frequentist $p$-values. 
We first investigate the calibration properties of the divided SPC when the conclusion of \cref{THM:SINGLESPC_RATE} holds. 
Let $\distUnif$ denote the uniform distribution on $[0,1]$, $\pvaldist$ be the distribution of a $\dspcK$ single SPC $\p$-value 
using $\dspcnumobs$ observations. 
Let $\pvalempdist$ denote the corresponding empirical distribution of $k$ such $p$-values computed on independent datasets. 

\begin{assumption}
	\label{assump:singleSPC_rate}
	There exist constants $C>0$ and $\bar\gamma \in (0,1]$ such that for all $\gamma < \bar\gamma$, $d_K\big(\pvaldist, \distUnif\big) < C\dspcnumobs^{-\gamma/2}$.
\end{assumption}
Under the hypotheses of \cref{THM:SINGLESPC_RATE}, \cref{assump:singleSPC_rate} holds with $\bar\gamma = s/(s+1)$. 
For sufficiently regular models, any $s \ge 2$ may be chosen, in which case $\bar\gamma$ can be arbitrarily close to $1$. 
\bthm
\label{THM:DIVIDEDSPC}
If \cref{assump:singleSPC_rate} holds and $\dspcK = \lfloor b\numobs^{\beta} \rfloor$ for some $b>0$ and  $0<\beta<\frac{\bar\gamma}{\bar\gamma +1}$, then 
for any $t > 0$,
\[
\lim_{\numobs \to \infty}\left|\Pr\left[\dspcK^{1/2}d_K\big(\pvalempdist, \distUnif\big) > t\right] - \left(1 - \distKS{t}\right)\right| = 0,
\]
where $\distKS{\cdot}$ is the CDF of the Kolmogorov distribution.
Hence, $\lim_{\numobs \to \infty}d_{K}(\dspcpval{\spcobsrv}, \distUnif) = 0$.
\ethm
The proof of \Cref{THM:DIVIDEDSPC}
is in Supplementary Materials \ref{appx:dspcproofs1}, which
suggests that $\dspcK$ cannot grow too fast compared to $\numobs$ if we wish to control the test size.
For example, if $s$ can be chosen arbitrarily large, then we may take $\dspcK = O(\numobs^{\beta})$ for any $\beta \in (0, 1/2)$. 
On the other hand, if we only assume $s \ge 2$, then we must take $\beta \in (0, 2/5)$. 

Having established the correct calibration of divided SPC $p$-values, we next show that when the model is misspecified the asymptotic power of divided SPC is one.

\bthm \label{THM:DSPC_POWER}
If $\liminf_{\dspcK \rightarrow \infty}d_K(\pvaldist, \distUnif) > 0$, %
then for any $t > 0$, 
\[
\lim\limits_{\dspcK \rightarrow \infty}\Pr\left[\sqrt{\dspcK}d_K\big(\pvalempdist, \distUnif\big) > t\right] = 1.
\]
\ethm
The proof of \cref{THM:DSPC_POWER} is in Supplementary Materials \ref{appx:dspcproofs2}. 
We note that when the model is misspecified and $\rho \ne 1$, \cref{THM:DIVIDEDSPC} shows that the distribution of SPC $\p$-values is nonuniform, so the 
hypotheses of \cref{THM:DSPC_POWER} hold. 

\section{Using Split Predictive Checks in Practice}
\label{sec:SPCinPractice}

		We next provide some guidance on how to apply split predictive checks in practice, which we verify in our simulation studies in \cref{sec:simulation}.

		\subsection{Using single SPCs}
		As discussed earlier, most of the assumptions required by  \cref{THM:SINGLESPC} are quite mild. %
		The only assumption that the analyst has to be careful with is \cref{assump:asymp_norm_stat}, which requires the choice of test statistic to be asymptotically normal. 

		\paragraph*{Choosing $\spcprop$.}
		The split proportion $\spcprop$ controls the relative weight placed on two forms of misspecification quantified by $\rho$: the uncertainty of the statistic $\statistic{}$ and the optimal parameter $\optparam$. 
		Choosing $\spcprop$ large assigns more weight to the comparison between
		$\trueasympsd$ and $\asympsd(\optparam)$ while a small $\spcprop$ emphasizes the mismatch between frequentist sampling-based uncertainty $\Ehessloglik{\optsym}^{-1}\mlevar_{\optsym}\Ehessloglik{\optsym}^{-1} $ and standard posterior uncertainty $\Ehessloglik{\optsym}^{-1}$ in the direction $\diffasympmean{}(\optparam)$. A canonical choice for $\spcprop$ is 0.5, which treats the well-calibration of the two types uncertainty as equally important. 
		
		In a small-sample regime, extreme values for $\spcprop$ are not recommended. Small split proportions $\spcprop$ may result in an inadequate amount of observed data, while choosing $\spcprop$ large may result in poor estimation of the statistic. 
		Hence, we recommend choosing a moderate value for $\spcprop$ (e.g., 0.5 or 0.7) when the dataset size is fairly small relative to the model complexity.
		
		\paragraph*{Accounting for model structure.}
		There are many ways to splitting the data for structured models such as those for time-series or spatial data. 
		The analyst must then take into account what aspects of the data the test statistic captures in order to choose an effective splitting strategy. 
		We discuss two canonical cases: hierarchical and time-series models. 
		
\begin{figure}[tp]
			\centering
			\subfloat[single cross-SPC]{\label{fig:single-cross}\includegraphics[width=40mm]{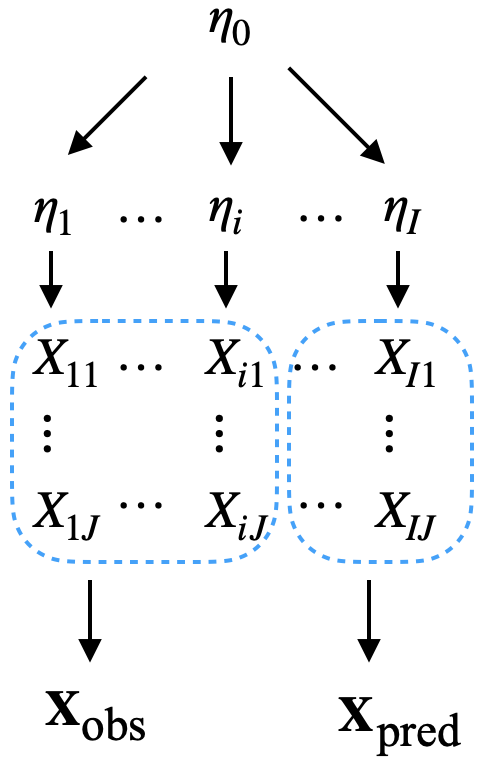}}
			\subfloat[single within-SPC]{\label{fig:single-within}\includegraphics[width=62.5mm]{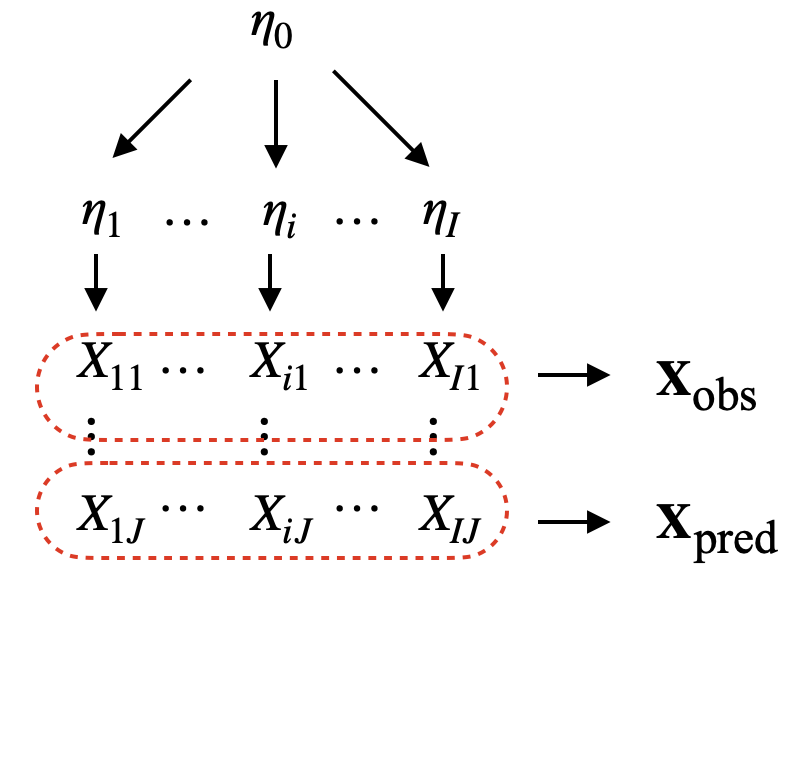}}
			\subfloat[within-divided cross-SPC]{\label{fig:within-divided-cross-SPC}\includegraphics[width=45mm, height=61.5mm]{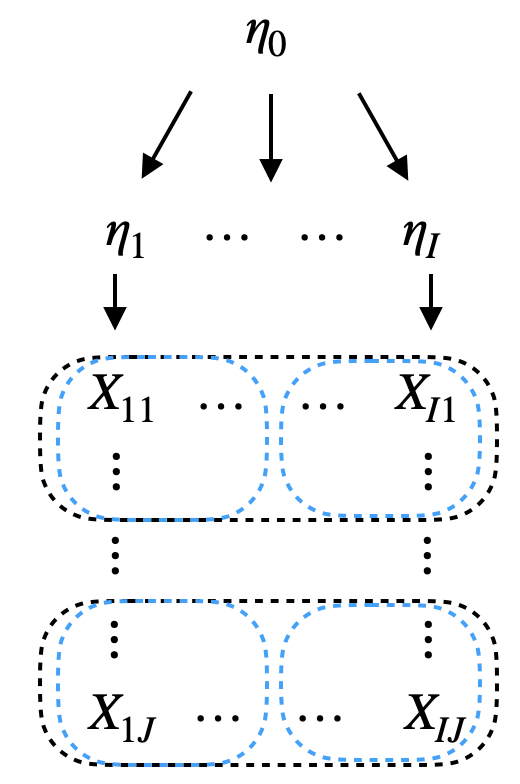}}\\
			\subfloat[within-divided within-SPC]{\label{fig:within-divided-within-SPC}\includegraphics[width=45mm, height=55mm]{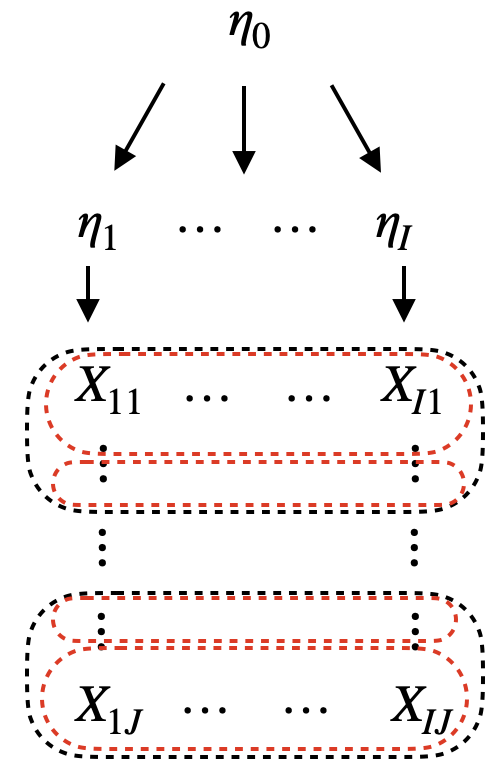}}
			\subfloat[cross-divided cross-SPC]{\label{fig:cross-divided-cross-SPC}	\includegraphics[width=55mm]{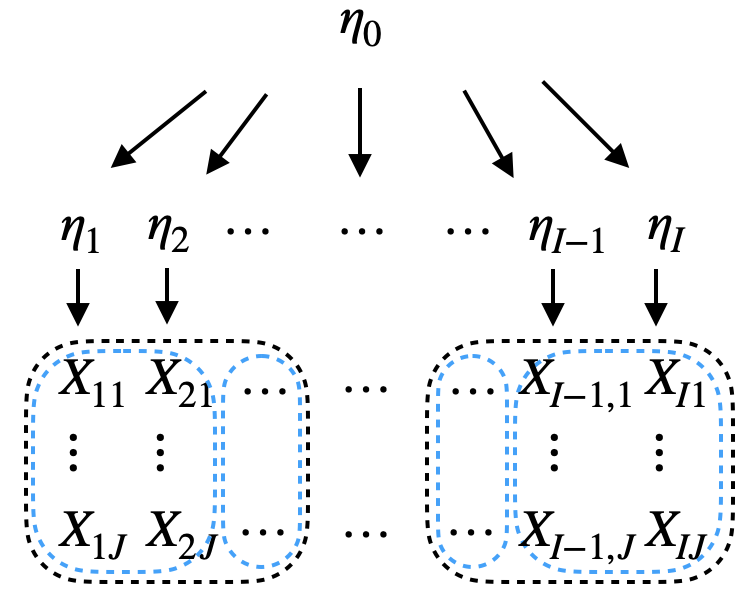}}
			\subfloat[cross-divided within-SPC]{\label{fig:cross-divided-within-SPC}	\includegraphics[width=55mm]{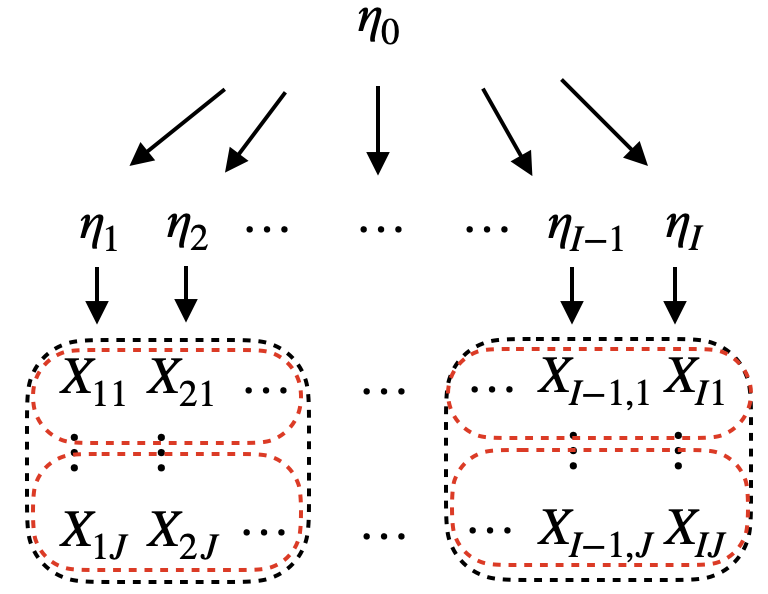}}
			\caption{A schematic diagram of splitting strategies for single SPCs and divided SPCs in a two-level hierarchical model. Blue dotted circles refer to a single cross-SPC test for the subset. Red dotted lines indicate a single within-SPC $\p$-value for each subset.}
			\label{fig:hier_diagram_SPCsplits}
		\end{figure}

		We first consider the case of a two-level hierarchical structure:
		\[
		\begin{split}
			\eta_{0} &\dist H \\
			\eta_{i} &\distiid G_{\eta_{0}} \quad (i = 1, \ldots, I), \\
			\obsrv{ij} &\distind F_{\eta_{i}} \quad (i = 1,\dots, I; j = 1,\dots,J_{i}).  \label{eq:two-level-hierarchical}
		\end{split}
		\] 
		When applying the single SPC to such a model, we have two ways to split the data: either across or within the groups, as shown in \cref{fig:single-cross,fig:single-within}. 
		We call the former \emph{single cross-SPC} and the latter \emph{single within-SPC}.  
		Which split is preferred depends on which level of misspecification is of interest. 
		For instance, if the analyst is interested in misspecification at the observation-level (that is, in the $F_{\eta_{i}}$) and has chosen an appropriate test statistic, then the single within-SPC 
		would be appropriate since it will compare data  from each group with the predictive distribution for that group. 
		On the other hand, if the analyst is interested in  group-level misspecification (that is, in $G_{\eta_{0}}$), then the single cross-SPC will compare data across different groups. 
		Similar considerations could be applied to more complex hierarchical structures. 

		For a time-series model, there are two splitting strategies to do the single split.
		In the \emph{extrapolated single SPC}, take the first $\lfloor \spcprop N \rfloor$ data points as $\spcobsrv$ and hold out the rest.
		In the \emph{interpolated single SPC}, for every $m \ll \numobs$ observations, take the first $\lfloor\spcprop m \rfloor$  as observed data and the reminder  as held out. 
		The extrapolated single SPC can be views as the ``extreme'' interpolated single SPC with $m = \numobs$. 
		But assuming $m \ll \numobs$, in the case of the interpolated SPC, both $\spcobsrv$ and $\spcnewrv$ contain information along the whole time-series.
		Which choice is appropriate will depend on how the analyst is using the model. 
		For short-term prediction or analysis of a fixed dataset, interpolation may be appropriate, while for medium-to-long-term prediction extrapolation might be 
		a better choice. 

		\subsection{Using divided SPCs}
		In addition to $\spcprop$, divided SPCs require choosing the number of splits $\dspcK$, while structured models also require additional considerations. 
		We address each in turn.

		\paragraph*{Choosing $\spcprop$ and $\dspcK$.}
		The choice of $\spcprop$ requires the same considerations as in the single SPC case. 
		However, now we must judge the dataset size by $\dspcnumobs$, the sample size used to compute the $\dspcK$ single $q$-SPC $p$-values. 
		The performance of divided SPCs could be significantly affected by the number of splits $\dspcK$. 
		In general, a larger $\dspcK$ will improve the power of the KS test but lead to $\dspcnumobs$ being smaller, which may result in single $q$-SPC $p$-values being closer to uniform. 
		We suggest using divided SPCs when $\numobs$ is very large, so choosing $\dspcK = \lfloor \numobs^{\beta} \rfloor$ with $\beta$ as large as possible is appropriate. 
		Hence, based on \cref{THM:DIVIDEDSPC} and the discussion that follows, for sufficiently regular models we suggest taking $\beta$ just below $1/2$ (e.g., $0.49$). 
		For less regular models, a safe choice is $\beta$ just below $2/5$ (e.g., $0.39$). 

		\paragraph*{Accounting for model structure.}
		Divided SPCs have two splitting steps: first divide the original data into $\dspcK$ subsets, then apply single splits to each of the subset.
		For the hierarchical model \cref{eq:two-level-hierarchical}, each split step can be across- or within-groups. 
		Therefore, we have four possibilities, which we illustrate in  \cref{fig:hier_diagram_SPCsplits}. 
		For $\mathsf{A}, \mathsf{B} \in \{ \text{cross}, \text{within} \}$, we refer to using the $\mathsf{A}$ split for the first step and the $\mathsf{B}$ split
		for the second step as the \emph{$\mathsf{A}$-divided $\mathsf{B}$-SPC}.
		For example, the \emph{cross-divided within-SPC} refers to doing cross-group splits to get subsets and single within-SPC for each subset.

		These four types of divided SPCs can effectively apply to different scenarios. Suppose a user aims to detect lower-level misspecification. From the analysis of single SPCs in hierarchical model, one should choose from cross-SPCs. Indeed, asymptotically cross-divided cross-SPCs and within-divided cross-SPC produce similar results. However, when the data size is small, we suggest choosing within-divided cross-SPC, which provides more lower-level information for each dataset compared to doing cross-group splitting twice. On the other hand, if the user is interested in finding mismatches in the group level, then with large dataset, both cross-divided within-SPCs and within-divided within-SPC should work well. We expect that for small data sets cross-divided within-SPC will usually be preferred. 
		
		The application of divided SPCs to time-series data works in a similar way as in hierarchical model. For each level of splitting in divided SPCs, we have two choices: extrapolation and interpolation. Interpolated splitting allows each subset to contain data across the whole time-series from the original dataset while extrapolated splits only include partial information for each subset. Depending on the purpose of the study, the user can choose different ways of splitting with SPCs.

		\section{Simulation Studies}
		\label{sec:simulation}

		To validate our theory from \cref{sec:singleSPC} and guidelines from \cref{sec:SPCinPractice}, 
		and investigate the finite-sample effects of SPC tuning parameters (the split proportion $\spcprop$ and, for the divided SPC, the number of folds $\dspcK$),
		we carry out two simulation studies, with a conjugate Poisson model and a Gaussian hierarchical model.  
		We compare SPCs to other general-purpose predictive checks: the widely used posterior predictive check (PPC) 
		and the population predictive check (POP-PC) since, like SPCs, it avoids double use of the data.
		
		\subsection{Poisson model} 
		\label{sec:Poisson model}
		
		We consider the conjugate Poisson model $\likdist{\param} = \distPoiss(\param)$, a Poisson distribution with rate parameter $\param$, and prior $\priordist = \distGam(0.1,0.2)$,
		a gamma distribution parameterized by the shape and rate.
		We investigate the behavior of the predictive checks for two data-generating distributions, the well-specified model $\obsdist = \distPoiss(2)$ 
		and the misspecified model $\obsdist = \distNegBin(2, 0.01)$, a negative binomial distribution parameterized by the mean and dispersion.
		We implement all methods using four statistics: (i) empirical mean $\statistic{\numobs}(\data) = \frac{1}{\numobs}\sum_{i = 1}^{\numobs}\obs{i}$, (ii) 2nd moment $\statistic{\numobs}(\data) = \frac{1}{\numobs}\sum_{i = 1}^{\numobs}\obs{i}^2$, (iii) 3rd moment $\statistic{\numobs}(\data) = \frac{1}{\numobs}\sum_{i = 1}^{\numobs}\obs{i}^3$ and (iv) mean squared error (MSE) $\statistic{\numobs}(\data, \param) = \frac{1}{\numobs}\sum_{i = 1}^{\numobs}(\obs{i} - \EE[\datarv_i\mid \param])^2$. 
		Similar results are also obtained in the case of the Gaussian location model, which we present in Supplementary Materials \ref{appx:Gaussian location model}.
		\begin{figure}[tp]
			\centering
			\subfloat[2nd moment]{\label{fig:pois_props:secmo_pw}\includegraphics[width=70mm]{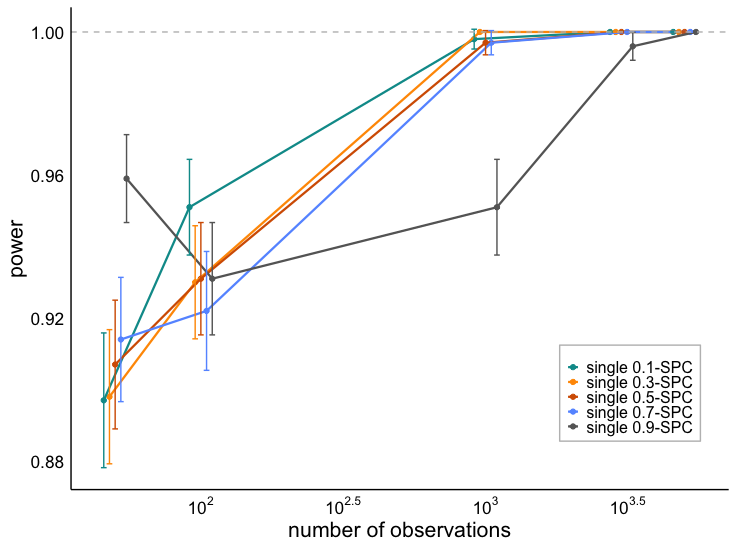}}
			\subfloat[ mean]{\label{fig:pois_props:mean_pw}\includegraphics[width=70mm]{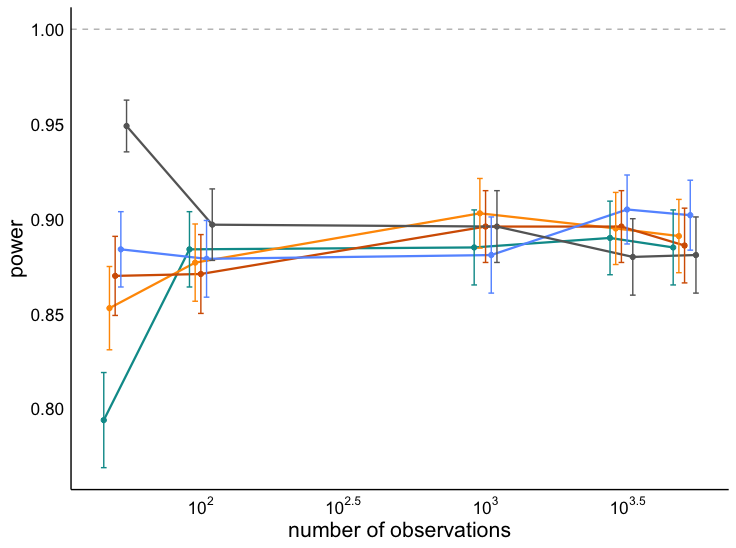}}
			\caption{Power plots for single SPC with different split proportions $\spcprop \in  \{0.1, 0.3, 0.5, 0.7, 0.9\}$ under conjugate Poisson model. The vertical bars represent the $95\%$ confidence intervals for the power estimates. 
				The dashed horizontal line in the test size plots indicates the significance level $\alpha = 0.05$ and the dashed line in the power plots is at 1. }
			\label{figur:pois_props}
		\end{figure}
		
		\paragraph*{Single SPCs.}
		\Cref{figur:pois_props} confirms that, as shown in \cref{THM:SINGLESPC}, 
		the choice of $\spcprop$ controls the asymptotic behavior of single SPCs when $\trueasympmean = \asympmean{}(\optparam)$ and has no impact when $\trueasympmean \neq \asympmean{}(\optparam)$. 
		For the small sample size regime, a moderate split proportion such as $\spcprop = 0.5$ is preferable to an extreme one.
		As shown in \cref{fig:pois_props:secmo_pw}, single SPC with $\spcprop = 0.9$ produces the smallest power when $\numobs = 1000$.
		\Cref{fig:pois_props:mean_pw} empirically verifies the limitations of single SPCs documented in \cref{THM:SINGLESPC}: for all single SPCs using the mean statistic and where mild-to-moderate misspecification occurs, the power stays around 0.9 and has no tendency to increase as the sample size grows. We present results for 3rd moment and MSE statistics in Supplementary Materials \ref{appx:poisson-single-spc}, which correspond to moderate misspecification case and thus single SPC achieves power 1 asymptotically .

\begin{figure}[tp]
			\centering
			\subfloat{\label{fig:pois_dspcprops:mean_pw}\includegraphics[width=67mm]{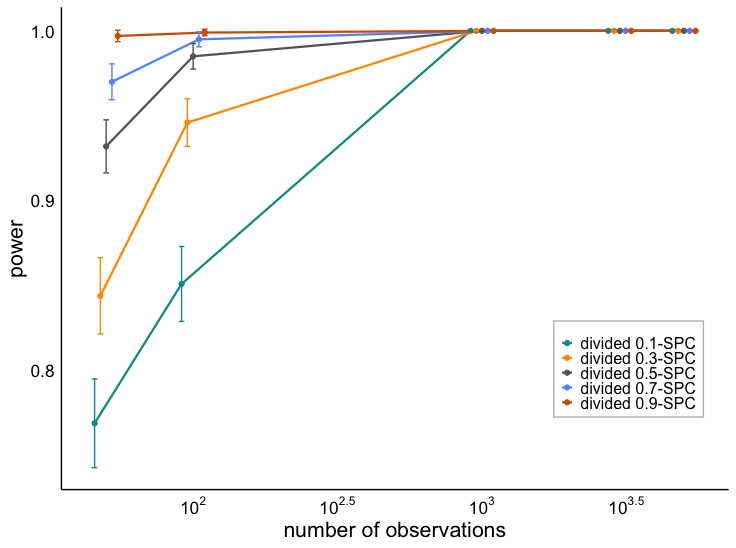}}
			\subfloat{\label{fig:pois_dspcKs:pw}\includegraphics[width=72mm]{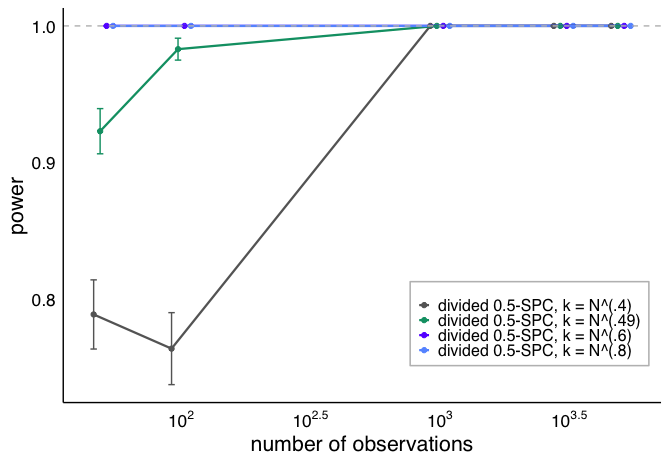}}
			
			\caption{\textbf{Left:} Power for divided SPC with mean statistic, $\dspcK = \numobs^{0.49}$, and varying $\spcprop \in  \{0.1, 0.3, 0.5, 0.7, 0.9\}$, under the conjugate Poisson model.
				\textbf{Right:}Power for divided SPC with 3rd moment statistic, $\spcprop= 0.5$, and varying the number of folds $\dspcK \in  \{\numobs^{0.4},\numobs^{0.49}, \numobs^{0.6}, \numobs^{0.8}\}$. See \cref{figur:pois_props} for further explanation and experimental details.} %
		\label{figur:pois_dspcprops}
	\end{figure}

\begin{figure}[tp]
		\centering
		\subfloat[]{\label{fig:pois_dspcKs:ts}\includegraphics[width=70mm]{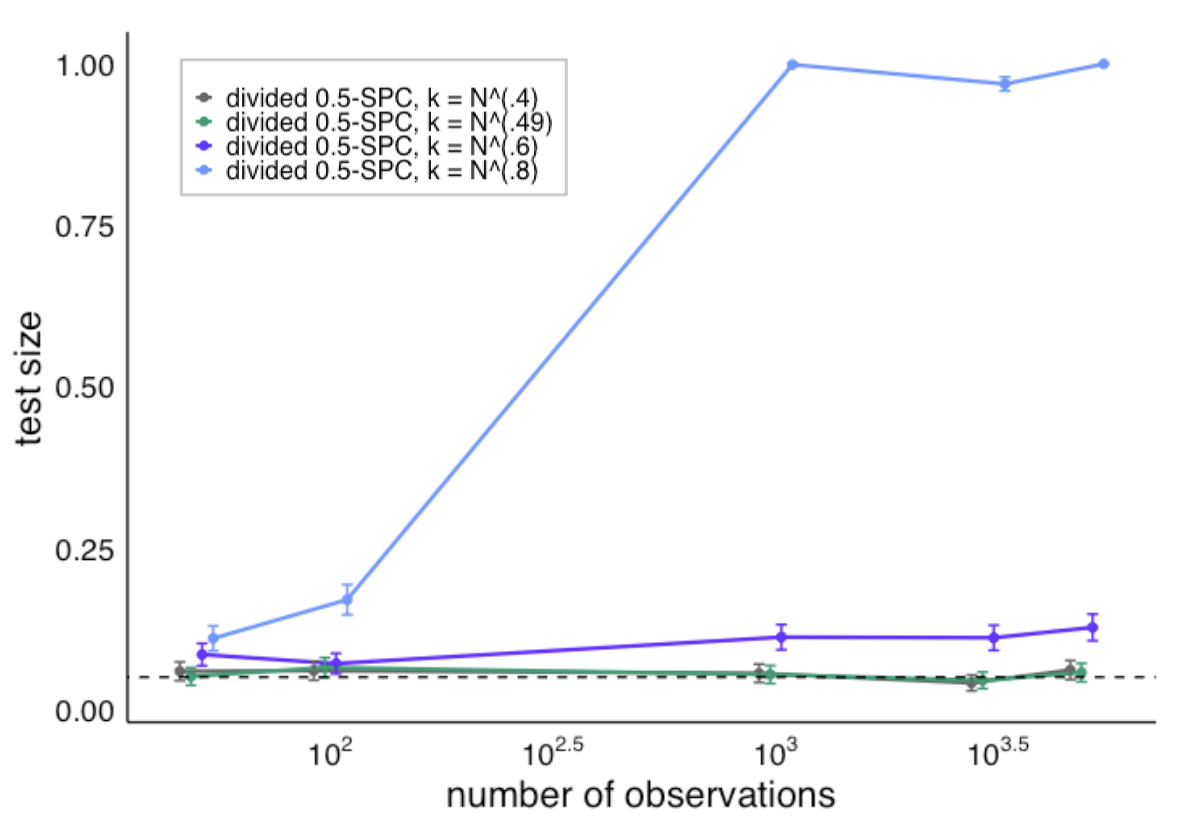}}
		\subfloat[]{\label{fig:poisson_ess}\includegraphics[width=70mm]{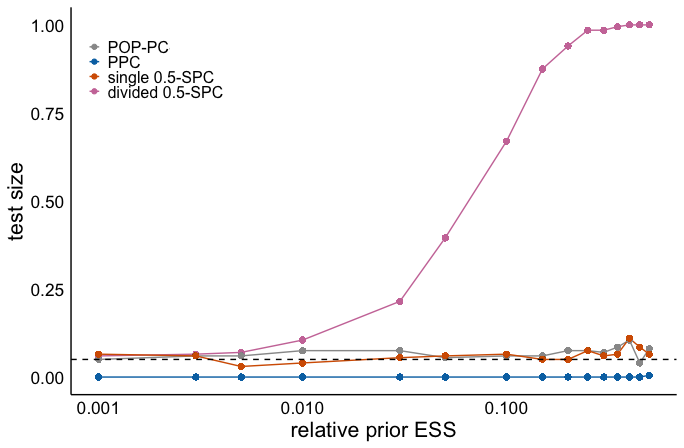}}
		
		\caption{\textbf{(a)} Test size for divided SPC with mean statistic, $\dspcK = \numobs^{0.49}$, and varying $\spcprop \in  \{0.1, 0.3, 0.5, 0.7, 0.9\}$, under the conjugate Poisson model. \textbf{(b)} Relative prior ESS versus test size with mean statistic for POP-PC, PPC, single $0.5$-SPC and divided $0.5$-SPC with $\dspcK = \numobs^{0.49}$ for under the conjugate Poisson model. 
			We set $\param_{\star} = 25$ for the data-generating distribution and adjust the prior parameters so that $\param_{\star}$ is at the 95th quantile of the prior. 
		}

	\end{figure}

		\paragraph*{Divided SPCs.}
		For divided SPCs, first we fix $\dspcK  = \numobs^{0.49}$ and vary $\spcprop$. 
		\cref{fig:pois_dspcprops:mean_pw} shows that the divided SPC has asymptotic power 1 and since $\rho = \trueasympsd/\asympsd(\optparam)$, the split proportion makes no difference in the power performance of divided SPC given a large sample size. However, when dealing with small-to-moderate sample sizes, one needs to be careful with the choice of split proportion for divided SPC. 
		\cref{fig:pois_dspcKs:ts} shows that with $\beta > 1/2$, the test size is not well controlled, which is in agreement with conclusion of \cref{THM:DIVIDEDSPC} that only guarantees the divided SPC is asymptotically well-calibrated when when $\beta \in (0, 1/2)$. 
		Similar results for 2nd and 3rd moments are presented in Supplementary Materials \ref{appx:poisson-divided-spc}.

	\paragraph*{Effect of number of folds $\dspcK$.}
	To investigate the effect of different scalings of $\dspcK$, we fix the split proportion $\spcprop = 0.5$ and consider $\dspcK = \numobs^{\beta}$ for different choices of $\beta$. 
	\cref{fig:pois_dspcKs:ts} shows that with $\beta > 1/2$, the test size is not well controlled, which is in agreement with conclusion of \cref{THM:DIVIDEDSPC} that only guarantees the divided SPC is asymptotically well-calibrated when when $\beta \in (0, 1/2)$.

	On the other hand, 
	choosing small $\beta$ results in an insufficient number of folds and thus degrades the power of the KS test. 
	We find that $\beta \approx 1/2$ provides the best power while $\beta \approx 0.4$ has smaller power. Results for other statistics are presented in \cref{figur:pois_dspcKs_appx} in Supplementary Materials.

\begin{figure}[tp]
		\centering
		\subfloat[ mean]{\label{fig:pois_compare_mean_mismatch}\includegraphics[width=70mm]{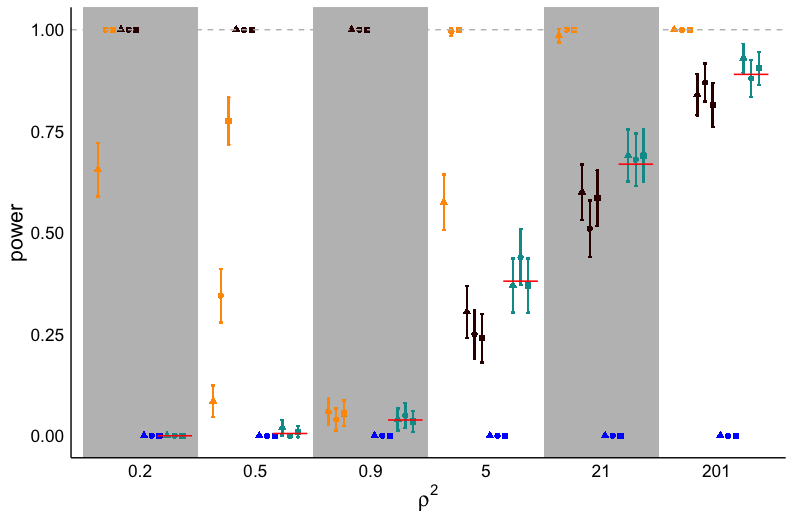}}
		\subfloat[ $2$nd moment]{\label{fig:pois_compare_secmo_mismatch}\includegraphics[width=70mm]{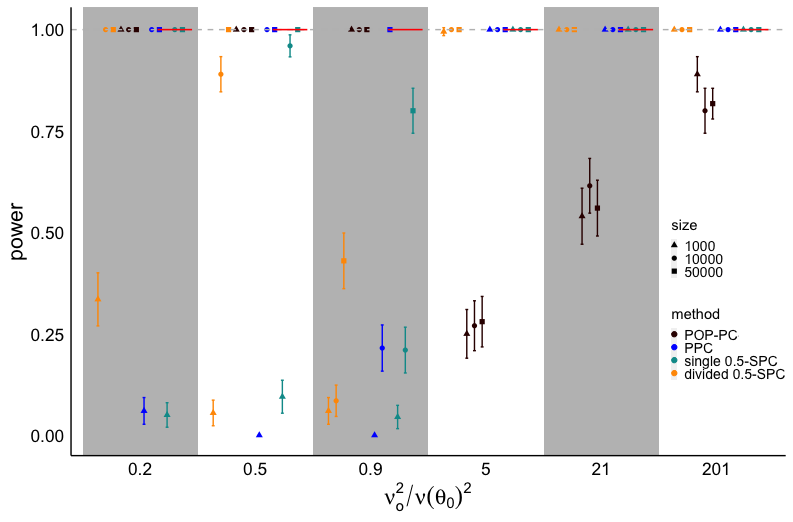}}
		\caption{The power plots of candidate checks given different levels of misspecification with conjugate Poisson model. 
			The red solid lines indicate theoretical asymptotic power of single SPC computed from \cref{THM:SINGLESPC}. All SPCs choose proportion as  $\spcprop = 0.5$ and for divided SPCs, the number of splits is set as $\dspcK = \numobs^{0.49}$.}
		\label{figur:Poisson_mismatch_comparison}
	\end{figure}

	\paragraph*{Effect of the degree of misspecification.}
	To investigate how the degree of misspecification affects each predictive check, we use the data-generating distributions 
	$\obsdist = \distNegBin(2,\tau)$ and $\obsdist = \distBinom(30,p)$ to create over- and under-dispersed data
	and compare the asymptotic power of each method given different types of misspecification. 
	Using a mean statistic in this case yields $\trueasympmean = \asympmean{}(\optparam)$, so the degree of misspecification is effectively characterized by $\rho = {\trueasympsd}/{\sigma(\optparam)}$
	from  \cref{THM:SINGLESPC}. 
	For the negative binomial model $\rho^2 = 1 + 2/\tau$  %
	and for the binomial model $\rho^{2} = 1 - p$. 
	Hence, varying $\tau \in \{0.01, 0.1, 0.5\}$ for the negative binomial model and $p \in \{0.1, 0.5, 0.8\}$ for the binomial model leads to the subtle-to-moderate misspecification scenarios $\rho^2 \in \{0.2, 0.5, 0.9, 5, 21, 201\}$.
	\cref{fig:pois_compare_mean_mismatch} shows that, as expected, the power of single SPC stabilizes near zero for $\rho^2 < 1$ but approaches one when $\rho^2 \gg 1$. 
	The power of divided SPC, on the other hand, reaches one for all cases except the nearly well-specified one of $\rho^2 = 0.9$.
	The power of PPC is zero in all cases.
	The power of POP-PC is inferior to single SPC when $\rho^{2} > 1$ but close to one when $\rho^{2} < 1$. 
	Switching to the 2nd moment statistic, $\trueasympmean \ne \asympmean{}(\optparam)$ so there is major misspecification. 
	As expected, \cref{fig:pois_compare_secmo_mismatch} shows that PPC, single and divided SPC all succeed to capture such major misspecification with power 1 given large datasets, 
	while POP-PC has power depending on how different $\asympmean{}(\optparam)$ and $\trueasympmean$ are when model is over-dispersed.
	On the whole, these results suggest that, when used together, single and divided SPC are superior to using either or both PPC and POP-PC. 
	
	\paragraph*{Effect of the prior.}
	Finally, we investigate the effect of the prior on the test size of the predictive checks.
	Conjugate exponential families lend themselves to the definition of a \emph{prior effective sample size (ESS)} (though see \citet{Reimherr.2021} for an extension to other models). 
	For a Poisson model with a $\distGam(\alpha, \beta)$ prior, the prior ESS is $\ess = \beta$.
	We measure the prior impact in terms of the \emph{relative prior ESS} is defined as $\unadjess \defined \frac{\ess}{\ess + \numobs_{*}}$,
	where $\numobs_{*}$ is the number of samples to construct a single posterior distribution.
	For PPC and POP-PC, $\numobs_{*} = \numobs$ while for the SPCs $\numobs_{*} = \spcNobs$. 
	We set $\numobs_{*} = 50$ and $\trueparam = 25$ for all methods, then 
	vary $\beta$ so that $\unadjess$ varies between $0.001$ and $0.5$ using the formula $\beta = \frac{\unadjess}{1-\unadjess}\numobs_{*}$.
	To mimic a realistic scenario, we chose $\alpha$ such that, for a given $\beta$, the true parameter $\trueparam$ lies at the $95$th quantile of the prior distribution. 
	\Cref{fig:poisson_ess} shows that for the mean statistics, single SPC and POP-PC fail to control the test size when $\adjess \ge 0.1$ while 
	divided SPC fails when $\adjess \ge 0.01$. %
	Results for 2nd moment when $\trueparam = 25$ and $\trueparam = 100$ are similar, which we present in \Cref{fig:poisson_ess_appx} in Supplementary Materials. 
	Hence, we do not recommend using single SPCs when $\adjess \ge 0.1$ or divided SPCs when $\adjess \ge 0.01$.

	\subsection{Gaussian hierarchical model} 
	\label{sec:simulation_hier}
	
	Next, we compare SPCs and the other predictive checks using the two-level Gaussian hierarchical model from \citet{Bayarri:2007},
	which corresponds to the model in \cref{eq:two-level-hierarchical} with $F_{\eta_{i}} = \mathcal{N}(\eta_{i}, 4)$, $G_{\eta_{0}} = \mathcal{N}(\mu_0, \sigma_0^2)$,
	and an improper prior $H$ on $\eta_{0} = (\mu_{0}, \sigma^{2}_{0})$ with density $h(\mu_{0}, \sigma^{2}_{0}) = \sigma_{0}^{-2}$.
	To illustrate the use of SPCs with different types of misspecification, we simulate data from four data-generating distributions in \cref{table:simulation-scenarios}. 

	\begin{table}[bt]
		\centering
		\caption{Four scenarios with 4 types of misspecification in a hierarchical setup.}
		\renewcommand{\arraystretch}{1}
		\begin{tabular}{ccc}
			\toprule
			\textbf{Scenario Name} & \textbf{$X_{ij}$ Distribution} & \textbf{$\eta_i$ Distribution} \\
			\midrule
			Well-specified   &  $\obsrv{ij} \mid \eta_{i} \dist \distNorm(\eta_{i}, 4)$  &$\eta_{i} \dist \distNorm (0, 1)$    \\
			Misspecified across groups  &  $\obsrv{ij}\mid \eta_{i}  \dist \distNorm(\eta_{i}, 4)$  &$\eta_{i} \dist \distGam(0.6, 0.2)$  \\
			Misspecified within groups  &  $\obsrv{ij}\mid \eta_{i} \dist \distNorm(\eta_{i}, 8)$  &$\eta_{i} \dist \distNorm (0, 1)$ \\
			Misspecified across and within groups   &  $\ln{\obsrv{ij}}\mid \eta_{i} \dist \distNorm(\eta_{i}, 4)$  &$\eta_{i} \dist \distNorm (0, 1)$ \\
			\bottomrule
		\end{tabular}
		\label{table:simulation-scenarios}
	\end{table}
	
	\paragraph*{Statistics for hierarchical model.}
	
\begin{figure}[tp]
		\centering
		\includegraphics[height=16mm]{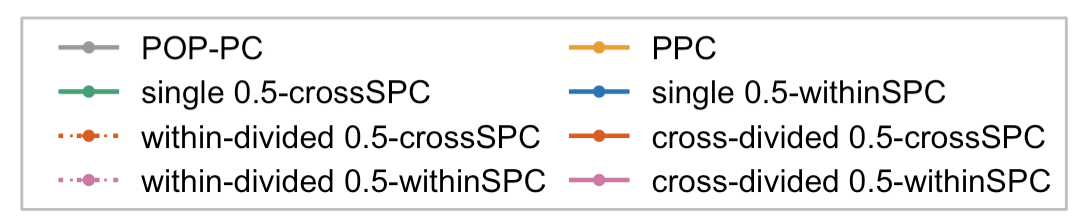}
		\subfloat[grand mean]{\label{fig:hier_ts_grmean}\includegraphics[width=60mm]{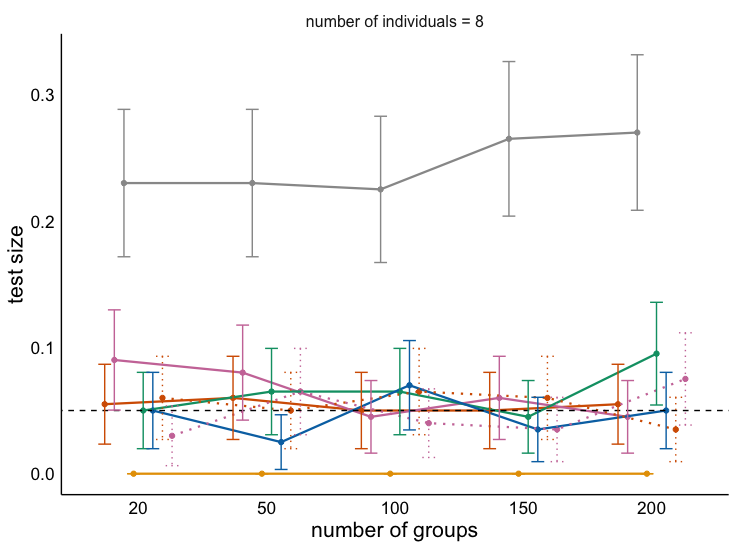}} 
		\subfloat[mean of 75th quantiles of groups]{\label{fig:hier_ts_meangrq75}\includegraphics[width=60mm]{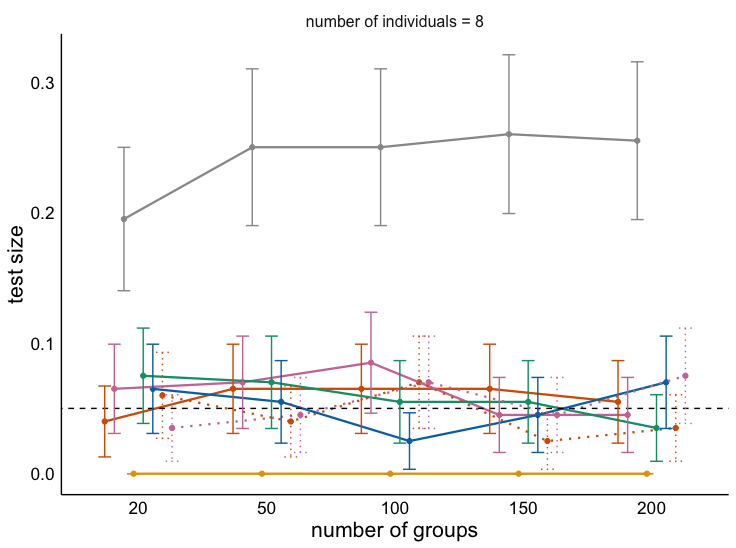}}
		\subfloat[75th quantile of group means]{\label{fig:hier_ts_q75means}\includegraphics[width=60mm]{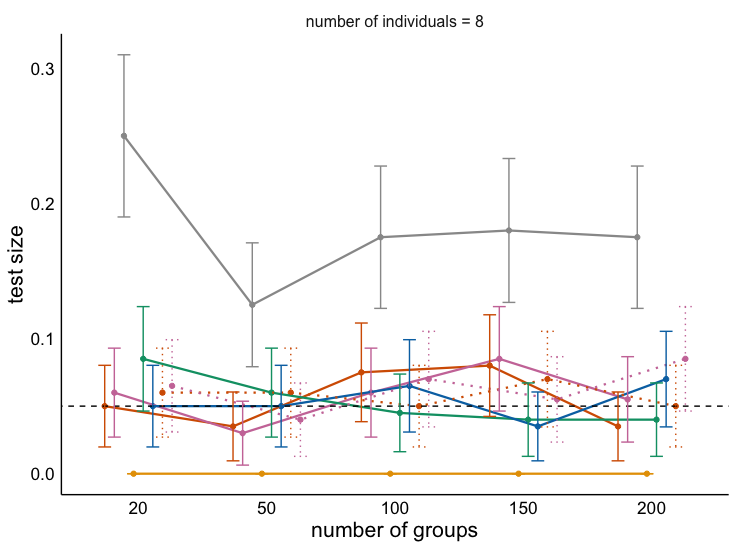}}
		\\
		\subfloat[Misspecified across groups]{\label{fig:hier_pw_J8_cr}\includegraphics[width=60mm]{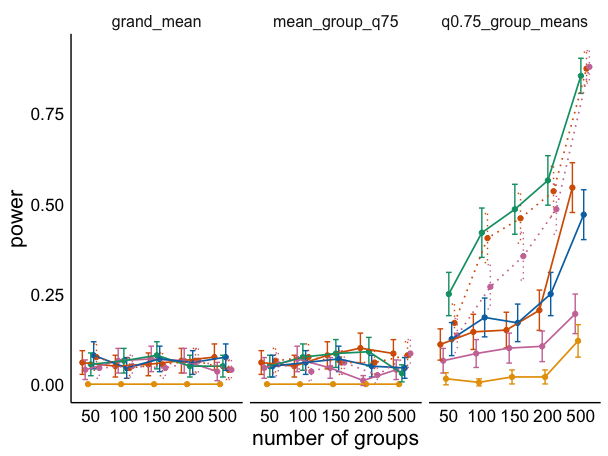}}
		\subfloat[Misspecified within groups]{\label{fig:hier_pw_J8_grvar}\includegraphics[width=60mm]{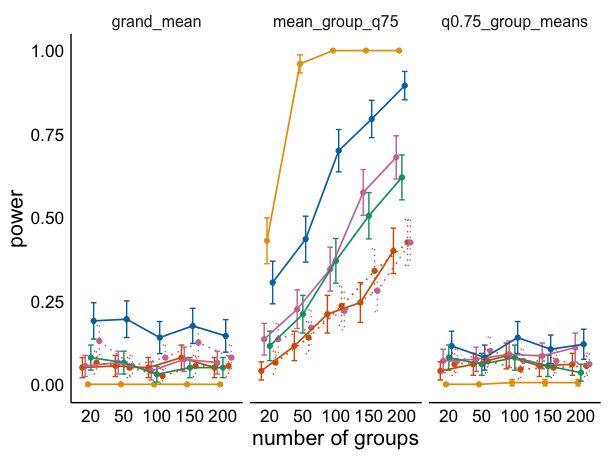}}
		\subfloat[Misspecified across and within groups]{\label{fig:hier_pw_J8_wilognorm}\includegraphics[width=60mm]{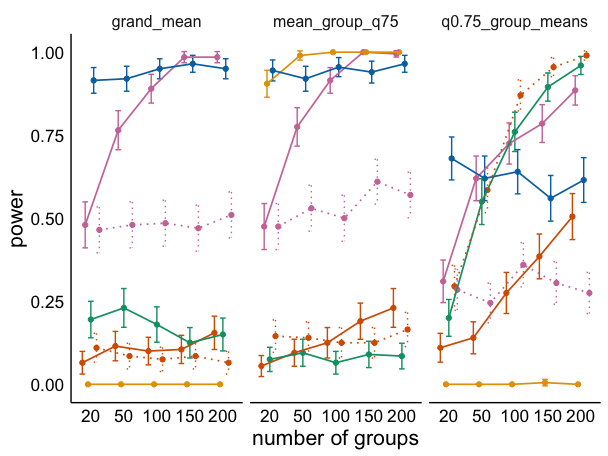}}
		
		\caption{ Test size \textbf{(a,b,c)} and power \textbf{(d,e,f)} plots for different checks and statistics given fixed number of individuals $ J= 8$ and increasing number of groups $I \in \{20, 50, 100, 150,200\}$ under Gaussian hierarchical model. Vertical bars represnet the $95$th confidence intervals for the estimation. The horizontal dashed line indicates significance level $\alpha = 0.05$. Dotted lines are divided SPC using within splitting strategy to get $\dspcK$ folds as discussed in \cref{sec:SPCinPractice}. }
		\label{fig:hier_J8}
	\end{figure}

	Let $Q(\data)$ denote the $75$th quantile of data $\data$ and $\bar{\obs{}}_i$ be the mean of the $i$th group of observations. 
	We employ three statistics in our simulations: 
	(i) the grand mean $\statistic{IJ}(\data) = \frac{1}{IJ}\sum_{i = 1}^{I}\sum_{j = 1}^{J}\obs{ij}$, 
	(ii) the mean of the $75$th quantiles of each group $\statistic{IJ}(\data) = \frac{1}{I}\sum_{i = 1}^I Q(\obs{i\cdot})$, and
	(iii) the $75$th quantile of the group means $\statistic{IJ}(\data) = Q(\bar{\obs{}}_1, \ldots,\bar{\obs{}}_I )$.
	For each candidate check, we estimate the test size by repeatedly generating data from well-specified model for $200$ times and run each candidate check in each iteration to get their corresponding $\p$-values. Powers are estimated with 200 repeated experiments for each scenario. 
	\Cref{fig:hier_J8} shows results for a fixed number of observations $J = 8$ per group and an increasing number of groups $I$. 
	Results and discussions for fixed $I = 20$ with increasing $J$ are presented in Supplementary Materials \ref{appx:gauss_hier_figures}.
	
	\paragraph*{Test size.}
	\Cref{fig:hier_ts_grmean,fig:hier_ts_meangrq75,fig:hier_ts_q75means} show that in the well-specified Scenario 1 for all statistics
	both single and divided SPCs correctly control test size and produce uniform $\p$-values (see also \cref{fig:hier_qq_grmean_dspcs} in Supplementary Materials). 
	PPCs have the smallest test size but the $\p$-values are not uniformly distributed (see also \cref{fig:hier_qq_grmean_ppc} in Supplementary Materials) and so are not well-calibrated.
	For all scenarios POP-PC  fails to control the test size. Hence, we omit POP-PC  in the remainder of the discussion.
	
	\paragraph*{Power.}
	As discussed in \cref{sec:SPCinPractice}, a hierarchical structure may produce different levels of misspecification for which only certain test statistics and checking methods will be effective.

	\texttt{Misspecified across groups} scenario is most effectively detected by the $75$th quantile of group means statistic.
	\Cref{fig:hier_pw_J8_cr} shows that, both single and divided SPCs dramatically outperform PPCs.
	Also, as suggested in \cref{sec:SPCinPractice}, cross-SPCs are best for testing for group-level misspecification
	and cross-divided cross-SPCs have smaller power than within-divided cross-SPCs because doing cross-splits twice in divided SPCs results in poor posterior estimation in a small data regime. 
	single cross-SPCs outperform single within-SPCs. 

	\texttt{Misspecified within groups} scenario introduces observation-level mismatch between the model and data. 
	In this case the mean of group $75$th quantiles statistic is most effective at detecting the misspecification. 
	\Cref{fig:hier_pw_J8_grvar} shows that PPCs outperform the SPCs, although, as suggested in \cref{sec:SPCinPractice}, the within-SPC methods are also quite effective, 
	with single within-SPC being the best SPC. 

	\texttt{Misspecified across and within groups} scenario introduces mismatches at both levels and all three statistics are capable of detecting the mismatches. 
	\Cref{fig:hier_pw_J8_wilognorm} shows that PPCs are only effective when used with the mean of group $75$th quantiles statistic while SPCs are effective 
	for all three statistics. 
	With the mean of group 75th quantiles statistic, which is used to detect the lower-level misspecification, and the grand mean, only the within-SPCs have good power. 
	Further, the cross-divided within-SPC outperforms the within-divided within-SPC due to the small data effects. Similar analysis applies to grand mean statistic. 
	For 75th quantile of group means, which captures the group-level misspecification, cross-SPCs outperform within-SPCs.

	\section{Experiments}
	\label{sec:experiments}

	In this section, we compare SPCs to PPCs on four real-data examples.
	We do not include POP-PCs in our comparison due to their lack of test size control in some of the simulation results.

	\subsection{Airline delays data: geometric model}
	
	We investigate the power of the single and divided SPCs compared to the PPC using the airline on-time data for all flights departing NYC in 2013\footnote{\url{https://www.transtats.bts.gov/DL_SelectFields.asp?Table_ID=236}. After filtering out missing data, we get a cleaned airline data with $\numobs = 327,346$.}.
	Suppose we want to model $y = \max(\texttt{arr\_delay} - 15, 0)$ the arrival delays (in minutes) over 15 minutes for all delayed flights. 
	We assume a geometric model
	$y \dist \distGeom(\theta)$ with conjugate prior $p \dist \distBeta(0.1, 0.2)$. 
	A comparison of observed data distribution and posterior predictive distribution is shown in Supplementary Materials \ref{appx:airlines_figs}, which shows that 
	model fails to capture the heavy right tail of the data distribution. Thus, we should expect statistics sensitive to right tail behavior to reject the assumed model.
	In particular, we consider the success rate statistic $\statistic{}(\datarv) \defined \numobs/\sum_{n=1}^{\numobs} X_n$ and the MSE statistic. 
	To estimate the power of all checks across different dataset sizes, we permute and segment the full dataset into $I \defined \lfloor \numobs/\numobs_{\text{sub}} \rfloor$
	disjoint subsets of equal size $\numobs_{\text{sub}} \in  \{ 500, 1000,2000, 3000, 5000\}$. 
	Each subset is regarded as an ``original observed data'' and all candidate checks are applied to produce one $\p$-value for each dataset, from which we can estimate the power at a fixed test size. 

\begin{figure}[tp]
		\centering
		\subfloat[Success rate]{\label{fig:airlines_pw_successrate}	\includegraphics[width=70mm]{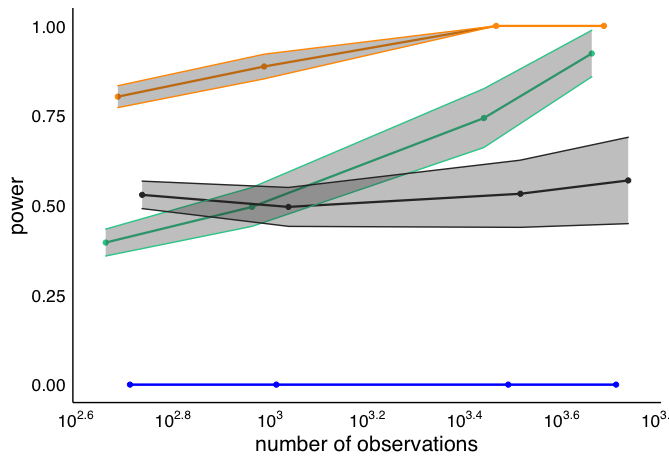}}
		\subfloat[MSE]{\label{fig:airlines_pw_mse}	\includegraphics[width=70mm]{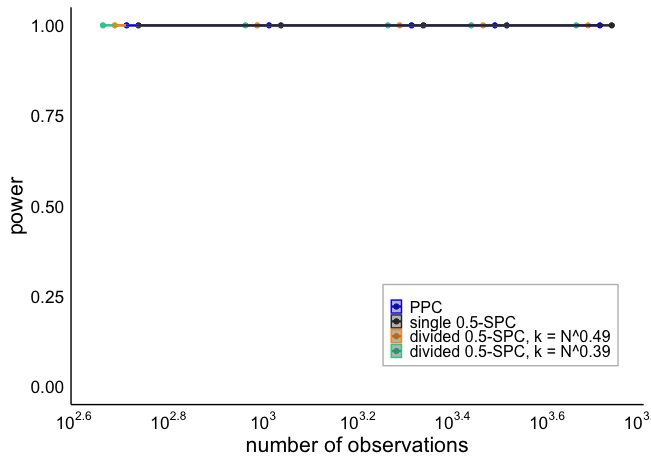}}
		\caption{The power estimates for PPC and SPCs with proportion $\spcprop = 0.5$ of NYC airline data using a geometric model. The power is estimated by segmenting the full dataset into subsets of equal sizes $\numobs_{\text{sub}} \in \{ 500, 1000,2000, 3000, 5000\}$ and computing $\p$-values for each subset. The shaded region represents $95\%$ confidence intervals for the power estimation.}
		\label{fig:airlines_nyc_data_power}
	\end{figure}

	\Cref{fig:airlines_nyc_data_power} shows that while all checks perform well for the MSE statistic, only the SPCs (with $\spcprop = 0.5$) have large power for the success rate statistic. 
	These results also establish the practical value of the divided SPC, which with $\dspcK = \numobs_{\text{sub}}^{0.49}$ rapidly reaches power of one 
	while PPC has power 0 and single SPC has approximately constant power of around 0.5 across sample sizes.
	Notably, divided SPC with a lower rate $\dspcK = \numobs_{\text{sub}}^{0.39}$ has increasing power but worse power than with $\dspcK = \numobs_{\text{sub}}^{0.49}$. %

	\subsection{Airline delays data: negative binomial regression}
	
	To illustrate the flexibility of SPCs for structured models compared to PPCs, we next consider a negative binomial regression model for the airline delays data with covariates 
	$\sin(\texttt{Month}/12)$, $\cos(\texttt{Month}/12)$, $\sin(\texttt{DayOfWeek}/7)$, $\cos(\texttt{DayOfWeek}/7)$, and $\texttt{Distance}$ (between airports). 
	As discussed in \cref{sec:SPCinPractice}, in a time-series model, there are two possible types of splitting, interpolated and extrapolated. 
	By plotting the observed data distribution and posterior predictive distribution, we can see that the negative binomial generalized linear model explains the data fairly well, 
	although the tail of the predictive distribution is heavier than the data. 
	Hence, it is not immediately clear which predictive checks pass. %
	From \cref{table: pvals_airline_data} we see that, in fact, interpolated single SPC and PPC $\p$-values are large, while the extrapolated single SPC $\p$-value indicates the model does not predict future data well. 
	On the other hand, the more sensitive divided SPC picks up on misspecification not just in the extrapolation setting but for interpolation as well. 
	These results suggest the success rate misspecification is fairly subtle. 
	Depending on the use case, these insights could guide whether further model elaboration is necessary. 
	\begin{table}[h]
		\centering
		\caption{Two-sided p-values of different split types in SPCs of $\spcprop = 0.5$ under negative binomial regression model with airline data and success rate statistic.}
		\renewcommand{\arraystretch}{1}
		\begin{tabular}{cc}
			\toprule
			\textbf{Method} &  \textbf{$p$-value} \\
			\midrule
			PPC   &  0.132  \\
			Interpolated single 0.5-SPC &  0.626	 \\
			Extrapolated single 0.5-SPC &  0.000	 \\
			Double-interpolated divided 0.5-SPC, $\dspcK = \numobs^{0.49}$&  0.000	 \\
			Interpolated divided extrapolated 0.5-SPC, $\dspcK = \numobs^{0.49}$&  0.001	 \\
			\bottomrule
		\end{tabular}
		\label{table: pvals_airline_data}
	\end{table}

\subsection{Birthday data}

\begin{figure}[tp]
	\centering
	\subfloat[Model 1]{\label{fig:birth_model1_lag}\includegraphics[width=85mm]{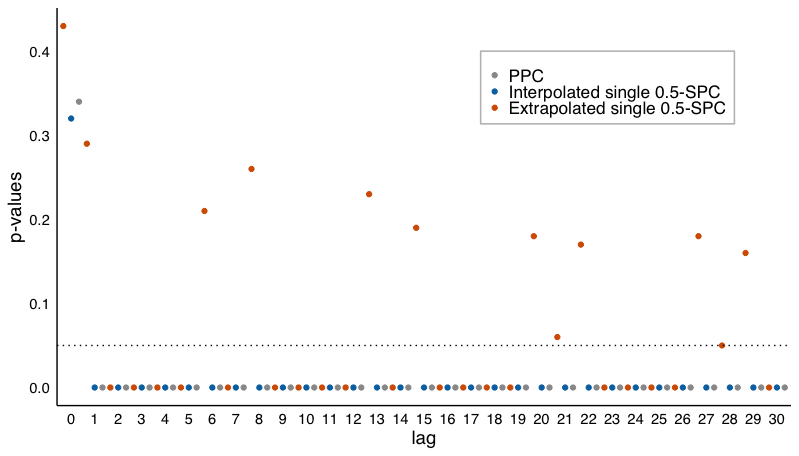}}
	\subfloat[Model 6]{\label{fig:birth_model6_lag}\includegraphics[width=85mm]{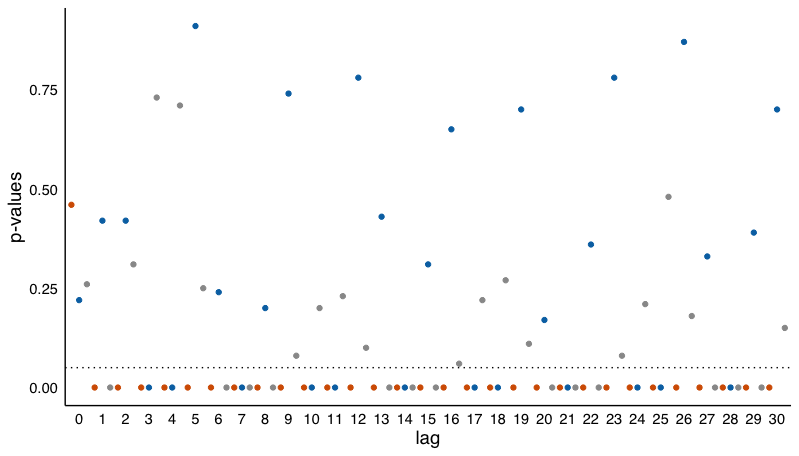}}
	\caption{Two-sided $p$-values for lag-0 to lag-30 auto-correlation coefficients under Models 1 and 6 for the birthday data. We compare interpolated and extrapolated single 0.5-SPCs to PPCs. The horizontal dotted line is the significance level of 0.05.}
	\label{fig:birth_model_lags}
\end{figure}

The birthday data consists of the number of births per day in the United Stats for the years 1969--1988.\footnote{Data source: National Vital Statistics System natality data, as
	provided by Google BigQuery and exported to cvs by Chris Mulligan (sum
	data http://chmullig.com/wp-content/uploads/2012/06/births.csv) and
	Robert Kern (whole time series
	http://www.mechanicalkern.com/static/birthdates-1968-1988.csv).}
We discuss two Gaussian process models from \citet[Chap.~21.2, pp.~507]{Gelman:2013}: Model 1 fits the long-term trend using an exponentiated quadratic kernel $k(x, x') = \sigma^2\exp(-\frac{\Vert x - x' \Vert^2}{2c^2})$; 
Model 6 uses a carefully designed kernel which contains seasonal, day-of-week, and day-of-year effects.  
To assess how well the model captures the correlation structure of the time-series data, we employ
lag 1 to 30 autocorrelation coefficients as statistics. 

As shown in \cref{fig:birth_model1_lag}, Model 1 fails all three checks. The extrapolated single SPC fails to reject for lags equal to $7i \pm 1$ for $i \in \{0, 1, 2, 3, 4\}$, which suggests
a potential improvement to Model 1 by adding day-of-week effects. 
With the improvement in Model 6, we can see in \cref{fig:birth_model6_lag} that the extrapolated single SPC rejects
for all lags greater than zero, while for many lags PPC and interpolated single SPC fail to reject. Large interpolated single SPC $\p$-values suggest that Model 6 can understand the autocorrelation on the number of births per day well, while extrapolated single SPC $\p$-values reveal a poor performance of Model 6 on predicting the autocorrelation effects for data. 

\section{Conclusion}

In this paper, we have proposed two types of split predictive checks (SPCs), single SPCs and divided SPCs, as asymptotically well-calibrated and well-powered alternatives to posterior predictive checks (PPCs).
Compared to to PPCs, our two types of SPCs offer complimentary strengths in terms of power, computational efficiency, and applicability to small and large datasets.  

\paragraph*{Power and degree of misspecification.}
To discuss the power of the three checks, recall that we roughly categorize misspecification into three types:  ``major'' (when $\trueasympmean$ and $\asympmean{}(\optparam)$ are very different), ``moderate'' (when $\trueasympmean$ and $\asympmean{}(\optparam)$ are  equal or nearly equal but $\trueasympsd/\asympsd(\optparam)$ is large) and ``mild'' (when $\trueasympmean \approx \asympmean{}(\optparam)$ and $\trueasympsd/\asympsd(\optparam)$ is not very large). 
This classification of misspecification in terms of both the accuracy of test statistic estimates
and the uncertainty of the statistic arose naturally from our theory.\footnote{There is some subtlety here. For single SPCs 
	in particular, the statistic uncertainty is also captured by
	the  parameter uncertainty \emph{in direction most relative to determining the statistic's value.}} %
Incorrect uncertainty quantification can lead to poorly calibrate model predictions. 
A relevant example the case of a sufficient test statistic, where the model will always be able 
to match the statistic, but may be overconfident about its true value. 
Following our classification, divided SPCs are the most sensitive, having asymptotic power of 1 even under mild misspecification. 
Single SPCs, on the other hand, have good sensitivity when the misspecification is moderate to large.
PPCs are only sensitive when misspecification is severe.

\paragraph*{Dataset size and computational efficiency.}
All three methods require similar computational effort. PPCs use only the already estimated posterior. 
Single SPCs require estimating the posterior of the observed split of the data, which will typically require about half the computational effort 
of the original posterior estimate since we recommend choosing $\spcprop \approx 0.5$.
While divided SPCs require one posterior computation for each subset, the dataset sizes are small. 
So, the total cost should be roughly equal to that of estimating the original posterior. 
As for different dataset sizes, divided SPCs require a moderate-to-large dataset since when the dataset is small, 
there is an insufficient amount of data in each subset of size $\dspcnumobs$. 
On the other hand, both PPCs and single SPCs remain effective in the small-data regime. 

\paragraph*{Choosing the right predictive check.} 
As a practical guide for SPCs, we have shown how SPCs provide additional flexibility in the context of structured models such as time-series and hierarchical models, which allows the data analyst to design the check to match how the model will be used. 
For hierarchical model checking, when the test statistics are affected by the misspecification within groups, the single within-SPC, cross/within-divided cross-SPC perform the best, while if the test statistics measure the effects across groups, then single cross-SPC and cross/within-divided within-SPC checks is the best choice detecting the misspecification across groups.
For time series setting, PPCs and interpolated single SPCs checks for the inference in the direction of statistics, while extrapolated single SPCs focuses on the prediction of the test statistics. Divided SPCs checks are sensitive to both inference and prediction behavior of the statistics.

\paragraph{Comparison to held-out predictive checks.} 
In independent and concurrent work, \citet{Moran:2024} similarly identify shortcomings with existing predictive checks, including  the POP-PC proposed in their original 2019 preprint (see Supplementary Materials B of the revised version of \citet{Moran:2024} for a detailed discussion of POP-PC). 
In a concurrent revision, they propose the \emph{held-out predictive check}, which is the same as the single SPC.
They show that it produces asymptotically uniform $p$-values when the model is correctly specified \citep[Theorem 1]{Moran:2024}, 
which is a special case of our \cref{THM:SINGLESPC}(1) when the model is well-specified. 
\citet{Moran:2024} also demonstrate empirically that post-hoc calibration approaches for PPCs do not resolve the ``double use of the data'' problem,
providing a complementary perspective to ours.

\paragraph*{Future work.} 
There are several interesting directions for future research. 
Our simulation study and theory for the normal location model suggest the calibration properties of SPCs extend to the case of realized discrepancies, but it would be worthwhile to develop a more comprehensive theory.
Such a theory could also cover discrepancies that have non-normal limiting distributions. 
Since SPCs can become unreliable when the prior is very influential, it could also be fruitful to develop a non-asymptotic theory of SPCs, which might lead to corrections to the $p$-values in the small-sample setting.

\subsection*{Acknowledgments}

Thanks to Aki Vehtari and Jeffrey Miller for helpful discussions, and to Jeffrey Negrea for help with conditions for uniform Donsker classes and sharing his preprocessed airline flight delay data. 
J.~Li and J.~H.~Huggins were supported by the National Institute of General Medical Sciences of the National Institutes of Health under grant number R01GM144963 as part of the Joint NSF/NIGMS Mathematical Biology Program. 
The content is solely the responsibility of the authors and does not necessarily represent the official views of the National Institutes of Health.

\bibliographystyle{imsart-nameyear}
\bibliography{Bibliography-MM-MC}

\newpage
\appendix

\section{Proofs}
\subsection{Technical lemmas}
\label{appx:lemmas-single}
We state some preliminary definitions and lemmas needed for the proofs of the single SPC-related results \cref{THM:SINGLESPC_RATE,THM:SINGLESPC}. 

To quantify the distance between probability measures, we rely on a number of different probability metrics: Kolmogorov distance, Wasserstein distance, total variation (TV) distance, and the bounded Lipschitz distance. 
It follows from the definitions $d_{TV,S} \le d_{TV}$ and $d_K \le d_{TV}$. 
The $s$-Wasserstein distance between distributions $\mu$ and $\nu$ defined on $\Omega \subseteq \reals^{d}$ 
is given by $W_{s}(\mu, \nu) \defined \inf_{\xi}\{\int |u - v|^{s}\xi(\dee u, \dee v)\}^{1/s}$, where the infimum
is over all distributions on $\Omega \times \Omega$ with marginals $\mu$ and $\nu$; 
that is, with $\xi(\cdot, \reals) = \mu$ and  $\xi(\reals, \cdot) = \omega$ \citep{Villani:2009}.
By \citet[Theorem 4.1]{Gibbs:2002}, we have
\begin{align}
	W_{s}(\mu,\nu)\leq \operatorname{diam}(\Omega)\,d_{TV}(\mu,\nu).
\end{align}
where $\operatorname{diam}(\Omega)\defined \sup\{|x - y|: x, y \in \Omega\}$.
If $d = 1$, by \citet[Theorem 1.2]{Bouchitte:2007}, we have
\[
d_{K}(\mu,\nu) \leq C_s W^{\frac{s}{1+s}}_{s}(\mu,\nu),
\]
where $C_s$ is a constant depending on $s \geq 1$. 

For a function $f : \reals^{d} \to \reals$, let $\norm{f}_{L} = \sup_{a \ne b}\frac{|f(a) - f(b)|}{\norm{a - b}_{2}}$
denote the Lipschitz constant of $f$ and let $\norm{f}_{\infty} = \sup_{a} |f(a)|$. 
Define norm $\BLnorm{\cdot}$ for all bounded and Lipschitz continuous functions $f$ as $\BLnorm{f}\defined \Vert f \Vert_{\infty} + \Vert f \Vert_{L}$.
The bounded Lipschitz distance is given by
\[
d_{BL}(X, Y) \defined \sup_{h : \BLnorm{h} \le 1}\left|\EE\{h(X)\} - \EE\{h(Y)\}\right|.
\]

Further, we can show that the Kolmogorov distance between two pairs of convolutions can be upper bounded by a combination of a bounded Lipschitz metric and Kolmogorov distance between single random variables. We present it as a lemma below.

\blem
\label{lem:bound_dK}
Suppose $X$, $Y$, $\tilde{X}$ and $\tilde{Y}$ are independent random variables. Assume that $F_{\tilde{Y}}$, the CDF of $\tilde{Y}$,
satisfies $\BLnorm{F_{\tilde{X}}}  < \infty$.
Then 
\[
d_K(\tilde{X} + \tilde{Y}, X +Y) \leq \BLnorm{F_{\tilde{X}}} d_{BL} (\tilde{Y}, Y) + d_{K}(\tilde{X}, X).
\]
\elem

\begin{proof}
	Let $F_{\tilde{X}}$, $F_{X}$, $ F_{Y}$ denotes the CDFs of, respectively, $\tilde{X}$, $X$ and $Y$. 
	By definition of convolutions and Kolmogorov distance,
	\begin{align}
		\lefteqn{d_K\left(\tilde{X} + \tilde{Y}, X + Y\right)}\\
		& = \sup_{t\in \mathbb{R}}\left|\int_{\mathbb{R}}F_{\tilde{X}}(t - y)\dee F_{\tilde{Y}}(y) - \int_{\mathbb{R}}F_{X}(t - y)\dee F_{Y}(y)  \right|\\
		& \leq \sup_{t\in \mathbb{R}}\left|\int_{\mathbb{R}}F_{\tilde{X}}(t - y)\dee F_{\tilde{Y}}(y) - \int_{\mathbb{R}}F_{\tilde{X}}(t - y)\dee F_{Y}(y) \right| + \sup_{t\in \mathbb{R}} \left|\int_{\mathbb{R}}F_{X}(t - y)-F_{\tilde{X}}(t - y)\dee F_{Y}(y)\right| \\
		& \leq  \sup_{t\in \mathbb{R}}\left|\EE\left[F_{\tilde{X}}(t - \tilde{Y}) -F_{\tilde{X}}(t - Y) \right]\right|+ \sup_{x \in \mathbb{R}}\left|F_{X}(x)- F_{\tilde{X}}(x)\right|\int_{\mathbb{R}}\dee F_{Y}(y)\\
		& = \sup_{t\in \mathbb{R}}\left|\EE\left[F_{\tilde{X}}(t - \tilde{Y}) -F_{\tilde{X}}(t - Y) \right]\right|+ d_K(\tilde{X}, X)\\
		& \leq \BLnorm{F_{\tilde{X}}}\sup_{f : \BLnorm{f} < \infty }\left|\EE\left[f(\tilde{Y}) -f(Y) \right]\right| + d_K(\tilde{X}, X)\\
		& = \BLnorm{F_{\tilde{X}}}d_{BL}(\tilde{Y}, Y)+ d_K(\tilde{X}, X).
	\end{align}
\end{proof}

\blem
\label{lem:A2}
Let $\paramseq{\numobs} \in \paramspace_{\numobs} \defined \{\param:\Vert\param - \optparam\Vert \leq CR_{\numobs}\}$ for some constant $C>0$ and a sequence $R_{\numobs} > 0$ depending on sample size $\numobs$. If \cref{assump2:asymp_mean_diff} holds, then there exists an absolute constant $c_2 > 0$ such that 
\begin{equation}
	\left|\frac{\numobs^{1/2}\{ \asympmean{}(\paramseq{\numobs})- \asympmean{}(\optparam)\}}{\asympsd(\optparam)} - \diffasympmean{}(\optparam)^\top\frac{\numobs^{1/2}\{\paramseq{\numobs}- \optparam\}}{\asympsd(\optparam)}\right|< \frac{c_2\numobs^{1/2}R_{\numobs}^2}{\asympsd(\optparam)} . 
\end{equation}
\elem
\begin{proof}
	By \cref{assump2:asymp_mean_diff}, we have $\Vert\ddot{\asympmean{}}(\optparam)\Vert_2 =  \lambda_{\max}(\ddot{\asympmean{}}(\optparam))< \infty$. Consider the first-order Taylor expansion of $\asympmean{}(\paramseq{\numobs})$ around $\optparam$, for some constant $c_2 > C \lambda_{\max}(\ddot{\asympmean{}}(\optparam))$ and $\paramstar \in \mathcal{U}(\optparam,\paramseq{\numobs}) \defined \{\param:\Vert \param - \optparam \Vert \leq  \Vert \paramseq{\numobs} - \optparam \Vert\}$, 
	\begin{align}
		\left|\asympmean{}(\paramseq{\numobs})- \asympmean{}(\optparam) + \diffasympmean{}(\optparam)(\paramseq{\numobs} - \optparam) \right|
		&= \left|\frac{(\paramstar - \optparam)^\top\ddot{\asympmean{}}(\optparam)(\paramstar - \optparam)}{2}\right| \\
		& \leq \frac{1}{2}\lambda_{\max}(\ddot{\asympmean{}}(\optparam))\Vert\paramstar - \optparam\Vert^2\\
		&\leq C\lambda_{\max}(\ddot{\asympmean{}}(\optparam))R_{\numobs}^2 \\
		&\leq c_2 R_{\numobs}^2,
		\label{eq:interpret_a2}
	\end{align}
	where the first inequality holds since $\ddot{\asympmean{}}(\optparam)$ is symmetric. The proof is complete after multiplying both sides by $\numobs^{1/2}/\asympsd(\optparam)$.
\end{proof}

\blem(Gaussian Concentration, \citet[Chap 2.5]{Vershynin:2018})
\label{lem:GaussConc}
Let $X \dist \mathcal{N}(0, v)$. Then, for all $t \geq 0$,
\begin{equation}
	\Pr\left[|X| \geq t\right] \leq 2 e^{-\frac{t^2}{2v}}.
\end{equation}
\elem

\blem
\label{lem:density_mass_1}
Let $\paramspace_{\numobs} \defined \{\param:\Vert\param-\optparam\Vert<C\log^{1/2}\numobs/
\numobs^{1/2}\}$. If \cref{assump:mle_and_bvm_rate} holds, then $\Pr_{\optparam}\left[\left|\postdistfull{\paramspace_{\numobs}}{\datarv_{\numobs}}- 1\right|  > c\numobs^{-1/2} \right] = \bigo(\numobs^{-s/2})$.
\elem
\begin{proof}
	It follows from \cref{lem:GaussConc} and \cref{assump:a:mle_convto_true_rate} that $\Pr_{\optparam}[|\distBVM{\paramspace_{\numobs}}{\datarv_{\numobs}}- 1| > C\numobs^{-1}] = \bigo(\numobs^{-s/2})$.
	Combining this inequality with \cref{assump:b:BVMrate} yields 
	\begin{align}
		\lefteqn{\Pr_{\optparam}\left[\left|\postdistfull{\paramspace_{\numobs}}{\datarv_{\numobs}}- 1\right|  > c\numobs^{-1/2} \right]} \\
		&\le \Pr_{\optparam}\left[ \big|\postdistfull{\paramspace_{\numobs}}{\datarv_{\numobs}} - \distBVM{\paramspace_{\numobs}}{\datarv_{\numobs}}\big| + \big|\distBVM{\paramspace_{\numobs}}{\datarv_{\numobs}} - 1\big| > c\numobs^{-1/2} \right] \\
		& \leq \Pr_{\optparam}\left[d_{TV,\paramspace_{\numobs}}\left(\postdistfull{\cdot}{\datarv_{\numobs}}, \distBVM{\cdot}{\datarv_{\numobs}}\right) >c\numobs^{-1/2} - C\numobs^{-1}\right]+ \bigo(\numobs^{-s/2})\\
		& \leq \Pr_{\optparam}\left[d_{TV,\paramspace_{\numobs}}\left(\postdistfull{\cdot}{\datarv_{\numobs}},  \distBVM{\cdot}{\datarv_{\numobs}}\right) > c_{\mathcal{K}}\numobs^{-1/2} \right] + \bigo(\numobs^{-s/2})\\
		& \leq \Pr_{\optparam}\left[d_{TV,\mathcal{K}}\left(\postdistfull{\cdot}{\datarv_{\numobs}},  \distBVM{\cdot}{\datarv_{\numobs}}\right) > c_{\mathcal{K}}\numobs^{-1/2} \right] + \bigo(\numobs^{-s/2})\\ 
		& =  \bigo(\numobs^{-s/2}), \label{eq:part1_densitymass1}
	\end{align}
	where the penultimate inequality follows by $d_{\text{TV}, \paramspace_{\numobs}} \leq d_{\text{TV}, \mathcal{K}}$ since $\paramspace_{\numobs} \subset \mathcal{K}$. 
\end{proof}

\blem
\label{lem:A3_asymsd_bounds}
If \cref{assump4:bdd_diff_sigma} holds, then for any $\param \in \paramspace_{\numobs}$ defined in \cref{lem:A2},  \[\left|\frac{\asympsd(\param)}{\asympsd(\optparam)}- 1 \right| \le \frac{C M R_{\numobs}}{\asympsd(\optparam)}.\]
\elem
\begin{proof}
	By Taylor's theorem, for any $\param \in \paramspace_{\numobs}$, there exists a $\paramstar \in \mathcal{U}(\optparam,\param) \defined \{\paramstar: \Vert \paramstar - \optparam\Vert \leq \Vert \param - \optparam\Vert \}$ such that
	\begin{equation}
		\asympsd(\param) = \asympsd(\optparam) + (\param - \optparam)^\top \diffasympvar{\paramstar}. \label{eq:taylor_asympvar}
	\end{equation}
	Since $\paramspace_{\numobs} \subseteq U_{\optparam}$, it follows from \cref{assump4:bdd_diff_sigma} together with \cref{eq:taylor_asympvar} that
	\begin{align}
		\left|\frac{\asympsd(\param)}{\asympsd(\optparam)}- 1 \right| 
		&= |(\param - \optparam)^\top \diffasympvar{\paramstar}|/\asympsd(\optparam) \\
		&\le \norm{\param - \optparam}_{2}\norm{\diffasympvar{\paramstar}}/\asympsd(\optparam)  \\
		&\le \frac{C M R_{\numobs}}{\asympsd(\optparam)}. \label{eq:part4_a3results}
	\end{align}
\end{proof}

\subsection{Proof of \protect\cref{THM:SINGLESPC}}

Consider the single SPC $\p$-value as a random variable
\begin{equation}
	\spcpvalue{\datarv} = \int_{\paramspace} \Pr\left[   \statistic{\spcNnew}(\spcreprv) \ge  \statistic{\spcNnew}(\spcnewrv) \given \param\right] \postdistfull{\dee\param}{\spcobsrv}. 
\end{equation}
Let $Z$ denote a standard normal random variable and $\phi(\cdot;\mu, \sigma^2)$ denote the PDF of normal random variables with mean $\mu$ and variance $\sigma^2$. 

Let $\densityBVM{\cdot} {\spcobsrv}\defined \phi(\cdot;\mle{}(\spcobsrv), \Ehessloglik{\optsym}^{-1}/\spcNobs)$ and $\distBVM{\cdot}{\spcobsrv}$ be the corresponding distribution.  Denote shrinking parameter space as $\paramspace_{\spcNobs}\defined \{\param:\Vert \param -\optparam \Vert \leq c {\spcNobs}^{-1/2}\}$. We emphasize on the difference between asymptotic means and standard deviations of $\statistic{\numobs}$ by defining $\asympmean{}(\param)$ and $\asympsd(\param)$ as under $P_{\param}$ and $\trueasympmean$ and $\trueasympsd$ as under $\obsdist$ respectively. We denote the ratio between asymptotic standard deviations under $\obsdist$ and $P_{\optparam}$ as $\sdratio \defined \trueasympsd/\asympsd_{}(\optparam)$.
For simplicity, we introduce some shorthand notation: 
\begin{align}
	&\pvalbvmdist{\datarv}\defined \int_{\paramspace} \Pr\left[\statistic{\spcNnew}(\spcreprv)\ge \statistic{\spcNnew}(\spcnewrv)\given \param\right] \distBVM{\dee\param}{\spcobsrv}\\
	&\pvalbvmdistball{\datarv}\defined \int_{\paramspace_{\spcNobs}} \Pr\left[ \statistic{\spcNnew}(\spcreprv)\ge \statistic{\spcNnew}(\spcnewrv)\given \param\right] \distBVM{\dee\param}{\spcobsrv}\\
	& s_{\statistic{}\asympmean{}}(\datarv;\param, \numobs) \defined  \numobs^{1/2}\{\statistic{\numobs}(\datarv)- \asympmean{}(\param)\}\\
	& s_{\statistic{}\trueasympmean}(\datarv;\numobs) \defined  \numobs^{1/2}\{\statistic{\numobs}(\datarv)- \trueasympmean\}\\
	&\xi(\param)\defined  \Pr\big[
	s_{\statistic{}\asympmean{}}(\spcreprv;\param, \spcNnew) +  \spcNnew^{1/2}\{\asympmean{}(\param) - \asympmean{}(\optparam)\}  \\
	&\hspace{4cm} \ge
	s_{\statistic{}\trueasympmean}(\spcreprv; \spcNnew)  + \spcNnew^{1/2}\{ \trueasympmean -  \asympmean{}(\optparam)\}\given  \param\big]\\
	&t(\param) \defined \frac{s_{\statistic{}\trueasympmean}(\spcnewrv;\spcNnew)}{\truesigma}\sdratio -  
	\frac{\spcNnew^{1/2}\diffasympmean{}(\optparam)\{\param - \optparam\}}{\asympsd(\optparam)}
	+ \frac{\spcNnew^{1/2}\{ \trueasympmean - \asympmean{}(\optparam)\}}{\asympsd(\optparam)}\\
	& Q(\spcnewrv,\spcobsrv)\defined \frac{1}{\tilde{\sigma}} \bigg(\frac{s_{\statistic{}\trueasympmean}(\spcnewrv;\spcNnew)}{\truesigma}\sdratio - \frac{\spcNnew^{1/2}}{\spcNobs^{1/2}}\frac{\spcNobs^{1/2}\diffasympmean{}(\optparam)\{ \mle{}(\spcobsrv) - \optparam\}}{\asympsd(\optparam)} \\
	& \hspace{4cm}
	+ \frac{\spcNnew^{1/2}\{ \trueasympmean - \asympmean{}(\optparam)\}}{\asympsd(\optparam)}\bigg).
\end{align}
Assume that $\spcNobs$ is sufficiently large such that $\paramspace_{\spcNobs}\subseteq \mcU_{\optsym,1} \cap \mcU_{\optsym,2}$. 
Letting $t = \Omega(\spcNobs^{-1/2})$ and $v = \bigo(\spcNobs^{-1})$ for \cref{lem:GaussConc} gives that $\distBVM{ \paramspace_{\spcNobs}}{\spcobsrv} \convP 1$. 
By \cref{assump:bvm}, \cref{lem:GaussConc} and the fact that $d_{TV}(\mu,\omega)\defined 2\sup_{f\in[0,1]}|\int f\dee\mu - \int f\dee\omega|$, we have 
\begin{align}
	\left|\spcpvalue{\datarv}  -  \pvalbvmdistball{\datarv}\right|
	& \le \left|\spcpvalue{\datarv} -  \pvalbvmdist{\datarv}\right|  + \left|	\pvalbvmdist{\datarv} -  \pvalbvmdistball{\datarv}\right|  
	\\ 
	&\le d_{TV}\left(\postdist{}(\cdot\mid\spcobsrv), \distBVM{\cdot}{\spcobsrv}\right) + \distBVM{ \paramspace^c_{\spcNobs}}{\spcobsrv}\\
	& \convP 0.\label{eq:pval_approx}
\end{align}
By centering both $\statistic{\spcNnew}(\spcreprv)$ and $\statistic{\spcNnew}(\spcnewrv)$, the integrand of $\pvalbvmdistball{\datarv} $ becomes
\begin{align}
	\lefteqn{ \Pr\left[\statistic{\spcNnew}(\spcreprv)\ge \statistic{\spcNnew}(\spcnewrv)\given \param\right]}\\
	& = \Pr\big[
	\spcNnew^{1/2}\{\statistic{\spcNnew}(\spcreprv)- \asympmean{}(\param)\}  +  \spcNnew^{1/2}\{\asympmean{}(\param) - \asympmean{}(\optparam)\} \\
	& \hspace{1.5cm} \ge \spcNnew^{1/2}\left\{\statistic{\spcNnew}(\spcnewrv) - \trueasympmean\right\} + \spcNnew^{1/2}\{ \trueasympmean -  \asympmean{}(\optparam)\}\given  \param\big]\\
	& = \Pr\big[
	s_{\statistic{}\asympmean{}}(\spcreprv;\param, \spcNnew) +  \spcNnew^{1/2}\{\asympmean{}(\param) - \asympmean{}(\optparam)\} \\
	& \hspace{1.5cm} \ge
	s_{\statistic{}\trueasympmean}(\spcreprv; \spcNnew)  + \spcNnew^{1/2}\{ \trueasympmean -  \asympmean{}(\optparam)\}\given  \param\big]\\
	& =: \xi(\param).\label{eq:rewrite_integrand}
\end{align}
Dividing by $1/\asympsd(\optparam)$ on both sides within the probability and applying \cref{lem:A2,lem:A3_asymsd_bounds} with $R_{\numobs} = C\spcNobs^{-1/2}$ yields  
\begin{align}
	\xi(\param)
	& \le  \Pr\big[
	s_{\statistic{}\asympmean{}}(\spcreprv;\param,\spcNnew)  +  \frac{\spcNnew^{1/2}}{\spcNobs^{1/2}}
	\left(\spcNobs^{1/2}\diffasympmean{}(\optparam)\{\param - \optparam\} + C/\spcNobs^{1/2} \right) \\
	& \hspace{1.5cm} > 
	s_{\statistic{}\trueasympmean}(\spcnewrv;\spcNnew)
	+ \spcNnew^{1/2}\{ \trueasympmean - \asympmean{}(\optparam)\}\given  \param \big]\\
	& =  \Pr\big[
	s_{\statistic{}\asympmean{}}(\spcreprv;\param,\spcNnew)  +  
	\spcNnew^{1/2}\diffasympmean{}(\optparam)\{\param - \optparam\} + C\spcNnew^{1/2}/\spcNobs  \\
	& \hspace{1.5cm} \ge
	s_{\statistic{}\trueasympmean}(\spcnewrv;\spcNnew)
	+ \spcNnew^{1/2}\{ \trueasympmean - \asympmean{}(\optparam)\}\given  \param \big]\\
	& =  \Pr\bigg[\frac{ s_{\statistic{}\asympmean{}}(\spcreprv;\param,\spcNnew)  }{\asympsd(\param)}\frac{\asympsd(\param)}{\asympsd(\optparam)}
	+  
	\frac{\spcNnew^{1/2}\diffasympmean{}(\optparam)\{\param - \optparam\}}{\asympsd(\optparam)} + \frac{C\spcNnew^{1/2}/\spcNobs}{\asympsd(\optparam)}  \\
	& \hspace{1.5cm} \ge \frac{s_{\statistic{}\trueasympmean}(\spcnewrv;\spcNnew)}{\truesigma}\frac{\truesigma}{\asympsd(\optparam)}
	+ \frac{\spcNnew^{1/2}\{ \trueasympmean - \asympmean{}(\optparam)\}}{\asympsd(\optparam)}\given  \param \bigg]\\
	& \le  \Pr\bigg[\frac{ s_{\statistic{}\asympmean{}}(\spcreprv;\param,\spcNnew)  }{\asympsd(\param)}\left(1 + \frac{CM}{\asympsd(\optparam)\spcNobs^{1/2}}\right)
	+  
	\frac{\spcNnew^{1/2}\diffasympmean{}(\optparam)\{\param - \optparam\}}{\asympsd(\optparam)}  \\
	& \hspace{1.5cm} + \frac{C\spcNnew^{1/2}/\spcNobs}{\asympsd(\optparam)}  \ge \frac{s_{\statistic{}\trueasympmean}(\spcnewrv;\spcNnew)}{\truesigma}\sdratio
	+ \frac{\spcNnew^{1/2}\{ \trueasympmean - \asympmean{}(\optparam)\}}{\asympsd(\optparam)}\given  \param \bigg]\\
	& =   1 - \Pr\bigg[\frac{ s_{\statistic{}\asympmean{}}(\spcreprv;\param,\spcNnew)  }{\asympsd(\param)}  \le  \frac{s_{\statistic{}\trueasympmean}(\spcnewrv;\spcNnew)}{\truesigma}\sdratio \\
	& \hspace{2cm}  -  
	\frac{\spcNnew^{1/2}\diffasympmean{}(\optparam)\{\param - \optparam\}}{\asympsd(\optparam)}
	+ \frac{\spcNnew^{1/2}\{ \trueasympmean - \asympmean{}(\optparam)\}}{\asympsd(\optparam)}  \given  \param \bigg] +  \littleoP(\spcNobs^{-1/2})\\
	& =  1  - \Pr\bigg[\frac{ s_{\statistic{}\asympmean{}}(\spcreprv;\param,\spcNnew)  }{\asympsd(\param)}  \le t(\param) \given \param\bigg]+  \littleoP(\spcNobs^{-1/2}). 
	\label{eq:bound_integrand}
\end{align}
By \cref{assump:asymp_norm_stat}, $\frac{ s_{\statistic{}\asympmean{}}(\spcreprv;\param,\spcNnew)}{\asympsd(\param)}$, we have 
\begin{align}
	\left|\Pr\bigg[\frac{ s_{\statistic{}\asympmean{}}(\spcreprv;\param,\spcNnew)  }{\asympsd(\param)} 
	\le t(\param) \given \param\bigg] -  \Phi\left(t(\param);\param\right)\right| \convP 0. \label{eq:conv_t}
\end{align}
Combining \cref{eq:conv_t,eq:bound_integrand,eq:rewrite_integrand} yields 
\begin{align}
	\left|\Pr\left[\statistic{\spcNnew}(\spcreprv)\ge \statistic{\spcNnew}(\spcnewrv)\given \param\right] - (1 - \Phi\left(t(\param);\param\right))\right|\convP 0,
\end{align}
and thus 
\begin{align}
	\left|  \pvalbvmdistball{\datarv} - \left(1- \int_{\paramspace_{\spcNobs}}  \Phi\left(t(\param);\param\right)\distBVM{\dee\param}{\spcobsrv}\right)\right|\convP 0, \label{eq:conv_pval_int}
\end{align}
Let $\EE_{\param}[\cdot]$ denote the expectation with respect to $\phi(\mle{}(\spcobsrv), \Ehessloglik{\optsym}^{-1}\mlevar_{\optsym}\Ehessloglik{\optsym}^{-1}/\spcNobs)$.
Let $Z$ be an independent standard normal random variable and define 
\begin{align}
	\tilde{Z}& \defined Z  +  \frac{\spcNnew^{1/2}}{\spcNobs^{1/2}}\frac{\spcNobs^{1/2}\diffasympmean{}(\optparam)\{\param - \mle{}(\spcobsrv)\}}{\asympsd(\optparam)} \dist \distNorm(0, \tilde{\sigma}^2),
\end{align}
where $\tilde{\sigma}^2  \defined  1 + \frac{\frac{\spcNnew}{\spcNobs}\diffasympmean{}(\optparam)^2 \Ehessloglik{\optsym}^{-1}}{\asympsd(\optparam)^2}$.
Consider 
\begin{align}
	\lefteqn{\int_{\paramspace_{\spcNobs}}\Phi\left(t(\param);\param\right)\distBVM{\dee\param}{\spcobsrv} }\\
	& = \EE_{\param}\left[\Pr_{Z}\bigg[Z \le \frac{s_{\statistic{}\trueasympmean}(\spcnewrv;\spcNnew)}{\truesigma}\sdratio -  
	\frac{\spcNnew^{1/2}\diffasympmean{}(\optparam)\{\param - \optparam\}}{\asympsd(\optparam)}
	+ \frac{\spcNnew^{1/2}\{ \trueasympmean - \asympmean{}(\optparam)\}}{\asympsd(\optparam)}\bigg]\right] \\
	& = \Pr_{Z,\param}\left[Z \le \frac{s_{\statistic{}\trueasympmean}(\spcnewrv;\spcNnew)}{\truesigma}\sdratio -  
	\frac{\spcNnew^{1/2}\diffasympmean{}(\optparam)\{\param - \optparam\}}{\asympsd(\optparam)}
	+ \frac{\spcNnew^{1/2}\{ \trueasympmean - \asympmean{}(\optparam)\}}{\asympsd(\optparam)}\right]\\
	& = \Pr_{Z,\param}\bigg[Z  +   
	\frac{\spcNnew^{1/2}}{\spcNobs^{1/2}}\frac{\spcNobs^{1/2}\diffasympmean{}(\optparam)\{\param - \mle{}(\spcobsrv)\}}{\asympsd(\optparam)} \\
	& \hspace{1.5cm} \le \frac{s_{\statistic{}\trueasympmean}(\spcnewrv;\spcNnew)}{\truesigma}\sdratio - \frac{\spcNnew^{1/2}}{\spcNobs^{1/2}}\frac{\spcNobs^{1/2}\diffasympmean{}(\optparam)\{ \mle{}(\spcobsrv) - \optparam\}}{\asympsd(\optparam)}
	+ \frac{\spcNnew^{1/2}\{ \trueasympmean - \asympmean{}(\optparam)\}}{\asympsd(\optparam)}\bigg]\\
	& = \Pr_{\tilde{Z}}\bigg[ \frac{\tilde{Z}}{\tilde{\sigma}} \le Q(\spcnewrv,\spcobsrv)\bigg] \\
	& = \Phi\left(Q(\spcnewrv,\spcobsrv)\right).\label{eq:pval_int_simplification}
\end{align}
Combining \cref{eq:pval_approx,eq:pval_int_simplification,eq:conv_pval_int} yields the asymptotic approximation
\begin{align}
	\spcpvalue{\datarv}  = 1 - \Phi\left(Q(\spcnewrv,\spcobsrv)\right) + \littleoP(1).\label{eq:Q_approx_pval}
\end{align}
\paragraph{Case 1: $\trueasympmean = \asympmean{}(\optparam)$.} Since $\spcobsrv$ and $\spcnewrv$ are independent, we have by \cref{assump:asymp_norm_stat,assump:MLE_conv} that 
\[
Q(\spcnewrv,\spcobsrv) \convD \distNorm(0, \rho^{2}). 
\]
Hence for any $\alpha \in (0,1)$, we have 
\begin{align}
	\Pr\left[\spcpvalue{\datarv}< \alpha\right]
	& = \Pr\left[ 1 - \Phi\left(Q(\spcnewrv,\spcobsrv)\right)< \alpha \right] + \littleoP(1)\\
	& = \Pr\left[ Q(\spcnewrv,\spcobsrv)> z_{1-\alpha} \right] + \littleoP(1)\\
	& = \Pr\left[\frac{Q(\spcnewrv,\spcobsrv)}{\rho} > \frac{z_{1- \alpha} }{\rho}\right] + \littleoP(1)\\
	& = 1 - \Phi\left(\frac{z_{ 1-\alpha} }{\rho}\right)+ \littleoP(1),
\end{align}
where $z_{\alpha}$ satisfies $\Pr[Z \le z_{\alpha}] = \alpha$.
\paragraph{Case 2: $\trueasympmean < \asympmean{}(\optparam)$.} 
Taking $\spcNnew \rightarrow \infty$ yields $\Pr(Q(\spcnewrv,\spcobsrv) > t) \rightarrow 0, \forall t \in \reals$ and thus $\spcpvalue{\datarv} \convP 1$.
Therefore, the tail probabilities of single SPC $\p$-values are immediate
\begin{align}
	\Pr\left[\spcpvalue{\datarv}> 1- \alpha\right]
	& = \Pr\left[ 1 - \Phi\left(Q(\spcnewrv,\spcobsrv)\right)> 1- \alpha \right] + \littleoP(1)\\
	& = 1 + \littleoP(1).
\end{align}
\paragraph{Case 3: $\trueasympmean > \asympmean{}(\optparam)$.} 
Similarly to Case 2, the single SPC $\p$-values converge to $0$ in probability and thus  
$\Pr\left[\spcpvalue{\datarv}< \alpha\right] \convP 1$. 

\subsection{Proof of \protect\cref{THM:SINGLESPC_RATE}}
\label{appx:proof-of-singleSPCthm}
\cref{THM:SINGLESPC_RATE} gives the rate of convergence results for single SPC $\p$-values to uniform distribution under a well-specified model. 
In addition to the regularity conditions in \cref{assump:regularity}, we require stronger versions of \cref{assump:asymp_norm_stat,assump:bvm_and_mle}.
First, we require the scaled test statistic satisfy a CLT at rate $\numobs^{-1/2}$ in Kolmogorov distance $d_{K}$.
\setcounter{assumption}{4}
\begin{assumption} \label{assump:asymp_norm_stat_rate} 
	There exists an open neighborhood $U_{\optsym,6} \ni \optparam$ and an absolute constant $c_1 > 0$ such that for $\param \in U_{\optsym,6}$ 
	and $Y_1, \dots, Y_{\numobs} \distiid \likdist{\param}$, 
	\begin{equation}
		\label{eq:A1_Kol}
		d_{K}\left(\frac{\numobs^{1/2}\{\statistic{\numobs}(\mathbf{Y}_{\numobs}) - \asympmean{}(\param)\}}{\asympsd(\param)}, Z\right) < c_1/\numobs^{1/2}, \tag{A1}
	\end{equation}
	where $Z$ is a standard normal random variable.		 
\end{assumption}
If the statistic takes the form of an average, $\statistic{\numobs}(\datarv) = \frac{1}{\numobs}\sum_{n = 1}^{\numobs}h(\obsrv{n})$,
the  assumption holds by the Berry--Esseen theorem as long as $\sup_{\param \in U_{\optsym,6}}\EE_{\param}[|h(\obsrv{1})|^3]  < \infty$ \citep[Theorem 6.1]{Gut:2013}. 
The assumption also holds for $U$-statistics $\statistic{\numobs}(\datarv) = \frac{2}{\numobs(\numobs-1)}\sum_{ i < j}h(\obsrv{i},\obsrv{j})$ 
as long as $\sup_{\param \in U_{\optsym,6}}\EE[|h(\obsrv{1},\obsrv{2})|^3]<\infty$ \citep{Alberink:1999}. 

We also require the maximum likelihood to be asymptotically normal in the bounded Lipschitz metric.
\begin{assumption}
	\label{assump:MLE_conv_rate}
	For $Z \dist \distNorm(0, I)$, the maximum likelihood estimator satisfies 
	\[
	d_{BL}(\numobs^{1/2}\mlevar_{\optsym}^{-1}\{\mle{}(\mathbf{X}_{\numobs}) - \optparam\}, Z) = \bigo(\numobs^{-1/2}).
	\]
\end{assumption}
\cref{assump:MLE_conv_rate} can be shown to hold using, for example, \citet[Lemma~2.1]{Anastasiou:2017}. 

\begin{proof}[Proof of \cref{THM:SINGLESPC_RATE}]
	We will first prove convergence in $s$-Wasserstein distance, which we will then use to control convergence in Kolmogorov distance. 
	
	Given data $\datarv$, the single SPC $\p$-value is defined as
	\begin{equation}
		\spcpvalue{\datarv} = \int_{\paramspace} \Pr\left[  \statistic{\spcNnew}(\spcreprv)\ge \statistic{\spcNnew}(\spcnewrv)\given \param\right] \postdistfull{\dee\param}{\spcobsrv}. 
	\end{equation}
	Let $Z$ denote a standard normal random variable and $\phi(\cdot;\mu, \sigma^2)$ denote the PDF of a normal random variables with mean $\mu$ and variance $\sigma^2$. Suppose $\mathcal{K}$ is a compact set containing $\optparam$
	and a shrinking parameter space is defined as $\paramspace_{\spcNobs}\defined \{\param:\Vert \param -\optparam \Vert \leq C \log^{1/2}\spcNobs/{\spcNobs}^{1/2}\}$. 
	Assume that $\spcNobs$ is sufficiently large that $\paramspace_{\spcNobs} \subseteq \mathcal{K} \cap \mcU_{\optsym,1} \cap \mcU_{\optsym,6}$.
	Denote $\densityBVM{\cdot} {\spcobsrv}\defined \phi(\cdot;\mle{}(\spcobsrv), \truemlevar/\spcNobs)$ and $\distBVM{\cdot}{\spcobsrv}$ the corresponding distribution. 
	For simplicity, we introduce some shorthand notation:
	\begin{align}
		&\pvaltoK{\datarv}\defined \int_{\paramspace_{\spcNobs}} \Pr[   \statistic{\spcNnew}(\spcreprv)\ge \statistic{\spcNnew}(\spcnewrv)\given  \param] \postdistfull{\dee\param}{\spcobsrv} \\
		&\spcpvalueapprox{\datarv} \defined \int_{\paramspace_{\spcNobs}} \Pr[\statistic{\spcNnew}(\spcreprv)\ge \statistic{\spcNnew}(\spcnewrv)\given   \param] \distBVM{\dee\param}{\spcobsrv}\\
		&s_{\statistic{} \asympmean{}}(\datarv,\param;\numobs) \defined \numobs^{1/2}\{\statistic{\numobs}(\datarv) - \asympmean{}(\param)\}\\
		&t(\param) \defined \spcNnew^{1/2}\{\statistic{\spcNnew}(\spcnewrv) - \asympmean{}(\optparam)\} - \spcNnew^{1/2}\{\asympmean{}(\param) - \asympmean{}(\optparam)\} \\
		&\tilde{Q}(\param) \defined \frac{s_{\statistic{} \asympmean{}}(\spcnewrv, \optparam; \spcNnew) - \diffasympmean{}(\optparam)^\top\spcNnew^{1/2}(\param - \optparam) }{\asympsd(\optparam)}\\
		& Q \defined \frac{s_{\statistic{} \asympmean{}}(\spcnewrv,\optparam;\spcNnew) - \diffasympmean{}(\optparam)^\top\spcNnew^{1/2}\{\mle{}(\spcobsrv)- \optparam\}}{(\asympsd^2(\optparam) + \frac{\spcNnew}{\spcNobs}\diffasympmean{}(\optparam)^\top\mlevar(\optparam)\diffasympmean{}(\optparam))^{1/2}}\\
		& \zeta(\param) \defined \Pr\left[s_{\statistic{} \asympmean{}}(\spcreprv,\param;\spcNnew) \ge t(\param)\mid \param\right].
	\end{align}
	Here $\pvaltoK{\datarv}$ approximates the single SPC $\p$-values by restricting the integral to $\paramspace_{\spcNobs}$ and then $\spcpvalueapprox{\datarv} $ defines a further approximation with standard posterior replaced by a limiting Gaussian density $\distBVM{\cdot}{\spcobsrv}$. We use $t(\param), \tilde{Q}(\param)$ and $Q$ in intermediate steps in our proof and that they are quantities approximating $\p$-values using assumptions listed above.
	For any real-valued function $g(\param)$, define the shorthands $\intBVMapprox  g(\param)\defined \int_{\paramspace_{\spcNobs}}g(\param)\distBVM{\dee \param}{\spcobsrv}$ and 
	$\intBVMapprox_{\paramspace}g(\param)\defined \int_{\paramspace}g(\param)\distBVM{\dee \param}{\spcobsrv}$.

	Let $\Phi(\cdot)$ be the CDF of a standard normal random variables. Applying triangle inequalities, we can divide the problem into several parts and bound each part in turn. 
	\begin{align}
		\lefteqn{W_s\left( \spcpvalue{\datarv},  U\right) = W_s\left(\spcpvalue{\datarv}, \Phi(Z)\right)} \\
		&\leq  W_s\left(\spcpvalue{\datarv}, \pvaltoK{\datarv}\right) + W_s\left(\pvaltoK{\datarv}, \spcpvalueapprox{\datarv}\right)\\ 
		&\quad + W_s\left(  \spcpvalueapprox{\datarv}, \intBVMapprox\zeta(\param)\right) + W_s\left(\intBVMapprox\zeta(\param), \intBVMapprox\Phi\left(\frac{t(\param)}{\asympsd(\param)}\right)\right) \\
		&\quad +  W_s\left(\intBVMapprox\Phi\left(\frac{t(\param)}{\asympsd(\param)}\right),\intBVMapprox\Phi\left(\frac{t(\param)}{\asympsd(\optparam)}\right)\right) + W_s\left(\intBVMapprox\Phi\left(\frac{t(\param)}{\asympsd(\optparam)}\right), \intBVMapprox\Phi\left(\tilde{Q}(\param)\right)\right)\\
		&\quad + W_s\left( \intBVMapprox\Phi\left(\tilde{Q}(\param)\right),  \Phi(Q)\right) + W_s(\Phi(Q),\Phi(Z)).
		\label{eq:divide_parts}
	\end{align}
	
	\textbf{Part I}:  $W_s\left(\spcpvalue{\datarv},  \pvaltoK{\datarv}\right) = \bigo(\spcNobs^{-1/2})$.
	
	We first show that the integral that defines SPC $\p$-values can be restricted to a shrinking ball while only introducing an error of order $\bigo(\spcNobs^{-1/2})$ in probability. 
	Fix constants $C > 0$ and $c > c_{\mathcal{K}} + C$. 
	It follows by the coupling characterization of Wasserstein distance that 
	\begin{align}
		&W_s^s\left(\spcpvalue{\datarv}, \pvaltoK{\datarv}\right) \\
		& \leq \EE\left|\spcpvalue{\datarv}- \pvaltoK{\datarv}\right|^s \\
		& = \EE\left|\int_{\paramspace \backslash \paramspace_{\spcNobs}}\Pr[\statistic{\spcNnew}(\spcreprv)\ge \statistic{\spcNnew}(\spcnewrv)\given   \param]\postdistfull{\dee\param}{\spcobsrv}\right|^s \\
		& \leq\EE\left|1 - \postdistfull{\paramspace_{\spcNobs}}{\spcobsrv}\right|^s \\
		&\leq c\spcNobs^{-s/2} + \Pr_{\optparam}[|\postdistfull{\paramspace_{\spcNobs}}{\spcobsrv}- 1| > c\spcNobs^{-1/2}]\\
		&\leq  c'\spcNobs^{-s/2} \label{eq:part1_term1}
	\end{align}
	where the last inequality follows by applying \cref{lem:density_mass_1} with $\spcobsrv$ of size $\spcNobs$.

	\textbf{Part II}: $W_s\left(\spcpvalue{\datarv},  \spcpvalueapprox{\datarv}\right) = \bigo(\spcNobs^{-1/2})$.

	Consider the equivalent definition of total variation distance, i.e., $d_{TV}(\mu,\nu) = 2\sup_f\{|\int f \dee\mu - \int f \dee\nu|: 0\leq f \leq 1\}$. A similar argument as used in \cref{eq:part1_term1} can be employed to get
	\begin{align}
		W_s^s\left(\pvaltoK{\datarv}, \spcpvalueapprox{\datarv}\right) 
		&\leq \EE\left|\pvaltoK{\datarv}-\spcpvalueapprox{\datarv}\right|^s\\
		& \leq \EE_{\param}\left[d^s_{TV,\paramspace_{\spcNobs}}(\postdistfull{\cdot}{\datarv}, \distBVM{\cdot}{\datarv})\right]\\
		& \leq c_{\mathcal{K}}\spcNobs^{-s/2} + \bigo(\spcNobs^{-s/2}).
	\end{align}
	Therefore,
	\begin{equation}
		W_s\left(\pvaltoK{\datarv}, \spcpvalueapprox{\datarv}\right) = \bigo(\spcNobs^{-1/2}).\label{eq:part1_term2_res}
	\end{equation}

	\textbf{Part III}: $W_s\left(  \spcpvalueapprox{\datarv},  \intBVMapprox\Phi\left(\frac{t(\param)}{\asympsd(\param)}\right) \right)= \bigo(\spcNobs^{-1/2})$.
	Starting from Part III, we mainly approximate the integrand of single SPC $\p$-values. 
	In this part, we first rewrite the integrand using our shorthand notations defined earlier. By definition, $ \spcpvalueapprox{\datarv} = \intBVMapprox\Pr[ \statistic{\spcNnew}(\spcreprv)\ge \statistic{\spcNnew}(\spcnewrv)\given   \param]$.
	But since
	\begin{align}
		\lefteqn{\Pr[\statistic{\spcNnew}(\spcreprv)\ge \statistic{\spcNnew}(\spcnewrv)\given   \param]}\\
		& = \Pr[
		\spcNnew^{1/2}\{\statistic{\spcNnew}(\spcreprv)- \asympmean{}(\optparam)\} -  \spcNnew^{1/2}\{\asympmean{}(\param) - \asympmean{}(\optparam)\}\\
		&\qquad \ge  \spcNnew^{1/2}\{\statistic{\spcNnew}(\spcreprv) - \asympmean{}(\optparam)\} -  \spcNnew^{1/2}\{\asympmean{}(\param) - \asympmean{}(\optparam)\} \given  \param]\\
		& = \Pr[ \spcNnew^{1/2}\{\statistic{\spcNnew}(\spcreprv) - \asympmean{}(\param)\}\ge t(\param)\given \param]\\
		& = \Pr[ s_{\statistic{}\asympmean{}}(\pcreprv,\param;\spcNnew)\ge t(\param)\given \param]\\
		& =\zeta(\param),
	\end{align}
	the desired equality follows. Now we only need to show that
	$
	d_W\left(\intBVMapprox\zeta(\param), \intBVMapprox\Phi\left(\frac{t(\param)}{\asympsd(\param)}\right)\right) = \bigo(\spcNobs^{-1/2})
	$.
	By definition of $\spcNobs \defined \spcobssize $ and $\spcNnew \defined \spcnewsize$, $\spcNnew^{1/2}/\spcNobs^{1/2} = \sqrt{\numobs/\spcNobs - 1} > c_0 > 0$.
	Since 
	\begin{align}
		\zeta(\param) 
		&=  \Pr\left[s_{\statistic{} \asympmean{}}(\spcreprv,\param;\spcNnew) \ge t(\param)\mid \param\right] 
		= \Pr\left[\frac{s_{\statistic{} \asympmean{}}(\spcreprv,\param;\spcNnew)}{\asympsd(\param)} \ge \frac{t(\param)}{\asympsd(\param)} \mid \param\right]
	\end{align}
	it follows from \cref{eq:A1_Kol} that for some constant $c_1 > 0$,
	\begin{equation}
		\sup\limits_{\param \in \paramspace_{\spcNobs}}\left|\Pr\left[\frac{s_{\statistic{} \asympmean{}}(\spcreprv,\param;\spcNnew)}{\asympsd(\param)} \ge \frac{t(\param)}{\asympsd(\param)} \mid \param\right] - \Phi\left(\frac{t(\param)}{\asympsd(\param)}\right)\right| < c_1/\spcNobs^{1/2}.\label{eq:part3_assump1}
	\end{equation}
	It then follows by the coupling characterization of Wasserstein distance and \cref{eq:part3_assump1} that 
	\begin{align}
		\lefteqn{W_s^s\left(\intBVMapprox\zeta(\param), \intBVMapprox\Phi\left(\frac{t(\param)}{\asympsd(\param)}\right)\right)}\\
		&\leq \EE\left|\intBVMapprox\zeta(\param) - \intBVMapprox\Phi\left(\frac{t(\param)}{\asympsd(\param)}\right)\right|^s \\
		& \leq \EE\left[\intBVMapprox\left| \Pr\left[\frac{s_{\statistic{} \asympmean{}}(\spcreprv,\param;\spcNnew)}{\asympsd(\param)} \ge \frac{t(\param)}{\asympsd(\param)}; \param\right]  - \Phi\left(\frac{t(\param)}{\asympsd(\param)}\right) \right|\right]^s\\
		&\leq \EE\left[\int_{\paramspace_{\spcNobs}}\sup\limits_{\param \in \paramspace_{\spcNobs}}\left| \Pr\left[\frac{s_{\statistic{} \asympmean{}}(\spcreprv,\param;\spcNnew)}{\asympsd(\param)} \ge \frac{t(\param)}{\asympsd(\param)}; \param\right]  - \Phi\left(\frac{t(\param)}{\asympsd(\param)}\right) \right|\distBVM{\dee \param}{\spcobsrv}\right]^s\\
		&\leq c_1/\spcNobs^{s/2}\EE\left[\distBVM{ \paramspace_{\spcNobs}}{\spcobsrv}\right]^s\\
		&\leq c_1/\spcNobs^{s/2}.
	\end{align}
	\textbf{Part IV}: 
	$
	W_s\left(\intBVMapprox\Phi\left(\frac{t(\param)}{\asympsd(\param)}\right),\intBVMapprox\Phi\left(\frac{t(\param)}{\asympsd(\optparam)}\right)\right) = \bigo(\log\spcNobs/\spcNobs^{1/2}).
	$

	We have
	\begin{align}
		\lefteqn{W_s^s\left(\intBVMapprox\Phi\left(\frac{t(\param)}{\asympsd(\param)}\right),\intBVMapprox\Phi\left(\frac{t(\param)}{\asympsd(\optparam)}\right)\right) }\\
		&\leq \EE\left[\intBVMapprox\left|\Phi\left(\frac{t(\param)}{\asympsd(\param)}\right) -\Phi\left(\frac{t(\param)}{\asympsd(\optparam)}\right)\right|\right]^s\\
		&\leq \Vert \Phi \Vert_{L}^s \EE\left[\int_{\paramspace_{\spcNobs}}\left|\frac{t(\param)}{\asympsd(\param)}\left(1 - \frac{\asympsd(\param)}{\asympsd(\optparam)}\right)\right|\distBVM{\dee\param}{\spcobsrv}\right]^s, \label{eq:part4_startup}
	\end{align}
	where $\Vert \Phi \Vert_{L}$ denotes the Lipschitz constant of $\Phi$. 
	By \cref{lem:A3_asymsd_bounds}, the distance can be further bounded as
	\begin{align}
		\lefteqn{W_s\left(\intBVMapprox\Phi\left(\frac{t(\param)}{\asympsd(\param)}\right),\intBVMapprox\Phi\left(\frac{t(\param)}{\asympsd(\optparam)}\right)\right)} \\
		& \leq  \frac{\Vert \Phi \Vert_{L} C M \log^{1/2}\spcNobs}{\asympsd(\optparam)\spcNobs^{1/2}} \EE\left[\int_{\paramspace_{\spcNobs}}\left|\frac{t(\param)}{\asympsd(\param)}\right|\distBVM{\dee\param}{\spcobsrv}\right].
		\label{eq:partiv_bound_relative_err}
	\end{align}
	Notice that $(\spcobsrv,\spcnewrv)$ are jointly independent. Applying \cref{lem:A2,lem:A3_asymsd_bounds}, we have
	\begin{align}
		\sup_{\param \in \paramspace_{\spcNobs}}	\EE\left[\frac{t(\param)}{\asympsd(\param)}\right] 
		& = \sup_{\param \in \paramspace_{\spcNobs}}\EE\left[\frac{\spcNnew^{1/2}\{\statistic{\spcNnew}(\spcnewrv) - \asympmean{}(\param)\}}{\asympsd(\param)}\right]\\
		& = \sup_{\param \in \paramspace_{\spcNobs}}\frac{\spcNnew^{1/2}\{\EE\left[\statistic{\spcNnew}(\spcnewrv)\right] - \asympmean{}(\param)\}}{\asympsd(\param)}\\
		& = \sup_{\param \in \paramspace_{\spcNobs}}\frac{\spcNnew^{1/2}\{\asympmean{}(\optparam) - \asympmean{}(\param)\}}{\asympsd(\optparam)}\frac{\asympsd(\optparam)}{\asympsd(\param)}\\
		&\leq \left(\sup_{\param \in \paramspace_{\spcNobs}}\diffasympmean{}(\optparam)\frac{\spcNnew^{1/2}\{\optparam - \param\}}{\asympsd(\optparam)} + c_2\log\spcNnew/\spcNnew^{1/2}\right)\\
		& \hspace{1.5in} \times \left(1 +\frac{C M \log^{1/2}\spcNobs}{\asympsd(\optparam)\spcNobs^{1/2}} \right) \\
		& \leq C_1\log^{1/2}\spcNnew + C_2\log\spcNnew/\spcNnew^{1/2}  \\
		& \hspace{1in} + C_3\log\spcNnew/\spcNnew^{1/2} +C_4 \log^{3/2}\spcNnew/\spcNnew\\
		& \leq \tilde{C}\log^{1/2}\spcNnew. 
		\label{eq:partiv_bound_exps}
	\end{align}
	It then follows by Fubini's theorem, and \cref{eq:partiv_bound_exps,eq:partiv_bound_relative_err} that
	\begin{align}
		\lefteqn{W_s\left(\intBVMapprox\Phi\left(\frac{t(\param)}{\asympsd(\param)}\right),\intBVMapprox\Phi\left(\frac{t(\param)}{\asympsd(\optparam)}\right)\right)} \\
		& \leq  \frac{\Vert \Phi \Vert_{L} C M \log^{1/2}\spcNobs}{\asympsd(\optparam)\spcNobs^{1/2}} \tilde{C}\log^{1/2}\spcNnew\\
		&= C'\log\spcNobs/\spcNobs^{1/2}. 
	\end{align}

	\textbf{Part V}: 
	$ W_s\left(\intBVMapprox\Phi\left(\frac{t(\param)}{\asympsd(\optparam)}\right), \intBVMapprox\Phi\left(\tilde{Q}(\param)\right)\right)
	= \bigo(\log \spcNobs/\spcNobs^{1/2})$
	
	By definition of $t(\param)$ and $\tilde{Q}(\param)$, we have
	\begin{equation}
		\left|\frac{t(\param)}{\asympsd(\optparam)} -\tilde{Q}(\param)\right| =  \left|\frac{\spcNnew^{1/2}\{ \asympmean{}(\param)- \asympmean{}(\optparam)\}}{\asympsd(\optparam)} - \diffasympmean{}(\optparam)^\top\frac{\spcNnew^{1/2}\{\param - \optparam\}}{\asympsd(\optparam)}\right|.
	\end{equation}
	Applying \cref{lem:A2,lem:A3_asymsd_bounds}, we conclude that for some constant $c_2 > 0$,
	\begin{align}
		\lefteqn	{d^s_W\left(\intBVMapprox\Phi\left(\frac{t(\param)}{\asympsd(\optparam)}\right), \intBVMapprox\Phi\left(\tilde{Q}(\param)\right)\right)}\\
		& \leq \EE\left|\intBVMapprox\Phi\left(\frac{t(\param)}{\asympsd(\optparam)}\right) -\Phi\left(\tilde{Q}(\param)\right)\right|^s\\
		& \leq \EE\left[\int_{\paramspace_{\spcNobs}}\left|\Phi\left(\frac{t(\param)}{\asympsd(\optparam)}\right) -\Phi\left(\tilde{Q}(\param)\right)\right|\distBVM{\dee\param}{\spcobsrv}\right]^s\\
		&\leq \Vert \phi \Vert^s_{\infty} \EE\left[\int_{\paramspace_{\spcNobs}}\left|\frac{t(\param)}{\asympsd(\optparam)} -\tilde{Q}(\param)\right|\distBVM{\dee\param}{\spcobsrv}\right]^s	\\
		&\leq \Vert \phi \Vert^s_{\infty}c_2 \log^s \spcNobs/\spcNobs^{s/2} \EE\left[\distBVM{\paramspace_{\spcNobs}}{\spcobsrv}\right]^s\\
		&\leq \Vert \phi \Vert^s_{\infty}c_2 \log^s \spcNobs/\spcNobs^{s/2}.
	\end{align}
	
	\textbf{Part VI}: 
	$
	W_s\left( \intBVMapprox\Phi\left(\tilde{Q}(\param)\right), \Phi(Q) \right) = \bigo(\spcNobs^{-1/2}).
	$

	Letting $\vartheta \dist \mathcal{N}(\mle{}(\spcobsrv), \mlevar_{\optsym}/\spcNobs)$, we have
	\begin{align}
		\intBVMapprox_{\paramspace}\Phi\left(\tilde{Q}(\param)\right) & = \int_{\paramspace}\Phi\left(\tilde{Q}(\param)\right) \distBVM{\dee\param}{\spcobsrv} \\
		& = \EE\left[\Phi(\tilde{Q}(\vartheta))\right]\\
		& = \EE\left\{\Pr\left[Z \leq \frac{s_{\statistic{}\asympmean{}}(\spcnewrv,\optparam;\spcNnew) - \diffasympmean{}(\optparam)^\top\spcNnew^{1/2}(\vartheta - \optparam)}{\asympsd(\optparam)}\right]\right\}\\
		& =\EE\left\{\Pr\left[\underbrace{Z\asympsd(\optparam) + \diffasympmean{}(\optparam)^\top\spcNnew^{1/2}(\vartheta - \optparam)}_{=: Z' \dist \mathcal{N}(\mu_{Z'}, \sigma_{Z'}^2)} \leq s_{\statistic{}\asympmean{}}(\spcnewrv,\optparam;\spcNnew)\right]\right\}\\
		& = \Pr\left[\frac{Z' - \mu_{Z'}}{\sigma_{Z'}} \leq \underbrace{\frac{s_{\statistic{}\asympmean{}}(\spcnewrv,\optparam;\spcNnew) - \mu_{Z'}}{\sigma_{Z'}}}_{=: Q}\right],
	\end{align}
	where $\mu_{Z'} \defined \diffasympmean{}(\optparam)^\top\spcNnew^{1/2}(\mle{}(\spcobsrv) - \optparam)$ and $\sigma^2_{Z'} \defined \asympsd^2(\optparam) + \frac{\spcNnew}{\spcNobs}\diffasympmean{}(\optparam)^\top\mlevar(\optparam)\diffasympmean{}(\optparam)$. It remains to show that $W_s\left( \intBVMapprox\Phi\left(\tilde{Q}(\param)\right), \intBVMapprox_{\paramspace}\Phi\left(\tilde{Q}(\param)\right) \right) = \bigo(\spcNobs^{-1/2})$. We apply \cref{lem:density_mass_1} and \cref{assump:b:BVMrate} to get
	\begin{align}
		\lefteqn{W^s_s\left( \intBVMapprox\Phi\left(\tilde{Q}(\param)\right), \intBVMapprox_{\paramspace}\Phi\left(\tilde{Q}(\param)\right) \right) } \\
		& \leq \EE\left[\left|\int_{\paramspace \backslash \paramspace_{\spcNobs}}\Phi(\tilde{Q}(\param))\right|\distBVM{\dee\param}{\spcobsrv}\right]^s\\
		&\leq \EE\left[\left|1 - \distBVM{\paramspace_{\spcNobs}}{\spcobsrv}\right|^s\right]\\
		& \leq C\spcNobs^{-s/2}.
	\end{align}
	
	\textbf{Part VII}: $W_s(\Phi(Q),\Phi(Z)) = \bigo(\spcNobs^{-1/2})$.
	
	Since $\Phi$ is a Lipschitz-continuous function and $\Phi(\cdot) \in [0,1]$, there exists some constant $C> 0$ such that
	\[
	W_s(\Phi(Q),\Phi(Z)) \le C d_K(\Phi(Q),\Phi(Z)) \le C \Vert\Phi\Vert_{L}d_K(Q,Z). \label{eq:dK_bound_Ws}
	\]
	Let $Z_1, Z_2 \dist N(0,1)$ be independent standard normal random variables. For simplicity, we use shorthand notations $\sigma_1^2 \defined \sigma_1^2(\optparam) = \frac{\spcNnew}{\spcNobs}\diffasympmean{}(\optparam)^\top\mlevar(\optparam)\diffasympmean{}(\optparam)$ and $\sigma_0^2 \defined \asympsd^2(\optparam)$. Observe that $s_{\statistic{}\asympmean{}}(\spcnewrv,\optparam;\spcNnew)$ and  $\spcNobs^{1/2}\{\mle{}(\spcobsrv)-\optparam)\}$ are jointly independent. Let $Z \defined \frac{\sigma_0Z_1 - \sigma_1Z_2}{(\sigma_0^2 + \sigma_1^2)^{1/2}} \dist \mathcal{N}(0,1)$ and $\tilde{Z}_1 \defined \frac{\sigma_0 Z_1}{(\asympsd_0^2  + \sigma_1^2)^{1/2}} \dist \distNorm(0,  \sigma_0^2/( \sigma_0^2 +  \sigma_1^2))$ with CDF $F_{\tilde{Z}}$. Let $\tilde{\sigma}_1^2$ be the variance of $\tilde{Z}_1$. By definition, $\BLnorm{F_{\tilde{Z}_1}} = \Vert F_{\tilde{Z}_1} \Vert_{\infty} + \Vert F_{\tilde{Z}_1} \Vert_{L} = 1 + 1/\sqrt{2\pi\tilde{\sigma}_1^2} < \infty$. It then follows by \cref{lem:bound_dK} that, 
	\begin{align}
		d_K\left(Q, Z\right)& = d_K\left(\frac{s_{\statistic{} \asympmean{}}(\spcnewrv,\optparam;\spcNnew) - \diffasympmean{}(\optparam)^\top\spcNnew^{1/2}\{\mle{}(\spcobsrv)- \optparam\}}{(\asympsd^2(\optparam) + \sigma_1^2(\optparam))^{1/2}}, \frac{\sigma_0Z_1 - \sigma_1Z_2}{(\sigma_0^2 + \sigma_1^2)^{1/2}} \right)\\
		& \leq  d_K\left(\frac{s_{\statistic{} \asympmean{}}(\spcnewrv,\optparam;\spcNnew) }{(\asympsd_0^2 + \sigma_1^2)^{1/2}}, \frac{\asympsd_0Z_1 }{(\asympsd_0^2+ \sigma_1^2)^{1/2}}\right) \\
		&\qquad + \BLnorm{F_{\tilde{Z}_1}}d_{\BL}\left(\frac{ \diffasympmean{}(\optparam)^\top\spcNnew^{1/2}\{\mle{}(\spcobsrv)- \optparam\}}{(\asympsd_0^2 + \sigma_1^2)^{1/2}}, \frac{\sigma_1 Z_2}{(\asympsd_0^2  + \sigma_1^2)^{1/2}}\right)\\
		& = d_K\left(\frac{s_{\statistic{} \asympmean{}}(\spcnewrv,\optparam;\spcNnew)}{\asympsd_0}, Z_1\right) \\
		& \hspace{1in}+ \BLnorm{F_{\tilde{Z}_1}}d_{\BL}\left(\frac{\diffasympmean{}(\optparam)^\top\spcNnew^{1/2}\{\mle{}(\spcobsrv)- \optparam\}}{\sigma_1},Z_2\right). \label{eq:dKbound}
	\end{align}
	By \cref{assump:asymp_norm_stat_rate,assump:a:mle_convto_true_rate,assump:MLE_conv_rate}, the two terms in \cref{eq:dKbound} are of order $\bigo(\spcNobs^{-1/2})$. It follows by the definition of $\spcNobs = \lceil \spcprop \numobs\rceil$ that $=\bigo(\spcNobs^{-1/2})= \bigo(\numobs^{-1/2})$. Combining \cref{eq:dK_bound_Ws} and above facts together, we obtain
	\[
	W_s(\Phi(Q),\Phi(Z)) \le C' d_K\left(Q, Z\right) \le \bigo(\numobs^{-1/2}).	\label{eq:upperbound_dK}
	\]	
	Hence, we get 
	$
	W_s(\Phi(Q),\Phi(Z))  = \bigo(\numobs^{-1/2}).
	$
	Now combining all parts from \textbf{I} to \textbf{VIII}, we have for sufficiently large constant $C>0$ and smoothing constant $s\ge 2$
	\[
	W_s\left(\spcpvalue{\datarv}, U\right) \leq C\log\numobs/\numobs^{1/2}.
	\]
	
	Therefore, assuming a smooth model which has large finite moments (large $s$) allows the rate for Kolmogorov distance to be close to $\bigo(\log\spcNobs/\spcNobs^{1/2})$.

	By \citep[Theorem 1.2]{Bouchitte:2007}, for constant $C_s > 0$ depending on $s \geq 2$ %
	\begin{align}
		d_K(\spcpvalue{\datarv}, U) \leq C_s W_s^{\frac{s}{1+s}}(\spcpvalue{\datarv}, U).
	\end{align}
	The proof is immediate by \cref{THM:SINGLESPC_RATE}.
\end{proof}

\subsection{Proof of \protect\cref{RMK:GAUSSIAN_MSE}}
\label{appx:MSE_gaussianmod}
Consider $\statistic{}(\datarv,\param) = \frac{1}{\numobs}\sum_{i = 1}^{\numobs}(\obsrv{i} - \param)^2$. 
For the Gaussian location model, we have MLE $\mle{}(\spcobsrv)  = \frac{1}{\spcNobs}\sum_{i = 1}^{\spcNobs}\spcobsobs{i} \dist \distNorm (\optparam, \truesigma^2/\spcNobs)$. 
Further, the scaled and centered statistics are asymptotically normal:
\begin{align}
	\sqrt{\spcNnew}\{\statistic{}(\spcreprv,\param) - \sigma^{2}\}
	& = \frac{1}{{\sqrt{\spcNnew}}}\left\{\sum_{i = 1}^{\spcNnew}(\spcrepobs{i} - \param)^2 - \sigma^{2}\spcNnew\right\}  \\
	&\dist \frac{\chi^2_{\spcNnew} - \spcNnew}{\sqrt{\spcNnew}}\cdot \sigma^2 \convD \distNorm(0, \sigma^4)  \label{eq:symp:MSE-pred}
\end{align}
and
\begin{align}
	{\sqrt{\spcNnew}\{\statistic{}(\spcnewrv,\optparam) - \truesigma^{2}\}} 
	& = \frac{1}{{\sqrt{\spcNnew}}}\left\{\sum_{i = 1}^{\spcNnew}(\spcnewobs{i} - \optparam)^2  - \truesigma^{2}\spcNnew\right\} \\
	& \dist \frac{\chi^2_{\spcNnew} - \spcNnew}{\sqrt{\spcNnew}}\cdot \truesigma^2 \convD \distNorm(0, \truesigma^4). \label{eq:symp:MSE-ho}
\end{align}
By definition of single SPC $\p$-value
\begin{align}
	\spcpvalue{\datarv} & =  1 - \int \Pr\left[\statistic{}(\spcreprv,\param)  \ge \statistic{}(\spcnewrv,\param)\mid \param \right]\postdist{}(\dee \param \mid \spcobsrv)\\
	& = 1 - \int \Pr\left[\sqrt{\spcNnew}\statistic{}(\spcreprv,\param)  \ge \sqrt{\spcNnew}\statistic{}(\spcnewrv,\param) \mid \param\right]\distBVM{\dee \param}{\spcobsrv} + \littleoP(1).\label{eq:pval}
\end{align}
Observe that $\sqrt{\spcNnew}\statistic{}(\spcnewrv,\param) $ is not centralized. We write
\[
\sqrt{\spcNnew}\statistic{}(\spcnewrv,\param) = \sqrt{\spcNnew}\statistic{}(\spcnewrv,\optparam) + \sqrt{\spcNnew}(\param - \optparam)^2 + 2(\param - \optparam)\frac{1}{\sqrt{\spcNnew}}\sum_{i = 1}^{\spcNnew}(\spcnewobs{i} - \optparam)^2. \label{eq:splitTnew}
\]
The first term in \cref{eq:splitTnew} follows a scaled $\chi^2_{\spcNnew}$ distribution. For the second term, we have
\begin{align}
	\sqrt{\spcNnew}(\param - \optparam)^2 & = \sqrt{\spcNnew}(\param - \mle{}(\spcobsrv))^2  + \sqrt{\spcNnew}(\mle{}(\spcobsrv) - \optparam)^2 \\
	& \qquad + 2\sqrt{\spcNnew}(\param - \mle{}(\spcobsrv))(\mle{}(\spcobsrv) - \optparam)\\
	& \dist \frac{\spcNnew}{\spcNobs}\frac{1}{\sqrt{\spcNnew}}(\sigma^2\chi_1^2 + \truesigma^2\chi_1^2 + 2 \sigma \truesigma\chi_1^2),
	\label{eq:splitT_secterm}
\end{align}
which converges to 0 in probability as $\numobs \to \infty$. Similarly for the last term in \cref{eq:splitTnew}, we have
\begin{align}
	2(\param - \optparam)\frac{1}{\sqrt{\spcNnew}}\sum_{i = 1}^{\spcNnew}(\spcnewobs{i} - \optparam) & = 2\bigg\{\sum_{i = 1}^{\spcNnew}\left(\param - \mle{}(\spcobsrv)\right)(\spcnewobs{i} - \optparam)\\
	& \qquad + \sum_{i = 1}^{\spcNnew}\left(  \mle{}(\spcobsrv) - \optparam \right)(\spcnewobs{i} - \optparam)\bigg\}\\
	& = \frac{2\truesigma}{\sqrt{\spcNnew\spcNobs}}\left(\sigma\chi^2_{\spcNnew} + \truesigma\chi^2_{\spcNnew}\right)\\
	& = \littleoP(1). \label{eq:splitT_thirdterm}
\end{align}
Substituting \cref{eq:splitT_secterm,eq:splitT_thirdterm} into \cref{eq:splitTnew} yields 
\[
\sqrt{\spcNnew}\statistic{}(\spcnewrv,\param) = \sqrt{\spcNnew}\statistic{}(\spcnewrv,\optparam) + \littleoP(1).
\]
Let $Z_1, Z_2$ be independent standard normal random variables and $\EE_{\param}[\cdot]$ denote the expectation with respect to the posterior distribution. 
By the DCT and \cref{eq:symp:MSE-ho,eq:symp:MSE-pred}, we can rewrite the $\p$-values in \cref{eq:pval} as 
\begin{align}
	\spcpvalue{\datarv} & =  1 - \EE_{\param}\left[\Pr\left[\sqrt{\spcNnew}\statistic{}(\spcreprv,\param)  \ge \sqrt{\spcNnew}\statistic{}(\spcnewrv,\param) \mid \param\right]\right] +  \littleoP(1)\\
	& =  1 - \EE_{\param}\bigg\{\Pr\bigg[\frac{\sqrt{\spcNnew}\{\statistic{}(\spcreprv,\param) -\sigma^2\}  }{\sigma^2}\ge \frac{\sqrt{\spcNnew}\{\statistic{}(\spcnewrv,\param) - \truesigma^2\} }{\truesigma^2}\frac{\truesigma^2}{\sigma^2} \\
	& \hspace{5cm} +\frac{\sqrt{\spcNnew}\left(\truesigma^2 - \sigma^2\right)}{\sigma^2} \mid \param\bigg]\bigg\} +  \littleoP(1)\\
	& = 1 - \Pr\left[Z_1 \ge \sdratio^2 Z_2 + \sqrt{\spcNnew}(\sdratio^2 - 1) \right] +  \littleoP(1)\\
	& =  \Phi\left(\sdratio^2 Z_2 + \sqrt{\spcNnew}(\sdratio^2 - 1)\right)+  \littleoP(1).
\end{align}                                                                                                             
Letting $\sdratio = 1$ we have $\spcpvalue{\datarv}  =  \Phi(Z_2) + \littleoP(1) \convD \distUnif$ and therefore 
\[
\Pr\left[\spcpvalue{\datarv} < \frac{\alpha}{2}\right] = \frac{\alpha}{2}.
\] 
For $\sdratio \neq 1$, followed by a similar argument in the proof of \cref{THM:SINGLESPC_RATE}, we have
\begin{align}
	\Pr\left[\spcpvalue{\datarv} < \frac{\alpha}{2}\right] & = \Pr\left[\Phi\left(\sdratio^2 Z_2 + \sqrt{\spcNnew}(\sdratio^2 - 1)\right)+  \littleoP(1)<  \frac{\alpha}{2}\right] \\
	& = \Pr\left[\frac{z_{ \frac{\alpha}{2}}- \sqrt{\spcNnew}(\sdratio^2 - 1) }{\sdratio^2}<  Z_2 + \littleoP(1)\right]\\
	& = 1 - \Phi\left(\frac{z_{ \frac{\alpha}{2}}- \sqrt{\spcNnew}(\sdratio^2 - 1)-  \littleoP(1) }{\sdratio^2}\right)\\
	& \overset{P}{\rightarrow} 1, 
\end{align}
where the last inequality follows by taking $\spcNnew \rightarrow \infty$ and applying DCT. Similarly, we can derive for $\sdratio \neq 1$
\begin{align}
	\Pr\left[\spcpvalue{\datarv} > 1 - \frac{\alpha}{2}\right]
	&  = \Phi\left(\frac{z_{\frac{\alpha}{2}} - \sqrt{\spcNnew}(\sdratio^2-1)- \littleoP(1)}{\sdratio^2} \right) \overset{P}{\rightarrow} 0.
\end{align}

\subsection{Proof of Theorem \protect\ref{THM:DIVIDEDSPC}}
\label{appx:dspcproofs1}

We first introduce some notation. 
Let $\empPn{n} \defined n^{-1}\sum_{i = 1}^{n}\delta_{\obsrv{i}}$ be the empirical measure of $\obsrv{1}, \ldots, \obsrv{n}$. The scaled empirical process is defined as
$\empGn{n}{} \defined \sqrt{n}(\empPn{n} - P)$. Given a collection of functions, the empirical process $\empPn{n}{}$ induces a map from $\fnclass$ to $\mathbb{R}$ defined as $f \mapsto \empPn{n}{f} \defined n^{-1/2}\sum_{i = 1}^nf(\obsrv{i})$ for some $f \in \fnclass$. Consider the collection of indicator functions $\indclass \defined \{\indicatorfn_{(-\infty, x]}, x\in \mathbb{R}\}$. The Kolmogorov distance can be rewritten as
\begin{equation}
	d_{K}(\empPn{n}{}, P) = \sup_{f\in \indclass} (\empPn{n}{} - P)f.
\end{equation}
Recall that divided SPCs generate $\p$-values by applying the Kolmogorov-Smirnov test to $\dspcK$ single SPC $\p$-values obtained from $\dspcK$ disjoint subsets. We require a uniform convergence on Kolmogorov distance in the proof of \cref{THM:DIVIDEDSPC}, which we present in the following lemma.
\blem(Uniform convergence on Kolmogorov distance)
\label{lem:unif_conv_KolDist}
If $\obsrv{1}, \obsrv{2}, \ldots$ i.i.d $\sim P$ a continuous distribution, then the empirical distribution uniformly converges to the true distribution:
\begin{equation}
	\sup_{P} 	d_{K}\left(\sup_{f\in\indclass}\empGn{n}{f} ,K \right) \rightarrow 0,
\end{equation}
where $K \dist\distKS{\cdot}$, the Kolmogov distribution with CDF $\distKS{t} \defined 1 - 2\sum_{i = 1}^{\infty} (-1)^{i-1} e^{-2i^2t}$. 
\elem
\begin{proof}
	Consider the class of indication functions $\indclass = \{\indicatorfn_{(-\infty, x]}, x \in \mathbb{R}\}$. It's easy to show that $\indclass$ has Vapnik-Chervonenkis (VC) dimension $\VCnum{\indclass}= 2$. Following the discussions of \citet[Theorem 14.13]{Ledoux:1991}, such VC class $\indclass$ belongs to uniform Donsker classes. It then follows from \citet[Definition 2.3 and Theorem 2.1]{Sheehy:1992} that an equivalent definition of uniform Donsker class is, for any $t > 0$
	\begin{equation}
		\label{eq:unifDonsker}
		\lim\limits_{n\rightarrow\infty}\sup_{P}\Pr^*\left\{\sup_{f\in\indclass}(\empGn{n}{} - \BB{n})(f)  > t\right\} = 0,
	\end{equation}
	where $\BB{n}$ is a sequence of coherent  $P$-Brownian bridge processes. The conclusion immediately follows by the fact that the supremum of Brownian bridge follows the Kolmogorov distribution \citep{Boukai:1990}.
\end{proof}

\begin{proof}[Proof of \cref{THM:DIVIDEDSPC}]

	Suppose $\dspcK = \numobs^{\beta}$ for $0<\beta<\frac{\gamma}{\gamma +1}$. By definition we have $\dspcnumobs = \lceil \frac{\numobs}{\dspcK}\rceil = \lceil \numobs^{1-\beta}\rceil$. It follows from \cref{assump:singleSPC_rate} that for constant $C' > C$ 
	\begin{align}
		\label{eq:Pk_F0}
		d_K\big(\pvaldist, \distUnif\big) < C\dspcnumobs^{-\gamma/2}  <  C'\numobs^{(\beta-1)\frac{\gamma}{2}}.
	\end{align}
	Consider the Kolmogorov distance between $\pvalempdist$ and $\distUnif$
	and fix some $t > 0$. 
	It follows from the triangle inequality that %
	\begin{align}
		{\Pr\left[\dspcK^{1/2}d_K\big(\pvalempdist, \distUnif\big) > t \right] }
		&\leq \Pr\left[\dspcK^{1/2}d_K\big(\pvaldist, \distUnif\big) + \dspcK^{1/2}d_K\big(\pvalempdist, \pvaldist\big) > t\right]\\
		& \leq \Pr\left[C'\numobs^{\beta/2} \numobs^{(\beta-1)\gamma/2}+ \dspcK^{1/2}d_K\big(\pvalempdist, \pvaldist\big) > t\right]\\
		& = \Pr\left[\dspcK^{1/2}d_K\big(\pvalempdist, \pvaldist\big)> s - C'\numobs^{\frac{1}{2}\left(\beta(1+\gamma)-\gamma\right)}\right].
		\label{eq:Ubound}
	\end{align}
	Note that $\beta < \frac{\gamma}{\gamma + 1}$ implies $\frac{1}{2}\left(\beta(1+\gamma)-\gamma\right) < 0$. By \cref{lem:unif_conv_KolDist} and taking $\numobs \to \infty$, we have the last line in \cref{eq:Ubound} converges to $1 - \distKS{t}$ in the Kolmogorov metric.
	On the other hand, by applying the triangle inequality, we have
	\begin{align}
		\Pr\left[\dspcK^{1/2}d_K\big(\pvalempdist, \distUnif\big) > t\right] & \geq     \Pr\left[\dspcK^{1/2}d_K\big(\pvalempdist, \pvaldist\big) - \dspcK^{1/2}d_K\big(\pvaldist, \distUnif\big) > t\right] \\ 
		& \geq\Pr\left[ \dspcK^{1/2}d_K\big(\pvalempdist, \pvaldist\big)> t + C'\numobs^{\frac{1}{2}\left(\beta(1+\gamma)-\gamma\right)}\right].
		\label{eq:Lbound}
	\end{align}
	Combining \cref{eq:Ubound,eq:Lbound} and \cref{lem:unif_conv_KolDist} yields 
	\[
	\left|\Pr\left[\dspcK^{1/2}d_K\big(\pvalempdist, \distUnif\big) > t\right] - \left(1 - \distKS{t}\right)\right|\overset{\dspcK \rightarrow \infty}{\longrightarrow} 0.
	\label{eq:dKconvtoKSdist}
	\]

\end{proof}

\subsection{Proof of Theorem \protect\ref{THM:DSPC_POWER}}
\label{appx:dspcproofs2}
\begin{proof}
	Fix any $t > 0$. 
	By assumption, there exists an $\epsilon > 0$ and $\bar{n}(\epsilon)$ such that for any $n_{\dspcK} > \bar{n}(\epsilon)$, 
	$
	d_K(\pvaldist, \distUnif) > \epsilon
	$.
	Together with the triangle inequality, we have 
	\begin{align}
		\Pr\left[\sqrt{\dspcK}d_K(\pvalempdist, \distUnif) > t \right] 
		& \geq \Pr\left[  \sqrt{\dspcK}d_K(\pvaldist, \distUnif)- \sqrt{\dspcK}d_K(\pvalempdist,\pvaldist) > t \right] \\
		&\geq \Pr\left[\sqrt{\dspcK}d_K(\pvalempdist,\pvaldist)< \sqrt{\dspcK}\epsilon - \qdistKS{\alpha}   \right].
		\label{eq:power_tri_ineq}
	\end{align}
	It follows from \cref{lem:unif_conv_KolDist} that %
	\begin{equation}
		\liminf\limits_{\dspcK \rightarrow \infty}\sup_{P}\Pr\left[\sqrt{\dspcK}d_K(\pvalempdist, P)  < \sqrt{\dspcK}\epsilon - t \right] = 1.
		\label{eq:power_lem3.3}
	\end{equation}
	Combining \cref{eq:power_tri_ineq,eq:power_lem3.3}  yields
	\begin{equation}
		\liminf\limits_{\dspcK \rightarrow \infty}\Pr\left[\sqrt{\dspcK}d_K\big(\pvalempdist, \distUnif\big) > t \right] = 1.
	\end{equation}
	Since $\liminf \le \lim$ and probabilities are upper bounded by 1, we have
	\begin{equation}
		\lim\limits_{\dspcK \rightarrow \infty}\Pr\left[\sqrt{\dspcK}d_K\big(\pvalempdist, \distUnif\big) > t \right] = 1.
	\end{equation}

\end{proof}

\section{Additional figures and tables}
\renewcommand{\thefigure}{\Alph{section}.\arabic{figure}}
We present additional figures and tables for the previous simulation study and experiments in this section. In Section\ref{appx:Gaussian location model}, we provide a simulation study of all candidate checks  under Gaussian location model. Some additional figures for previous Poisson model simulation study are displayed in Section \ref{appx:pois_add_figures}. For additional figures of Gaussian hierarchical simulation study, see Section \ref{appx:gauss_hier_figures}. We finally include some additional figures for the airline delays experiment in Section \ref{appx:airlines_figs}.
\subsection{Gaussian location model}
\label{appx:Gaussian location model}
Similar to the discussions in the Poisson model study in \cref{sec:Poisson model}, we investigate all the candidate checks with a Gaussian location model. Consider
\begin{equation}
	\label{mod:gaussian}
	\begin{split}
		&\obsrv{} \dist \distNorm(\param, 1)\\
		&\param \dist \distNorm(0, 10000).
	\end{split}
\end{equation}
We design two generative models to indicate two extreme cases of model mismatches. The assumed model in (\ref{mod:gaussian}) is well-specified when the data is generated from $P_{o}: \obsrv{} \dist \distNorm(0, 1)$ and misspecified if we generate from $P_{o}: \obsrv{} \dist \distNorm(0, 15^2)$. We implement all checking methods with the following statistics: (i) empirical mean $\statistic{\numobs}(\data_{\numobs}) \defined \frac{1}{\numobs}\sum_{i = 1}^{\numobs}\obs{i}$, (ii) 2nd moment $\defined \frac{1}{\numobs}\sum_{i = 1}^{\numobs}\obs{i}^2$, (iii) $75$th quantile $\defined \arg\min_{\statistic{}}|\frac{1}{\numobs}\sum_{i = 1}^{\numobs}\indicatorfn_{\{\obs{i} \le \statistic{}\}} - 0.75|$ and (iv) mean squared error (MSE) $\defined \frac{1}{\numobs}\sum_{i = 1}^{\numobs}(\obs{i} - \EE[\datarv_i\mid \param])^2$. All statistics are asymptotically normal distributed as required by \cref{assump:asymp_norm_stat_rate}. 

\begin{figure}[tp]
\centering
\subfloat[ mean]{\label{fig:gauss_props:mean_ts}\includegraphics[width=65mm]{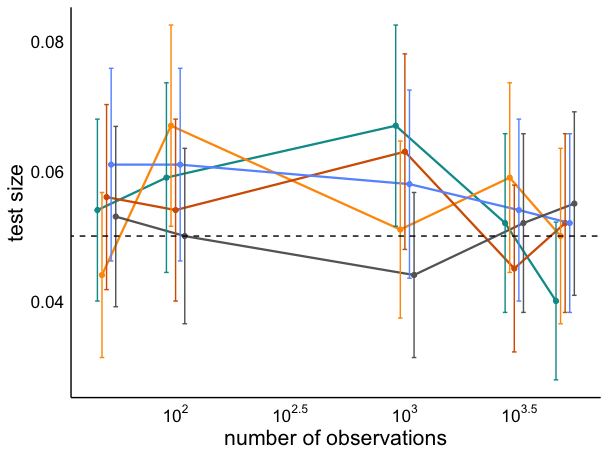}}
\subfloat[ mean]{\label{fig:gauss_props:mean_pw}\includegraphics[width=65mm]{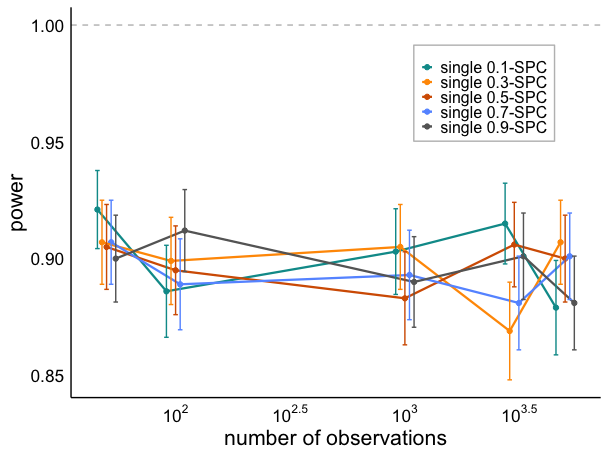}}
\\
\subfloat[2nd moment]{\label{fig:gauss_props:secmo_ts}\includegraphics[width=65mm]{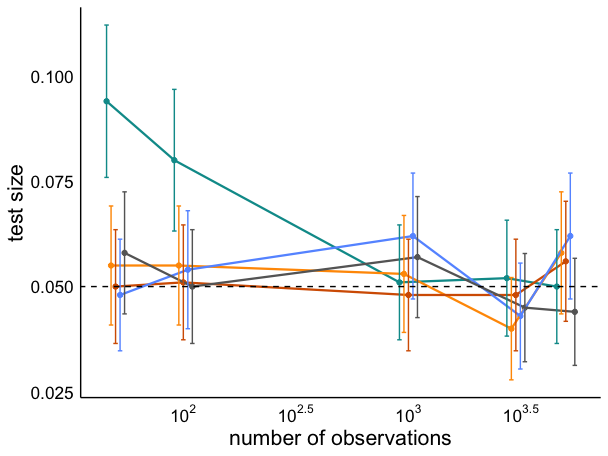}}
\subfloat[2nd moment]{\label{fig:gauss_props:secmo_pw}\includegraphics[width=65mm]{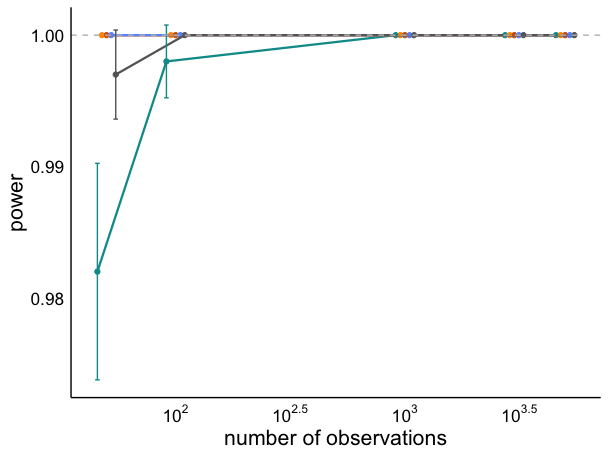}}
\caption{Test size and power plots in the log scale for single SPC with different split proportions $\spcprop \in  \{0.1, 0.3, 0.5, 0.7, 0.9\}$, under the Gaussian location model. The vertical bars refer to the confidence interval of each estimate. All estimates of test size and power are computed by 1000 replicate observed data drawn from the generative model. We display experiments for $\numobs \in \{50, 100, 1000, 3000, 5000\}$ to show how test size and power change as number of observations increases. The dashed horizontal line in test size plots indicates the significance level $\alpha = 0.05$ and the dashed line in power plot shows the optimal power 1.}
\label{figur:gauss_props}
\end{figure}

\paragraph{Single SPCs} \cref{figur:gauss_props} verifies the results in \cref{THM:SINGLESPC} that when $\trueasympmean = \asympmean{}(\optparam)$, with mean statistic, the asymptotic power of single SPC depends on $\rho^2 = \frac{\spcprop\truesigma^2 + (1-\spcprop)\truesigma^2/\spcNnew}{\spcprop\asympsd(\optparam)^2 + (1-\spcprop)\asympsd(\optparam)^2/\spcNnew} = \truesigma^2/\asympsd(\optparam)^2$. The asymptotic power is independent of $\spcprop$ in this case. For 2nd moment statisitic, single SPCs with all proportions successfully capture the major mismatches and $\spcprop$ does not affect the performance when $\trueasympmean \neq \asympmean{}(\optparam)$.

\begin{figure}[tp]
\centering
\subfloat[75th quantile]{\label{gauss_props:q75_ts}\includegraphics[width=70mm]{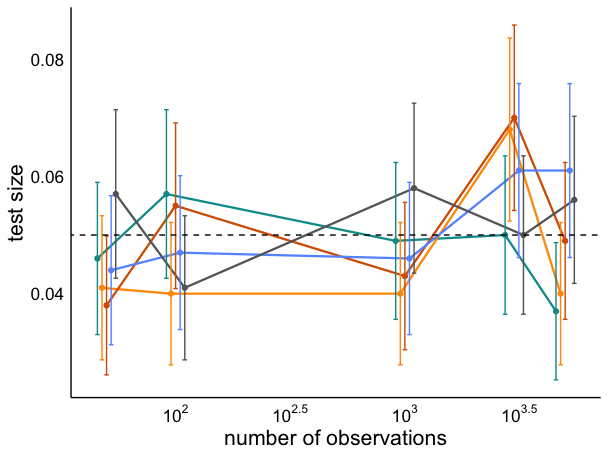}}
\subfloat[75th quantile]{\label{gauss_props:q75_pw}\includegraphics[width=70mm]{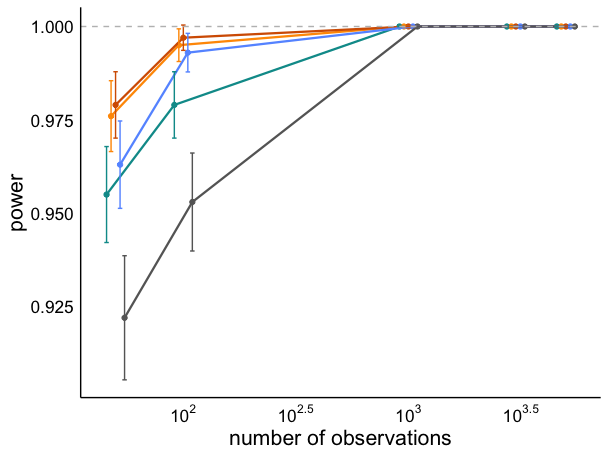}}
\\
\subfloat[MSE]{\label{gauss_props:mse_ts}\includegraphics[width=70mm]{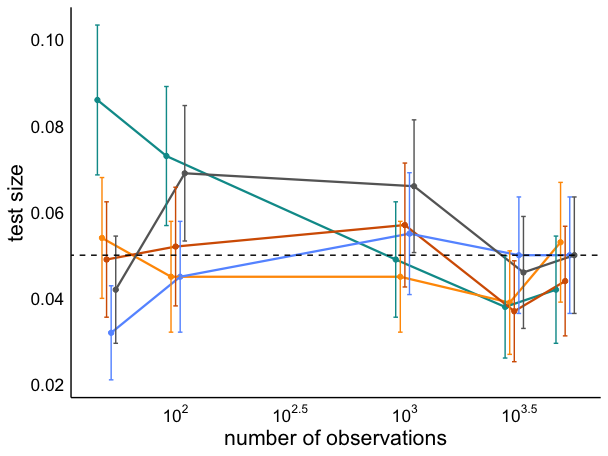}}
\subfloat[MSE]{\label{gauss_props:mse_pw}\includegraphics[width=70mm]{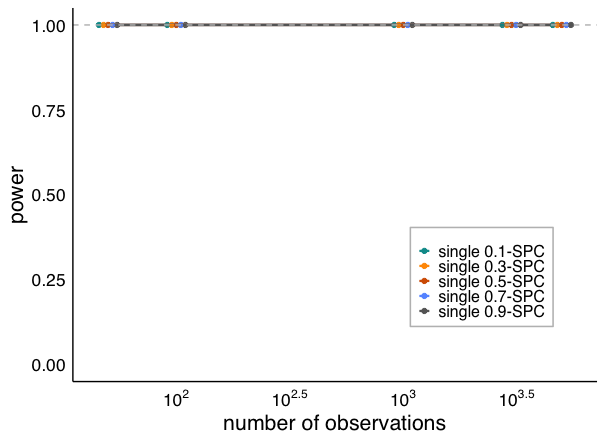}}
\caption{Test size and power plots (in log scale) for single SPC with different split proportions $\spcprop \in  \{0.1, 0.3, 0.5, 0.7, 0.9\}$. See caption for \cref{figur:gauss_props} for further explanation. }
\label{fig:gauss_props_appx}
\end{figure}

However, for small sample regime, a small split proportion such as $\spcprop = 0.1$ is not recommended due to a limited number of split observed data in computing posterior distributions. In \cref{fig:gauss_props:secmo_ts}, when $\numobs = 50$, the single SPC with $\spcprop = 0.1$ computes the posterior distribution with only $\spcNobs = 5$ samples, which tends to produce overconfident $\p$-values. \cref{fig:gauss_props_appx} shows similar results for $75$th quantile and MSE statistic with different proportions of single SPC.

\begin{figure}[tp]
\centering
\subfloat[ mean]{\label{fig:gauss_dspcprops:mean_ts}\includegraphics[width=65mm]{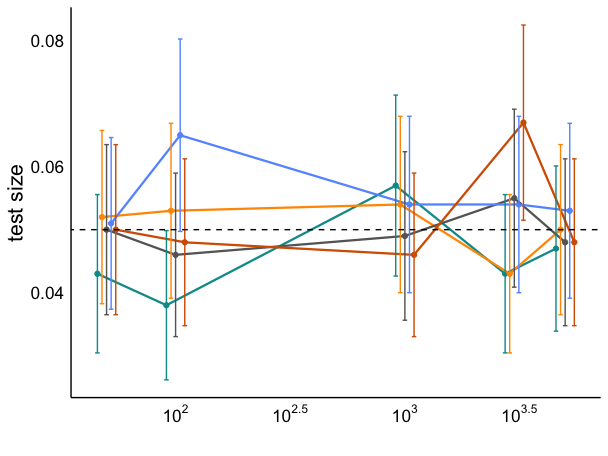}}
\subfloat[ mean]{\label{fig:gauss_dspcprops:mean_pw}\includegraphics[width=65mm]{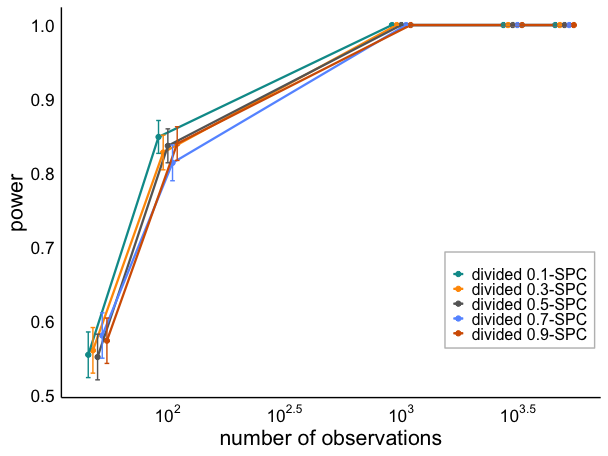}}
\\
\subfloat[2nd moment]{\label{fig:gauss_dspcprops:secmo_ts}\includegraphics[width=65mm]{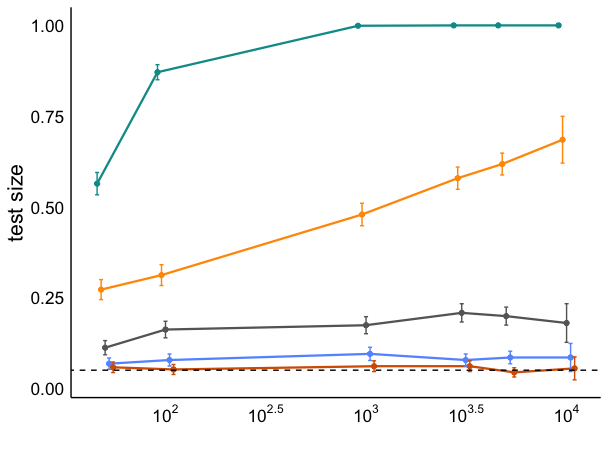}}
\subfloat[2nd moment]{\label{fig:gauss_dspcprops:secmo_pw}\includegraphics[width=65mm]{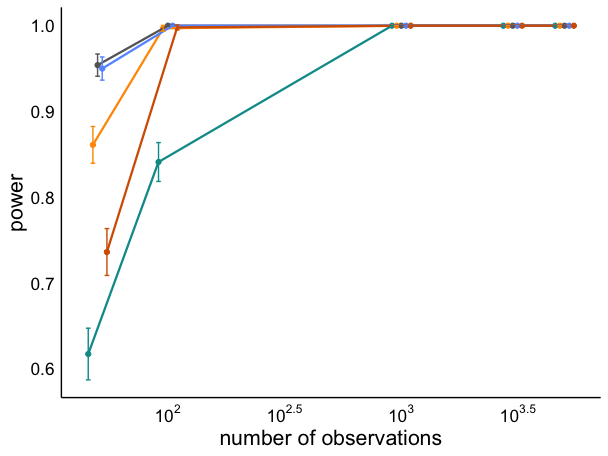}}\\
\subfloat[ 75th quantile]{\label{fig:gauss_dspcprops:q75_ts}\includegraphics[width=65mm]{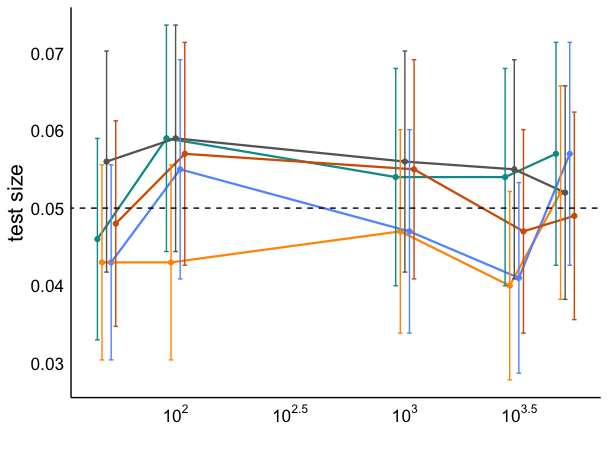}}
\subfloat[ 75th quantile]{\label{fig:gauss_dspcprops:q75_pw}\includegraphics[width=65mm]{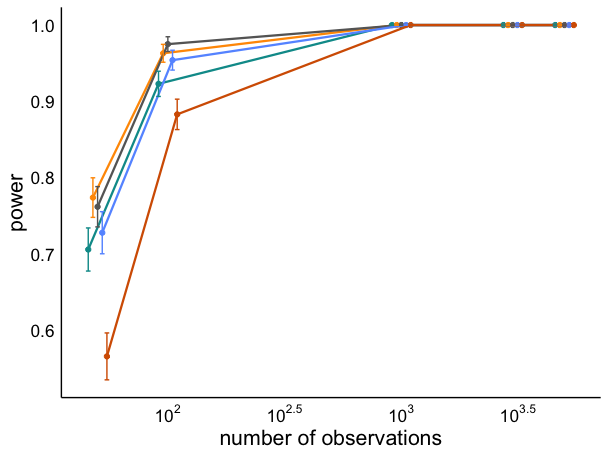}}
\caption{Test size and power plots in the log scale for divided SPC with identical $\dspcK = \numobs^{0.49}$, varying split proportions $\spcprop \in  \{0.1, 0.3, 0.5, 0.7, 0.9\}$, under the Gaussian location model. We include experiments for $\numobs \in \{50, 100, 1000, 3000, 5000\}$. }
\label{figur:gauss_dspcprops}
\end{figure}

\paragraph{Divided SPCs}
The choices of split proportion $\spcprop$ are of significance to the performance of divided SPCs in a small-data regime, and takes the same role with large datasets as in single SPC does. Given a sample of size $\numobs = 5000$, \cref{fig:gauss_dspcprops:secmo_ts} shows that the divided SPC with $\spcprop = 0.1, 0.3$ blows up the Type I error. The reason is that the data used to compute the posterior distribution only contains $\spcNobs = 7$ observations for $\spcprop = 0.1$ and $\spcNobs = 23$ for $\spcprop = 0.3$. Hence, we recommend choosing moderate values for proportions such as $\spcprop \in \{0.5, 0.7\} $ for divided SPC when the data set is not large. 

\begin{figure}[tp]
\centering
\subfloat[ 2nd moment]{\label{fig:gauss_dspcKs:ts}\includegraphics[width=65mm]{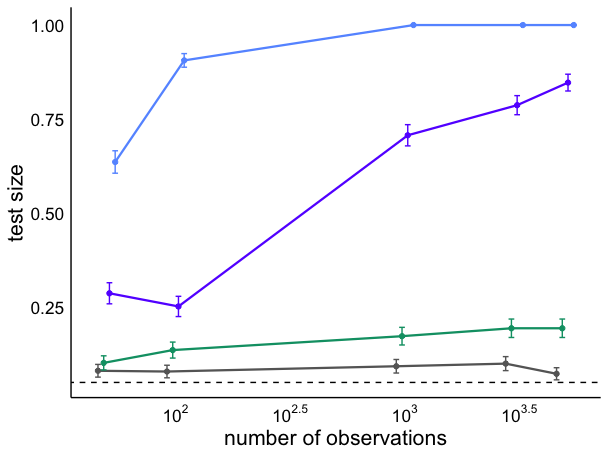}} 
\subfloat[ 2nd moment]{\label{fig:gauss_dspcKs:pw}\includegraphics[width=65mm]{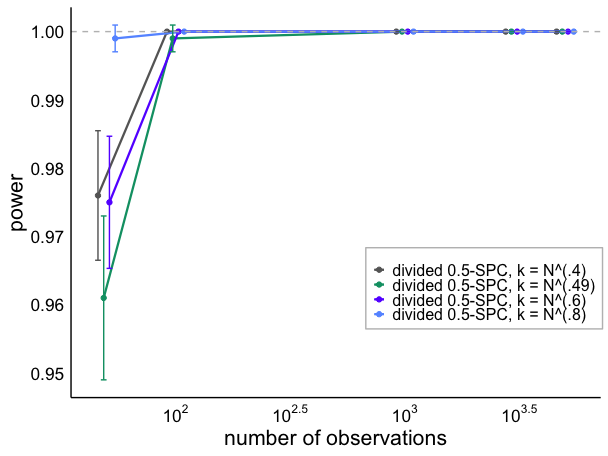}}\\
\subfloat[ 75th quantile]{\label{fig:gauss_dspcKs_ts_q75}\includegraphics[width=65mm]{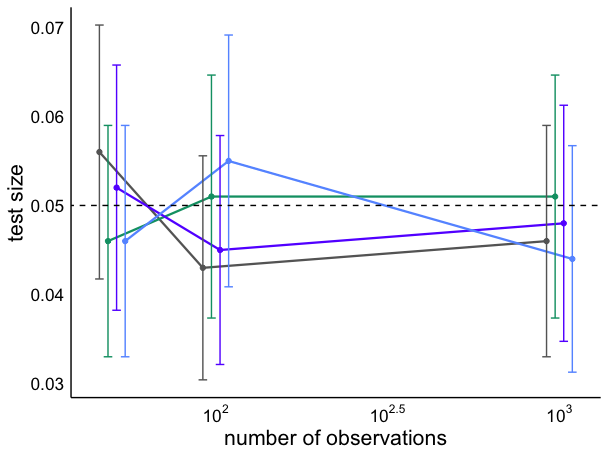}}
\subfloat[ 75th quantile]{\label{fig:gauss_dspcKs_pw_q75}\includegraphics[width=65mm]{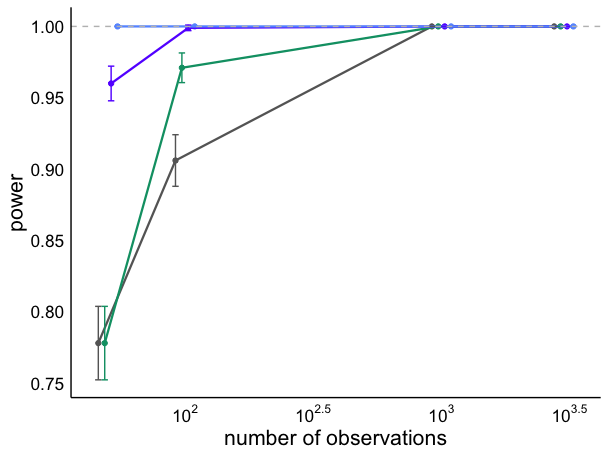}}\\
\subfloat[mean]{\label{fig:gauss_dspcKs_ts_mean}\includegraphics[width=65mm]{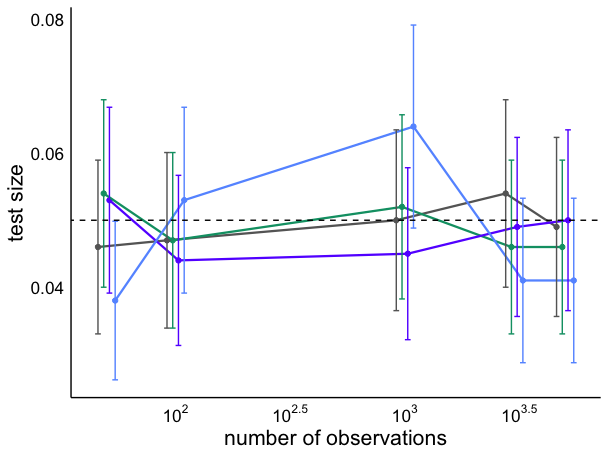}}
\subfloat[mean]{\label{fig:gauss_dspcKs_pw_mean}\includegraphics[width=65mm]{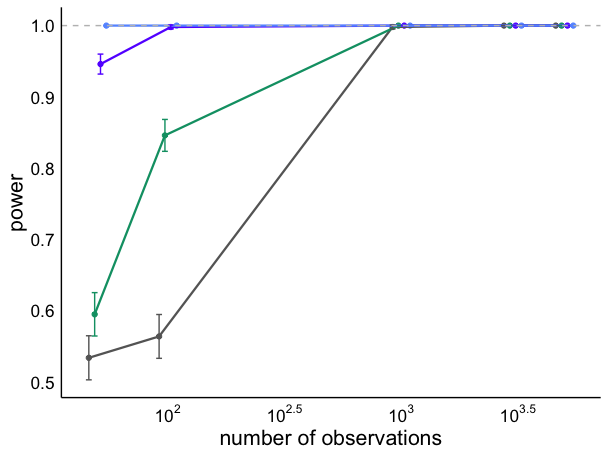}}
\caption{Test size and power plots in the log scale with 2nd moment and 75th quantile statistics for divided SPC under the Gaussian location model. Divided SPC $\p$-values are computed with identical split proportions $\spcprop= 0.5$ with varying number of folds $\dspcK \in  \{\numobs^{0.4},\numobs^{0.49}, \numobs^{0.6}, \numobs^{0.8}\}$. We include experiments for $\numobs \in \{50, 100, 1000, 3000, 5000\}$. }
\label{figur:gauss_dspcKs}
\end{figure}

\paragraph{Effective number of folds $\dspcK$} We show different choices of $\dspcK$ with various statistics in \cref{figur:gauss_dspcKs}.  
In \cref{fig:gauss_dspcKs:ts}, divided SPC with 2nd moment statistic and rate $\beta > 0.5$ results in large Type I errors, which coincides with \cref{THM:DIVIDEDSPC}. On the other hand, choosing rates too small may produce inefficient powers in small-data scenario, as in \cref{fig:gauss_dspcKs_pw_q75}. 

\begin{figure}[tp]
\centering
\subfloat[ mean]{\label{fig:gauss_compare_mean_ts}\includegraphics[width=60mm]{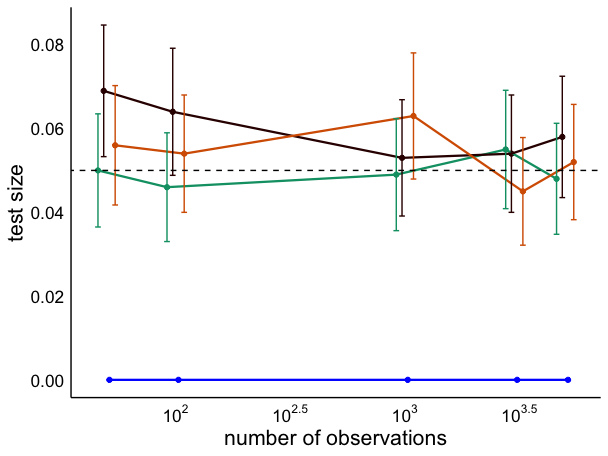}}
\subfloat[ mean]{\label{fig:gauss_compare_mean_pw}\includegraphics[width=60mm]{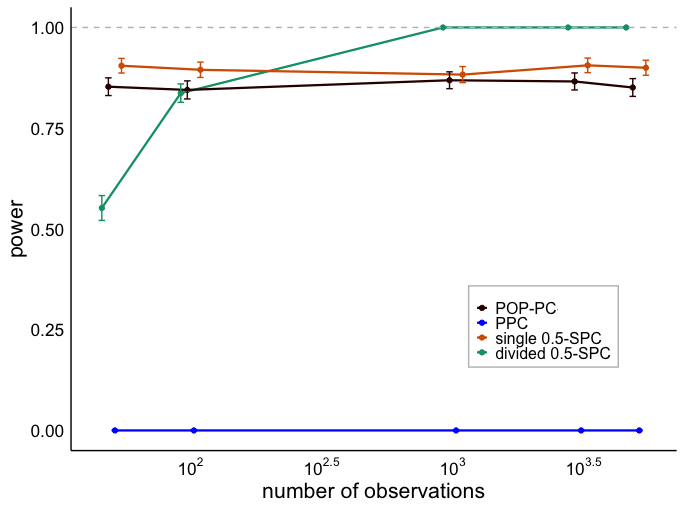}}
\\
\subfloat[75th quantile]{\label{fig:gauss_compare_q75_ts}\includegraphics[width=60mm]{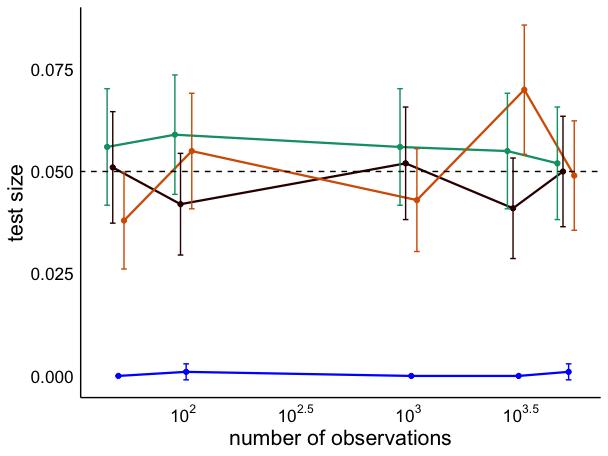}}
\subfloat[75th quantile]{\label{fig:gauss_compare_q75_pw}\includegraphics[width=60mm]{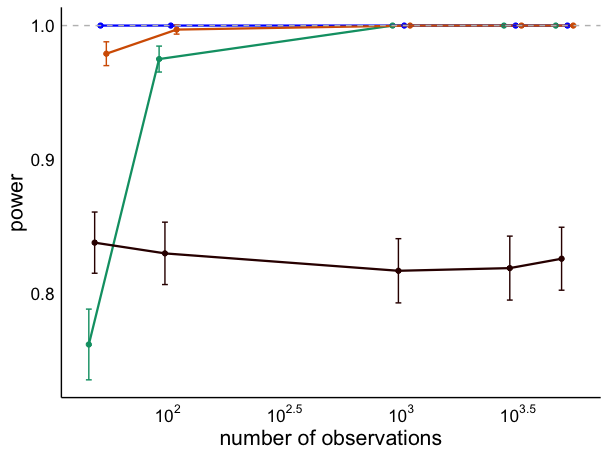}}
\\
\subfloat[2nd moment]{\label{fig:gauss_compare_secmo_ts}\includegraphics[width=60mm]{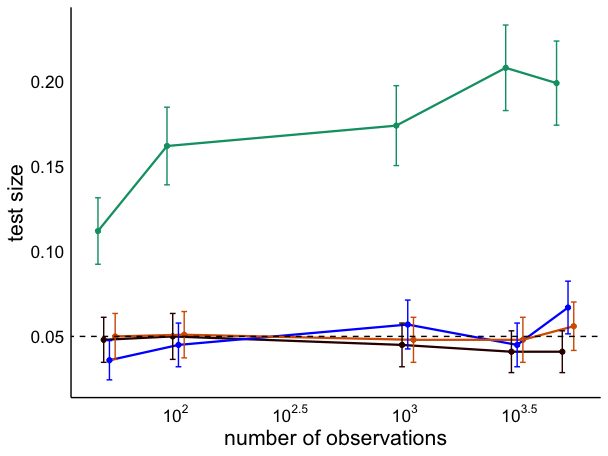}}
\subfloat[2nd moment]{\label{fig:gauss_compare_secmo_pw}\includegraphics[width=60mm]{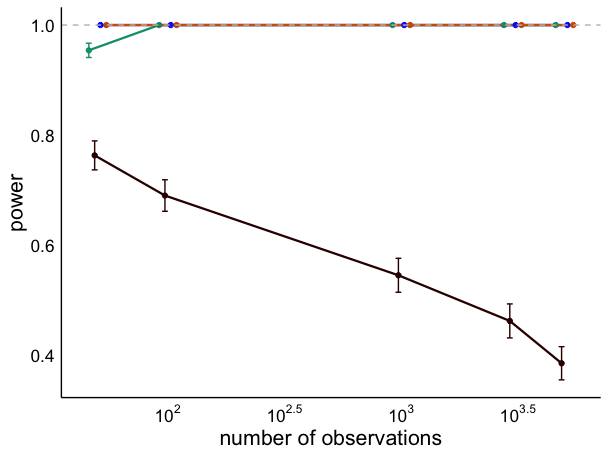}}
\caption{Test size and power plots in log scale for different checking methods and statistics under Gaussian location model. We display experiments for $\numobs \in \{50, 100, 1000, 3000, 5000\}$. All SPCs choose proportion $\spcprop = 0.5$ and for divided SPCs, the number of splits is set as $\dspcK = \numobs^{0.49}$.}
\label{figur:Gaussian_mod_comparison}
\end{figure}

\begin{figure}[tp]
\centering
\subfloat[mean]{
	\label{figur:gauss_mean_qq}
	\includegraphics[width=60mm]{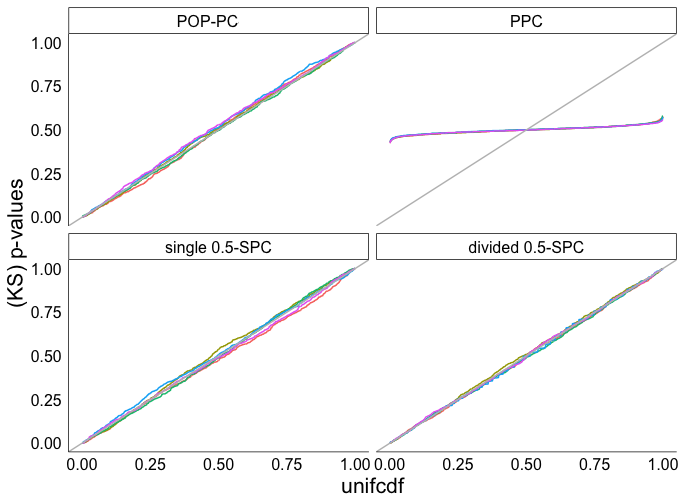}
} \qquad
\subfloat[75th quantile]{\label{figur:gauss_q75_qq}\includegraphics[width=60mm]{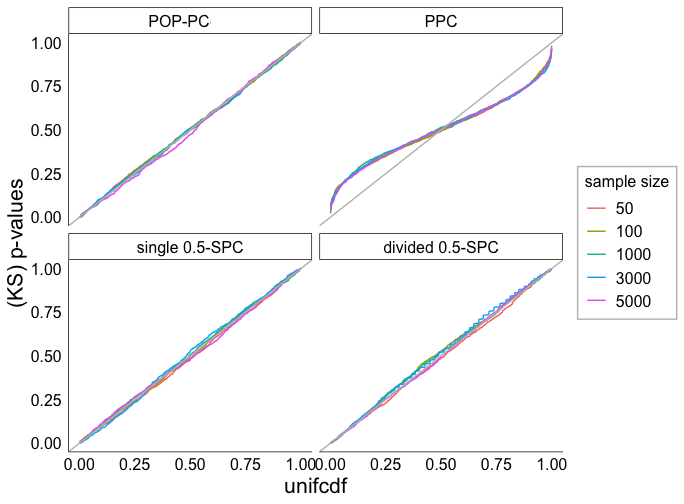}}
\caption{Q-Q plots for POP-PC, PPC, single $0.5$-SPC and divided $0.5$-SPC methods for mean and 75th quantile statistic under Gaussian location model. Diagonal gray lines are reference line for uniform Q-Q plot.  We include sample sizes $\numobs \in \{50, 100, 1000,3000, 5000\}$. }
\label{fig:gaussian_qqplots}
\end{figure}

\paragraph{Comparison of PPC, POP-PC  and SPCs} 
In \cref{figur:Gaussian_mod_comparison}, divided SPC outperforms all methods in power with mean statistic. 
PPC produces 0 power, while both POP-PC and single 0.9-SPC has power stabilized at a value not equal to 1. All four candidate methods succeed to achieve power 1 with the 75th quantile and 
2nd moment statistics except that POP-PC has a sudden drop on power given a large sample in 2nd moment case.
For the test size comparison, all methods have the test size under control. However, by plotting the Q-Q plots for each methods in \cref{fig:gaussian_qqplots}, we can conclude that PPC fails to produce frequentist $\p$-values under the mean and 75th quantile statistic given a well-specified null model.

\subsection{Simulation study for Poisson model of \cref{sec:Poisson model}}
\label{appx:pois_add_figures}
We present some additional figures for Poisson model of \cref{sec:Poisson model} here.
\subsubsection{Single SPCs}
\label{appx:poisson-single-spc}
\begin{figure}[tp]
	\centering
	\subfloat[ mean]{\label{fig:pois_props:mean_ts}\includegraphics[width=70mm]{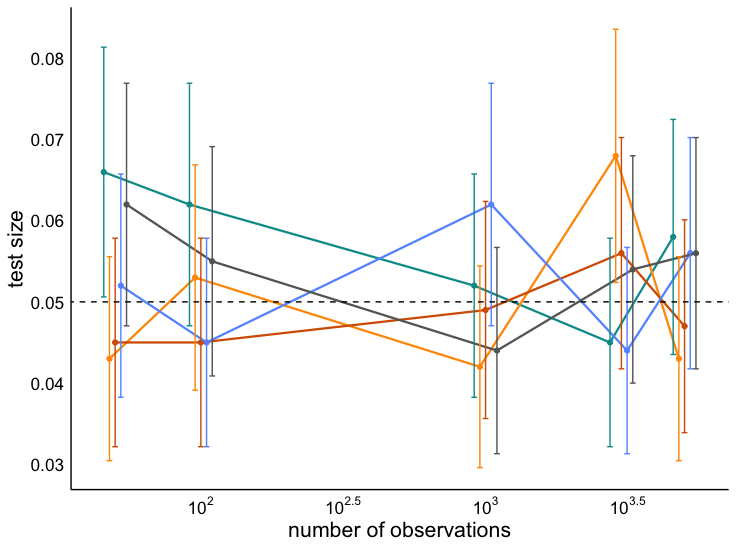}}
	\subfloat[2nd moment]{\label{fig:pois_props:secmo_ts}\includegraphics[width=70mm]{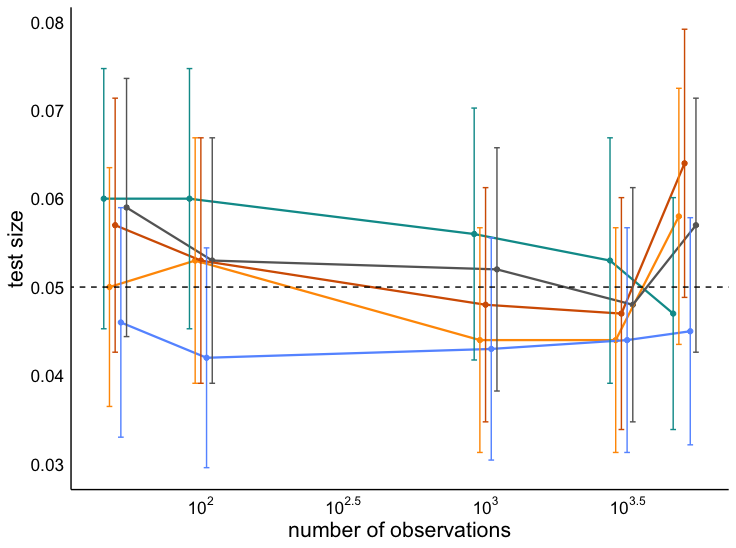}}
	\caption{Test size plots for single SPC with different split proportions $\spcprop \in  \{0.1, 0.3, 0.5, 0.7, 0.9\}$, under conjugate Poisson model. The vertical bars represent the $95\%$ confidence intervals for the test size estimates. Each test size and power are estimated with $1000$ single SPC $\p$-values that are computed with $1000$ datasets of size $\numobs \in \{ 50, 100, 1000, 3000, 5000\}$.
		The dashed horizontal line in the test size plots indicates the significance level $\alpha = 0.05$ and the dashed line in the power plots is at 1. }
	\label{appx:pois_props}
\end{figure}

As shown in \cref{appx:pois_props}, a small split proportion such as $\spcprop = 0.1$ results in a slightly larger Type I error. Hence, small split proportions are not recommended for single SPCs due to a limited number of split observed data in computing posterior distributions.

\begin{figure}[tp]
\centering
\subfloat[3rd moment]{\label{pois_props:q75_ts}\includegraphics[width=70mm]{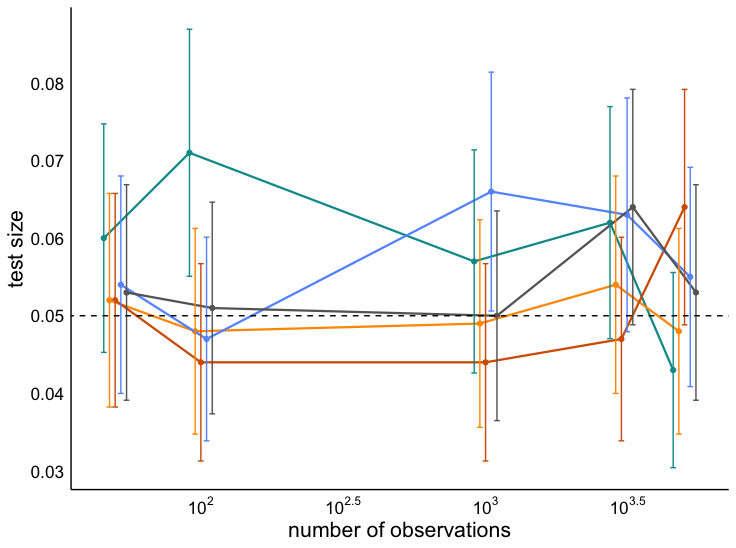}}
\subfloat[3rd moment]{\label{pois_props:q75_pw}\includegraphics[width=70mm]{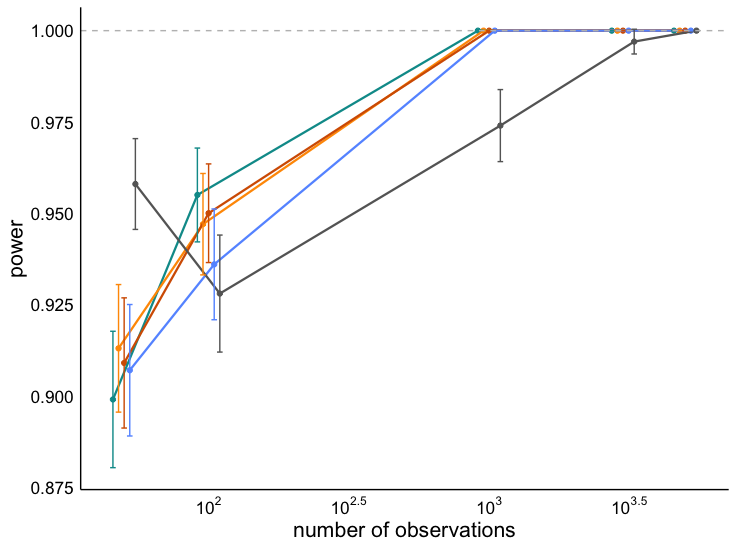}}
\\
\subfloat[MSE]{\label{pois_props:mse_ts}\includegraphics[width=70mm]{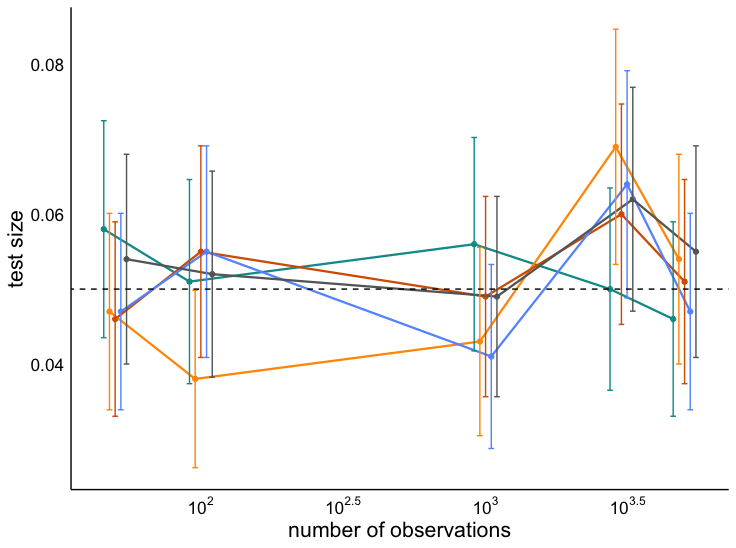}}
\subfloat[MSE]{\label{pois_props:mse_pw}\includegraphics[width=70mm]{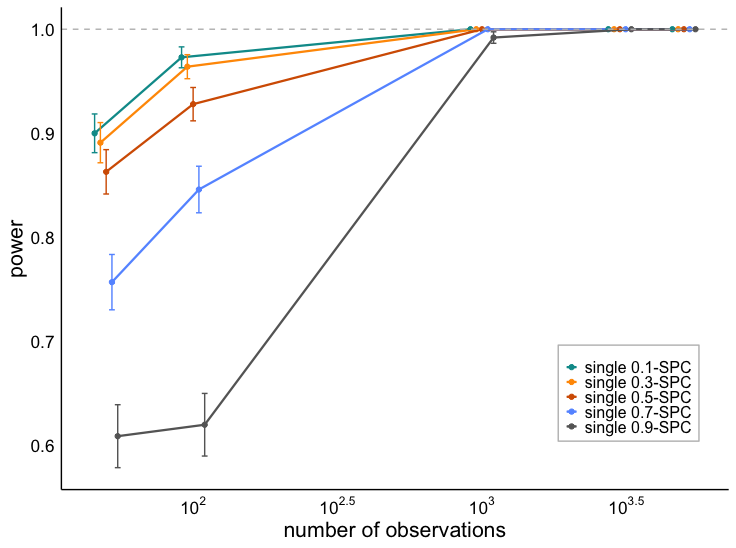}}
\caption{Test size and power plots in log scale for single SPC with different split proportions $\spcprop \in  \{0.1, 0.3, 0.5, 0.7, 0.9\}$, under the conjugate Poisson model. See caption for \cref{figur:pois_props} for further explanation. }
\label{fig:pois_props_appx}
\end{figure}

Results for 3rd moment and MSE statistic with different proportions of single SPC are displayed in \cref{fig:pois_props_appx}. The power plot with MSE in \cref{figur:pois_props} shows another limitation of single SPC that with small sample, a large split proportion $\spcprop$ may result in a slow rate of convergence in power. 

\subsubsection{Divided SPCs}
\label{appx:poisson-divided-spc}
\begin{figure}[tp]
\centering
\subfloat[2nd moment]{\label{fig:pois_dspcprops:secmo_ts}\includegraphics[width=70mm]{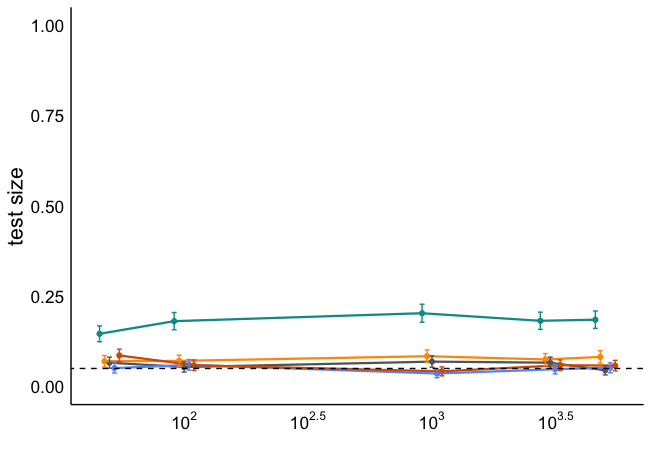}}
\subfloat[2nd moment]{\label{fig:pois_dspcprops:secmo_pw}\includegraphics[width=70mm]{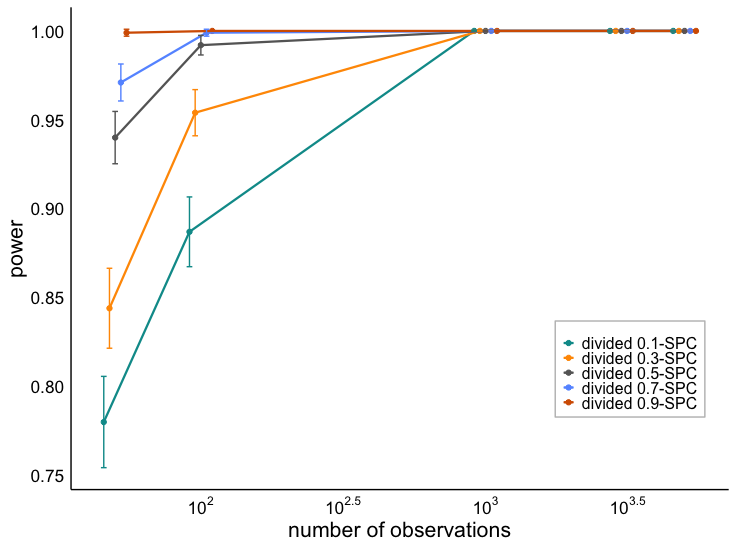}}\\
\subfloat[ 3rd moment]{\label{fig:pois_dspcprops:thirdmo_ts}\includegraphics[width=70mm]{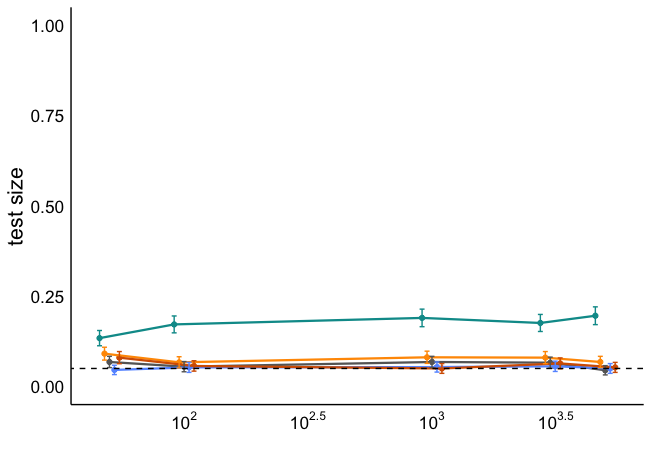}}
\subfloat[ 3rd moment]{\label{fig:pois_dspcprops:thirdmo_pw}\includegraphics[width=70mm]{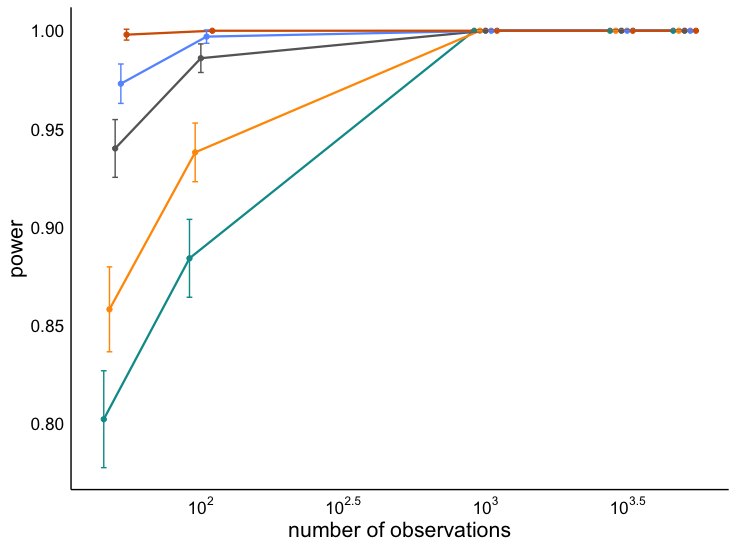}}
\caption{Test size and power plots in the log scale for divided SPC with identical $\dspcK = \numobs^{0.49}$ but different split proportions $\spcprop \in  \{0.1, 0.3, 0.5, 0.7, 0.9\}$, under the conjugate Poisson model. }
\label{figur:pois_dspcprops_appx}
\end{figure}

\begin{figure}[tp]
\centering
\subfloat[ 2nd moment]{\label{fig:pois_dspcKs_ts_secmo}\includegraphics[width=70mm]{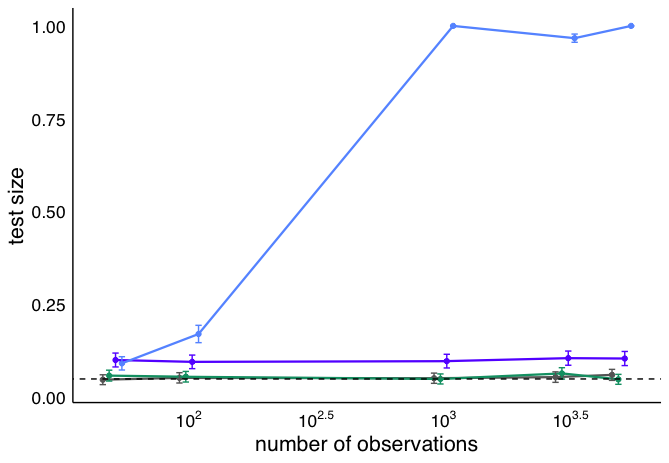}}
\subfloat[ 2nd moment]{\label{fig:pois_dspcKs_pw_secmo}\includegraphics[width=70mm]{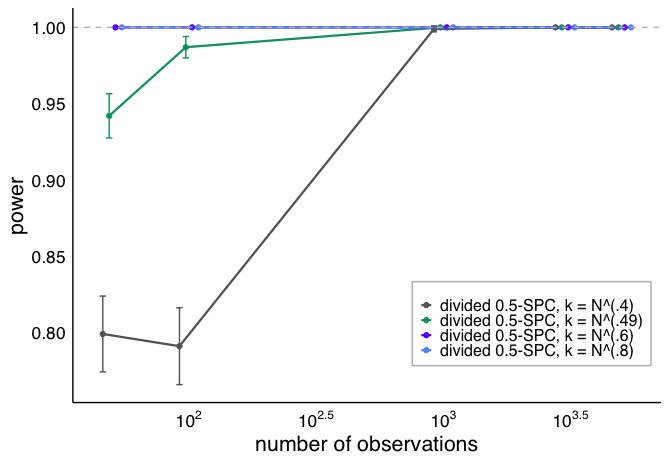}}
\\
\subfloat[ Mean]{\label{fig:pois_dspcKs_ts_mean}\includegraphics[width=70mm]{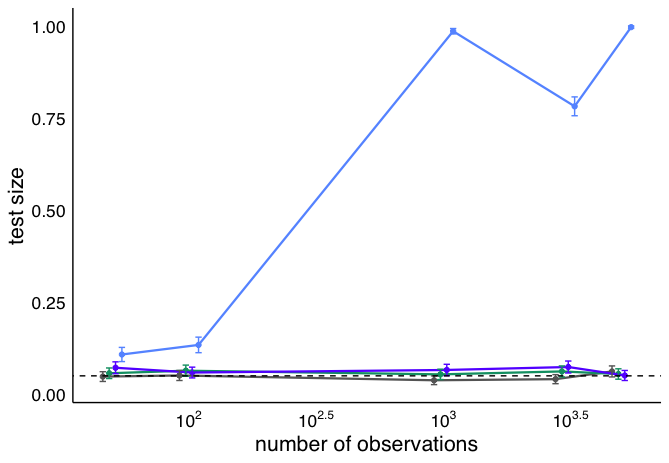}}
\subfloat[ Mean]{\label{fig:pois_dspcKs_pw_mean}\includegraphics[width=70mm]{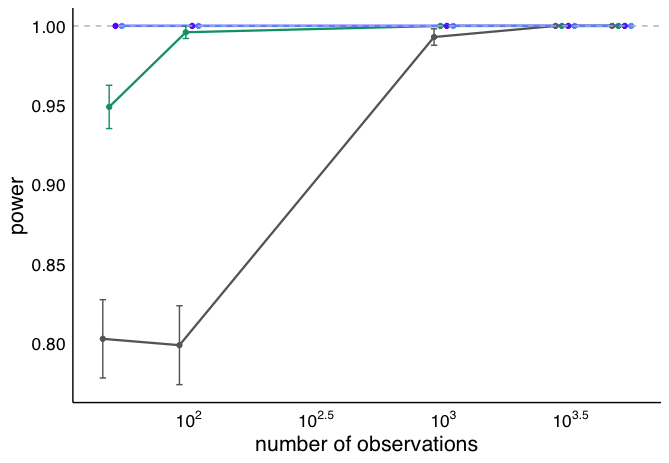}}
\caption{Test size and power plots in the log scale for divided SPC with identical split proportion $\spcprop = 0.5$ but various $\dspcK = \numobs^{0.49}$, under the conjugate Poisson model. See caption for \cref{figur:pois_dspcprops} for further explanation.}
\label{figur:pois_dspcKs_appx}
\end{figure}

The comparison of divided SPC with different proportions given statistics 2nd moment and 3rd moment is shown in \cref{figur:pois_dspcprops_appx}. Divided SPC with different choices of $\dspcK$ using 2nd moment and mean are shown in \cref{figur:pois_dspcKs_appx}. Both demonstrate similar results as in \cref{figur:pois_dspcprops}. 

\begin{figure}[tp]
	\centering
	\includegraphics[width=105mm]{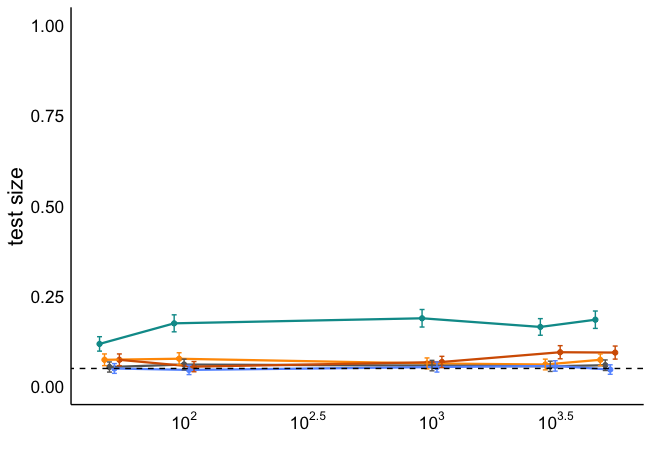}
	\caption{Test size for divided SPC with mean statistic, $\dspcK = \numobs^{0.49}$, and varying $\spcprop \in  \{0.1, 0.3, 0.5, 0.7, 0.9\}$, under the conjugate Poisson model.
		See \cref{figur:pois_props} for further explanation and experimental details.} %
\label{fig:pois_dspcprops:mean_ts}
\end{figure}

As shown in \cref{fig:pois_dspcprops:mean_ts}, the divided SPC with both $\spcprop = 0.1$ and $\spcprop = 0.9$ fail to control the test size well even when $\numobs = 5000$.  
Therefore, even though the power increases with $\spcprop$, in the small-to-moderate sample-size regime the best choice for the proportion is $\spcprop = 0.5$. 

\subsubsection{Comparison to PPC and POP-PC.}
\begin{figure}[tp]
\centering
\subfloat[ mean]{\label{fig:pois_compare_mean_ts}\includegraphics[width=60mm]{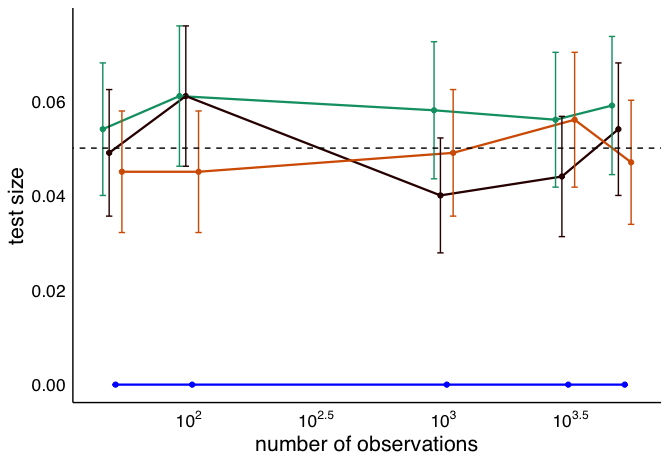}}
\subfloat[ mean]{\label{fig:pois_compare_mean_pw}\includegraphics[width=60mm]{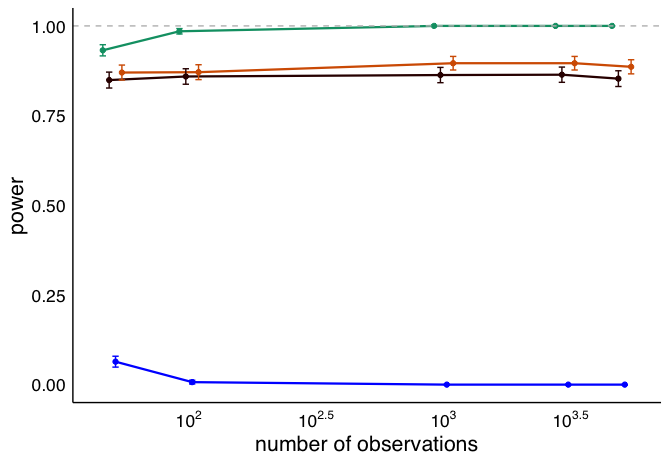}}
\\
\subfloat[2nd moment]{\label{fig:pois_compare_secmo_ts}\includegraphics[width=60mm]{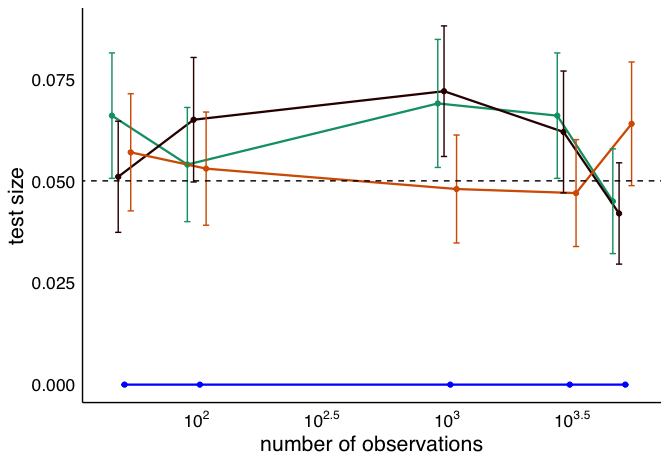}}
\subfloat[2nd moment]{\label{fig:pois_compare_secmo_pw}\includegraphics[width=60mm]{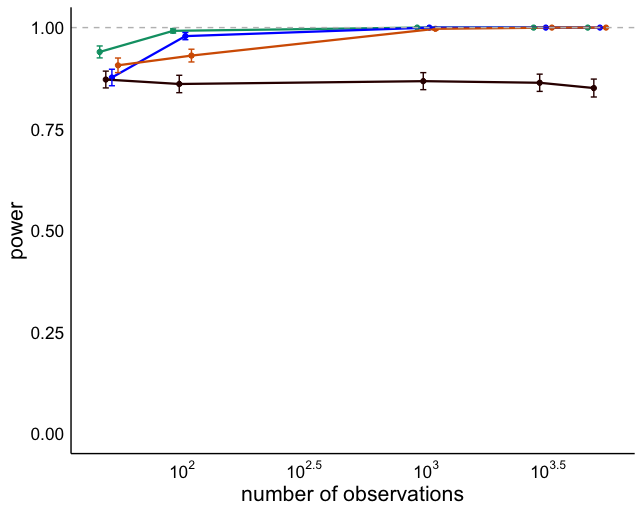}}
\\
\subfloat[3rd moment]{\label{fig:pois_compare_thirdmo_ts}\includegraphics[width=60mm]{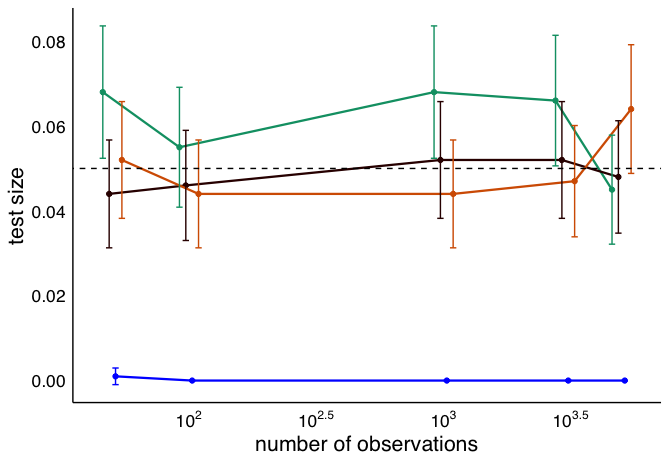}}
\subfloat[3rd moment]{\label{fig:pois_compare_thirdmo_pw}\includegraphics[width=60mm]{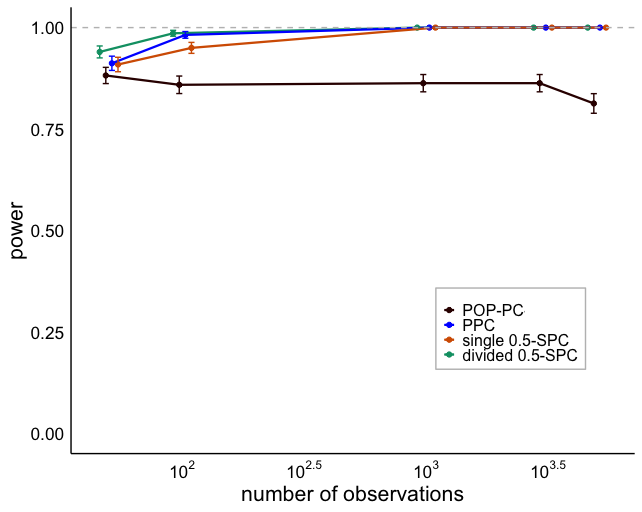}}
\caption{Test size and power plots in log scale for different checking methods and statistics under the conjugate Poisson model. We display experiments for $\numobs \in \{50, 100, 1000, 3000, 5000\}$. All SPCs choose proportion as  $\spcprop = 0.5$ and for divided SPCs, the number of splits is set as $\dspcK = \numobs^{0.49}$.}
\label{figur:Poisson_mod_comparison}
\end{figure}

\begin{figure}[tp]
\centering

\subfloat[2nd moment]{\label{figur:pois_secmo_qq}\includegraphics[width=60mm]{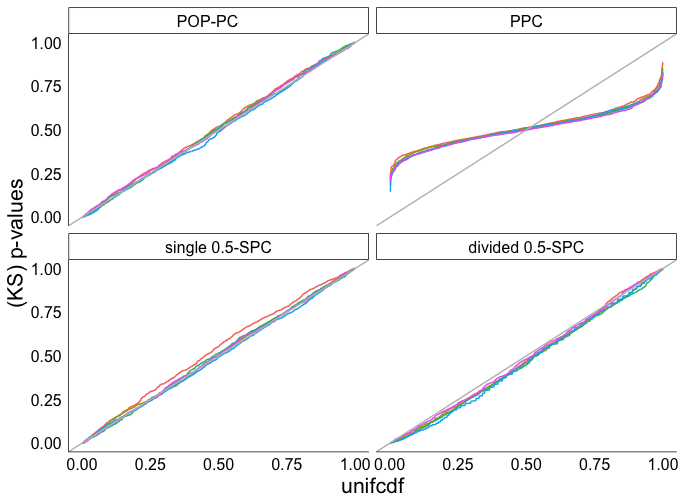}} \qquad
\subfloat[3rd moment]{\label{figur:pois_thirdmo_qq}\includegraphics[width=60mm]{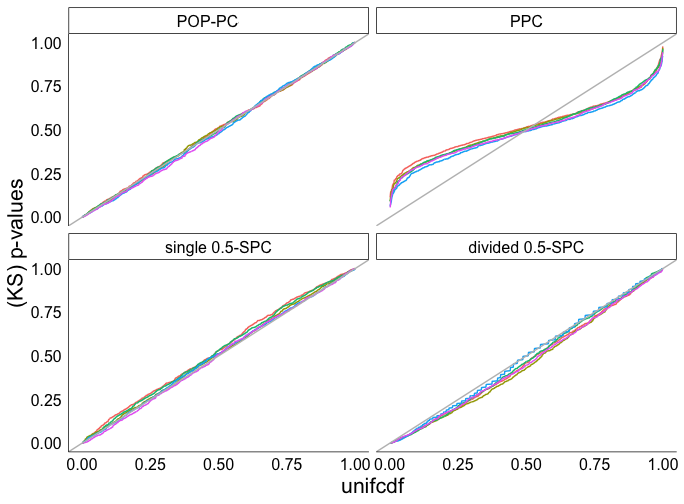}}\\
\subfloat[mean]{	\includegraphics[width=75mm]{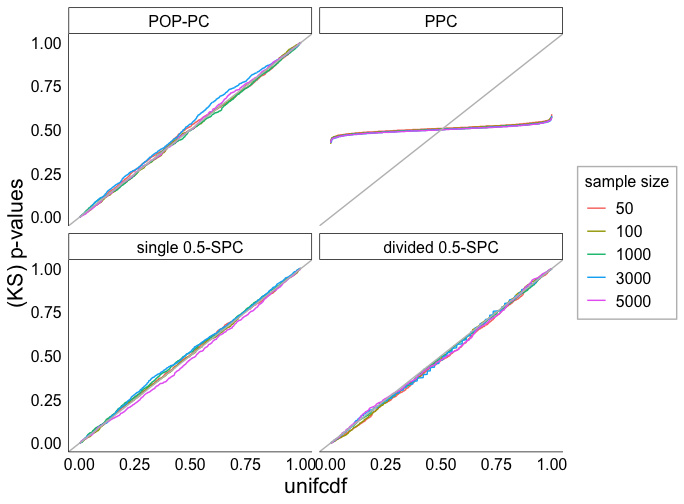}}
\caption{Q-Q plots for POP-PC, PPC, single $0.5$-SPC and divided $0.5$-SPC methods for  mean, 2nd moment and 3rd moment statistic under the conjugate Poisson model. We include sample sizes $\numobs \in \{50, 100, 1000,3000, 5000\}$. }
\label{fig:poisson_qqplots_appx}
\end{figure}

We compare the single $0.5$-SPC and divided $0.5$-SPC with $\dspcK = \numobs^{0.49}$ to the PPC and POP-PC. 
\Cref{fig:pois_compare_mean_pw} demonstrates divided SPC is very sensitive to misspecification when the sample size is large. 
In this case, the PPC has power near zero and the power of both the POP-PC and single 0.5-SPC stabilizes below one. 
\Cref{fig:pois_compare_secmo_pw,fig:pois_compare_thirdmo_pw} show that for the 2nd moment and 3rd moment statistics, 
all methods have power approaching one, except POP-PC. %
\cref{fig:pois_compare_mean_ts,fig:pois_compare_secmo_ts,fig:pois_compare_thirdmo_ts} show that all candidate checks control the test size well. However, even with the smallest test size,  PPC fails to produce frequentist $\p$-values for all three statistics, 
while POP-PC, single SPC, and divided SPC all have $\p$-values that are close to uniform under the well-specified case (\cref{fig:poisson_qqplots_appx}).
The Q-Q plots of SPCs with PPC and POP-PC using mean, 2nd moment and $3$rd moment under a well-specified model are presented in \cref{fig:poisson_qqplots_appx}.

\subsection{Additional figures for Gaussian Hierarchical simulation study of \cref{sec:simulation_hier}}
\label{appx:gauss_hier_figures}
\begin{figure}[tp]
\centering
\subfloat[mean of $75$th quantiles]{\label{fig:hier_qq_meanq75_ppc}\includegraphics[width=55mm]{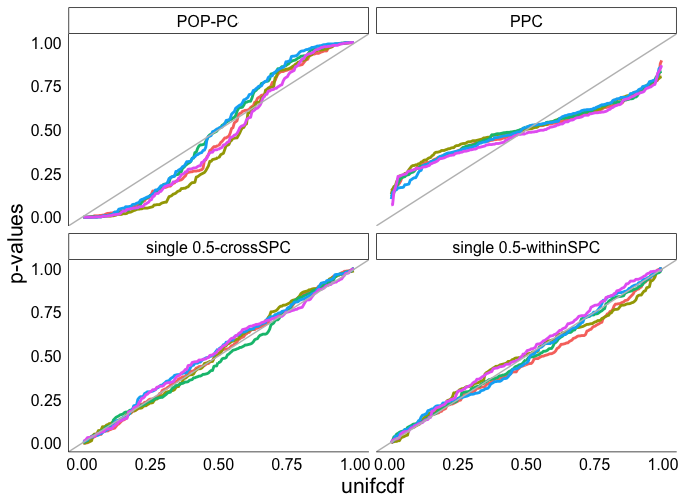}}
\subfloat[mean of $75$th quantiles]{\label{fig:hier_qq_meanq75_dspcs}\includegraphics[width=55mm]{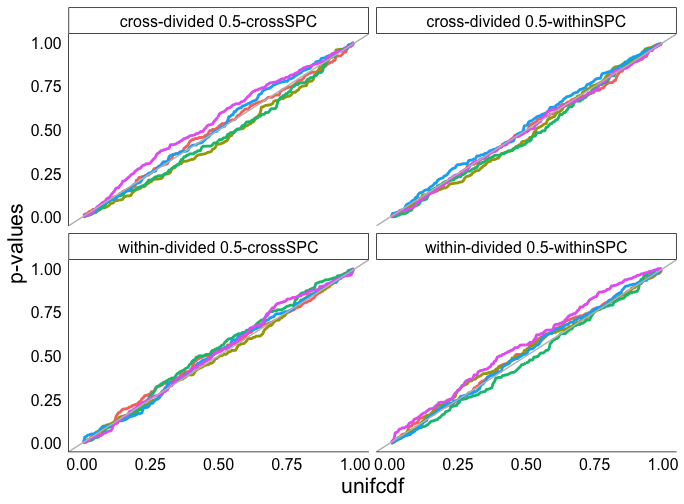}} \\
\subfloat[$75$th quantile of group means]{\label{fig:hier_qq_q75means_ppc}\includegraphics[width=55mm]{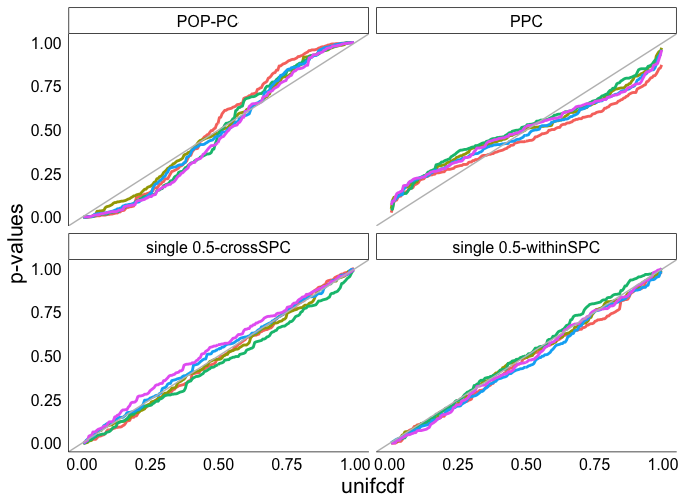}}
\subfloat[$75$th quantile of group means]{\label{fig:hier_qq_q75means_dspcs}\includegraphics[width=55mm]{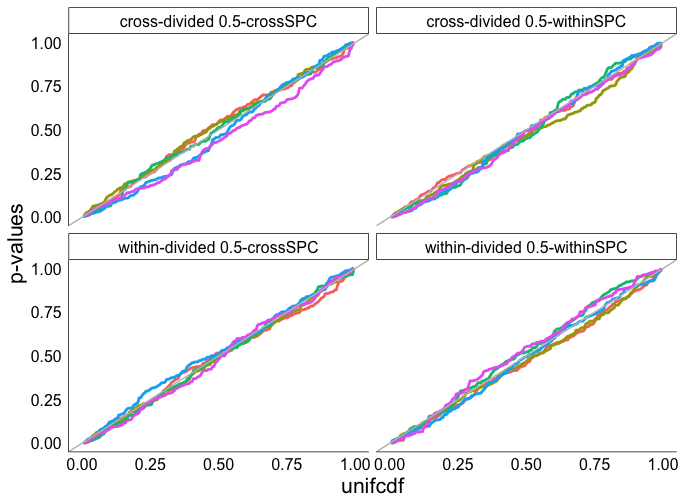}}\\
\subfloat[grand mean]{\label{fig:hier_qq_grmean_ppc}\includegraphics[width=55mm]{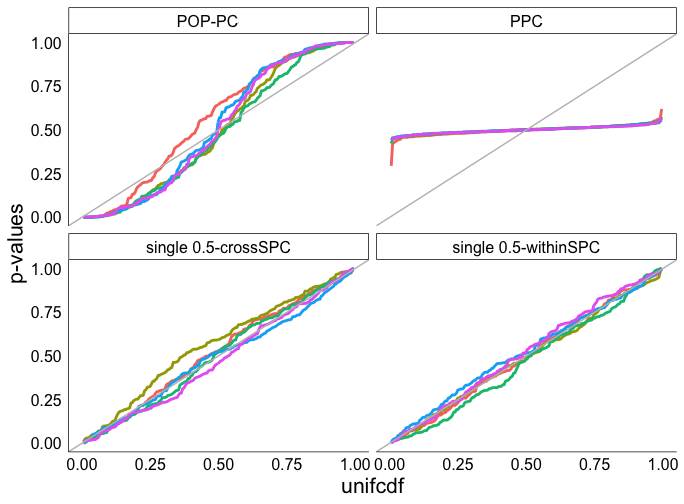}}
\subfloat[grand mean]{\label{fig:hier_qq_grmean_dspcs}\includegraphics[width=55mm]{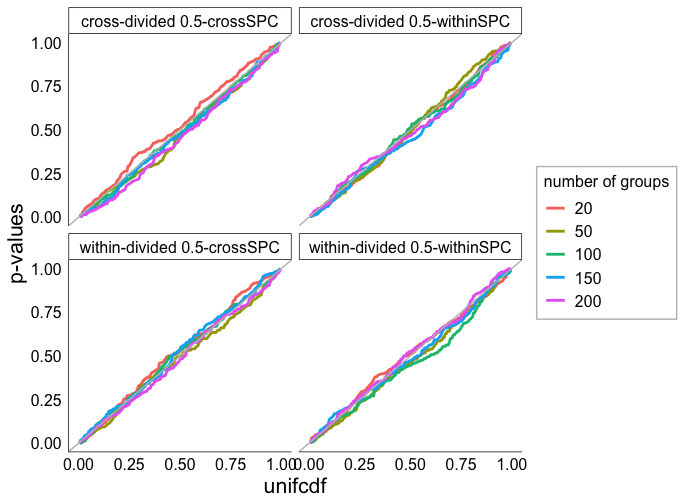}} 
\caption{The Q-Q plots of all checks under the well-specified model with grand mean, mean of $75$th quantiles and $75$th quantile of group means statistic, under the Gaussian hierarchical model. }
\label{fig:hier_qqplot_appx}
\end{figure}

We include the Q-Q plots of all candidate checks and statistics in \cref{fig:hier_qqplot_appx}. It turns out all types of SPCs produce well-calibrated $\p$-values while PPC and POP-PC  fail to generate uniform $\p$-values with all statistics used in the study. 

\begin{figure}[tp]
\centering
\subfloat[Misspecified across groups]{\label{fig:hier_pw_I20_cr}\includegraphics[width=65mm]{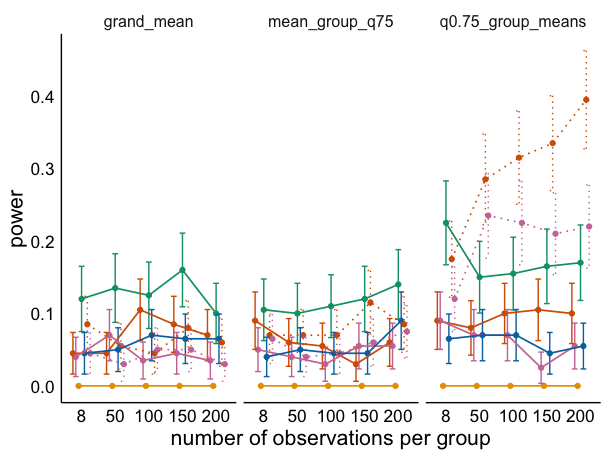}}
\qquad
\subfloat[Misspecified within groups]{\label{fig:hier_pw_I20_grvar}\includegraphics[width=65mm]{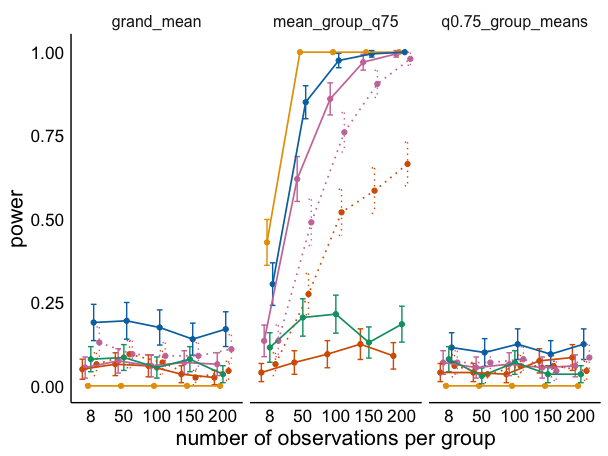}}
\\
\subfloat[Misspecified across and within groups]{\label{fig:hier_pw_I20_wilognorm}\includegraphics[width=80mm]{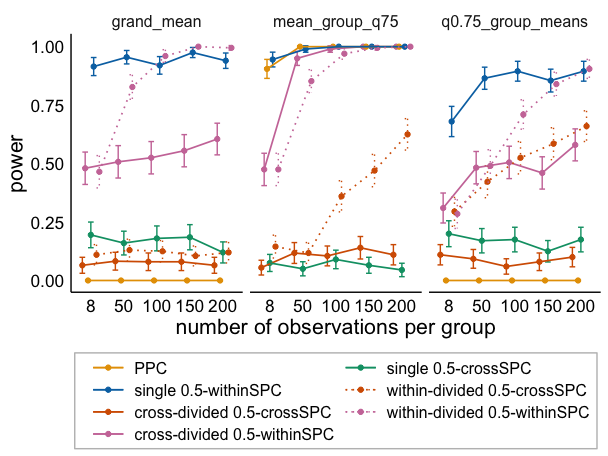}}
\caption{Power plot for different checks and statistics given different types of misspecification with fixed the number of groups $I = 20$. We increase the number of individuals per group to investigate how the power changes accordingly in different scenarios. Powers are estimated with 200 repeated experiments for each scenario. Vertical bars show the 95th confidence interval for each power estimate. }
\label{fig:hier_pw_I20}
\end{figure}

We show results of all methods given fixed number of groups $I = 20$ and increasing number of individuals per group $J$ in \cref{fig:hier_pw_I20}. 
For group-level mismatches, \cref{fig:hier_pw_I20_cr} shows that increasing number of observations per group may not help in detecting such misspecification even with proper statistics. However, with lower-level misspecified model, we see in \cref{fig:hier_pw_I20_grvar} that power of PPC and SPCs with within-splits tend to increase as the number of observations per group increases. A large sample size for each group also provides more information to checking methods when the model is misspecified in both levels as shown in \cref{fig:hier_pw_I20_wilognorm}. 

\subsection{Real-data experiment: light speed data}

\begin{figure}[tp]
\centering
\label{fig:light_dist}\includegraphics[width=105mm]{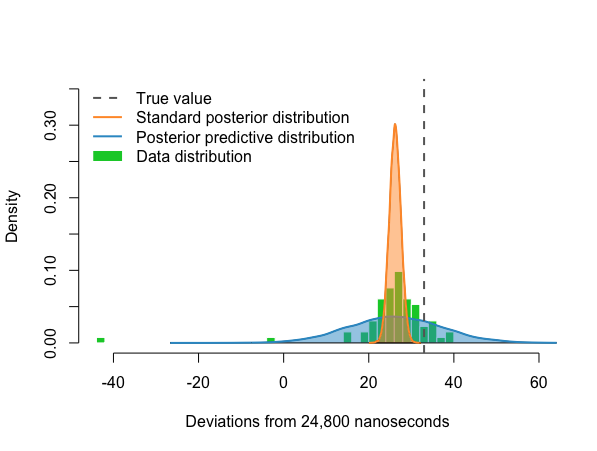}
\caption{The histogram of light data, the standard posterior and posterior prediction density curves. The vertical dashed line is the ``true value'' based on the currently accepted value of speed of light, $\nu = 33.0$. }
\end{figure}

The \textsf{light speed} data is discussed in both \citet{Gelman:2013} and \citet{Robins:2000}. 
The data were collected by Simon Newcomb (1882), who measured the 
time that light takes to travel 7442 meters at sea level. 
The observations are recorded as deviations from 24800 nanoseconds and $\numobs = 66$.
We consider a normal model $\likdist{\param} = \distNorm(\mu, \sigma^{2})$ with an improper prior density $\pi_{0}(\mu, \sigma^2) = \sigma^{-2}$. 
A comparison of the data distribution and the posterior predictive distribution is displayed in \cref{fig:light_dist}, which shows that the model 
does not capture the heavy left tail of the data and that the 95\% posterior credible interval does not contain the true value. 
We consider five statistics: mean, standard deviation, MSE, $5$th and $95$th quantiles.
We use $\spcprop = 0.5$ to ensure the statistics (particularly the quantiles) can be reasonably estimated. 
Two-sided $\p$-values computed for each method and statistic are given in \cref{table:pvals_light_data}.
Both the PPC and single SPC yield small $p$-values for the left quantile.
The single SPC yields a small $p$-value for the standard deviation and MSE while the PPC results in a small $p$-value for the right quantile.
Thus, we find that the single SPC captures complementary (but overlapping) forms of misspecification compared to the PPC. 
On the other hand, as expected in this small-data regime, the divided SPCs yield $p$-values far from zero.

\begin{table}[tp]
\centering
\caption{Two-sided $p$-values for light data.}
\begin{tabular}{cccccc}
	\toprule
	\textbf{Method} & \textbf{5th quantile} & \textbf{Mean} & \textbf{95th quantile} & \textbf{Std Dev} & \textbf{MSE} \\
	\midrule
	PPC   &  0.004   &0.996&  0.010 & 0.944& 0.894 \\
	single 0.5-SPC&  0.012   & 0.608   & 0.762 & 0.000& 0.000 \\
	divided 0.5-SPC, $\dspcK = \lfloor \numobs^{0.49}\rfloor $ &   0.592 & 0.533 & 0.259 & 0.239&0.200\\
	\bottomrule
\end{tabular}
\label{table:pvals_light_data}
\end{table}

\subsection{Additional figures for airline delays example}
\label{appx:airlines_figs}
\begin{figure}[tp]
\centering
\subfloat[Success rate]{\label{appx_fig:airlines_pw_sr}\includegraphics[width=65mm]{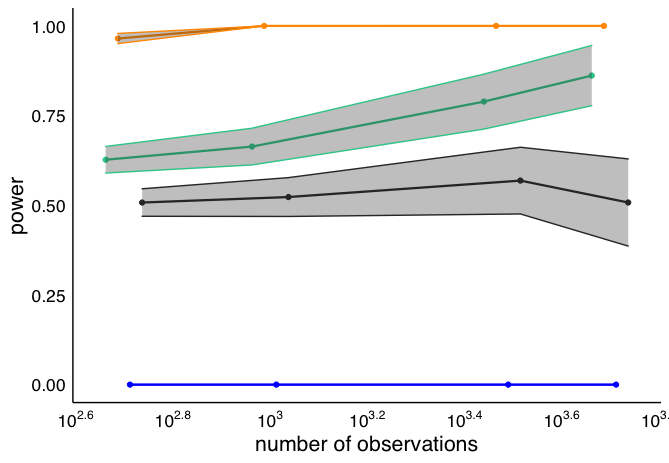}}
\subfloat[MSE]{\label{appx_fig:airlines_pw_mse}\includegraphics[width=65mm]{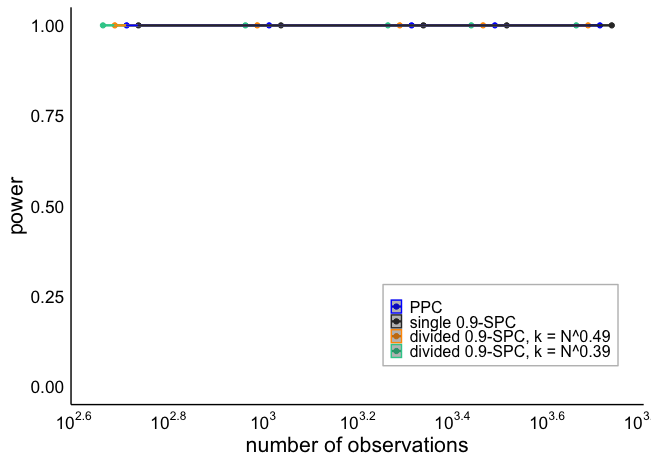}}
\caption{The power estimates for PPC and SPCs with proportion $\spcprop = 0.9$ with NYC airline data given a geometric model and success rate statistic. See caption for \cref{fig:airlines_nyc_data_power} for further explanation.}
\label{fig:airlines_nyc_data_power_appx}
\end{figure}

We present the comparison of SPCs with $\spcprop = 0.9$, PPC and POP-PC with  airlines delay data in \cref{fig:airlines_nyc_data_power_appx}. The  analysis is similar to that of \cref{fig:airlines_nyc_data_power} in \cref{sec:experiments}.

\begin{figure}[tp]
\centering
\label{fig:airlines_dist_hist}\includegraphics[width=105mm]{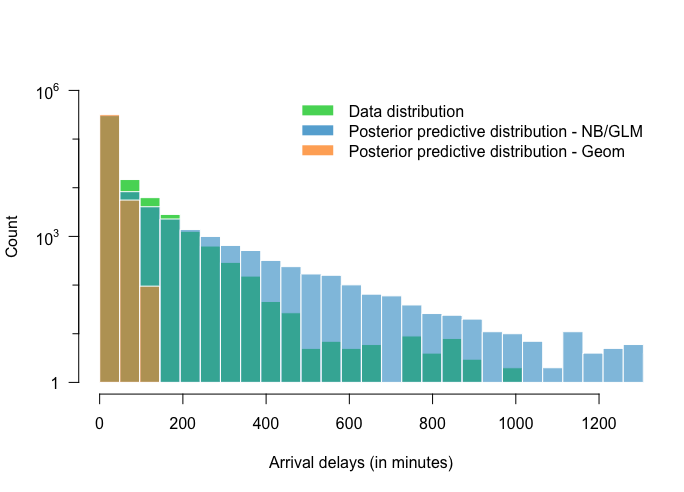}
\caption{The histograms of airline delays data and $\numobs = 327,346$ posterior predictive draws under simple geometric model (Geom) and negative binomial generalized linear model (NB/GLM) with counts in $\log$ scale.}
\end{figure}

\clearpage

\begin{figure}[tp]
\centering
\subfloat[mean, $\param_{\star} = 100$ ]{\label{figur:pois_ess_mean}\includegraphics[width=60mm]{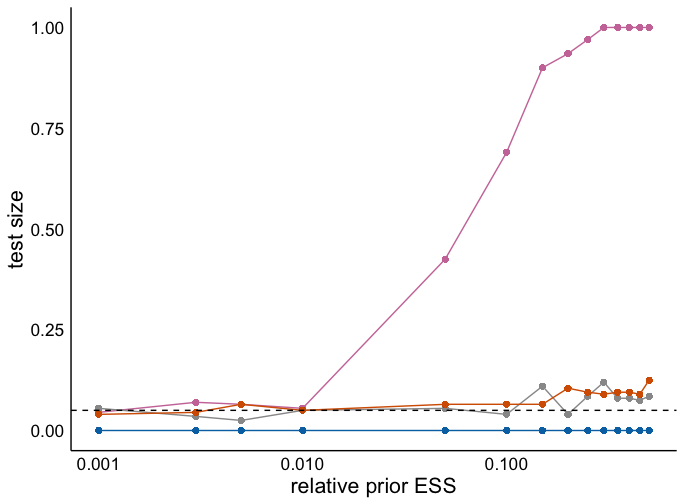}}\qquad
\subfloat[2nd moment,  $\param_{\star} = 100$]{\label{fig:pois_ess_secmo}\includegraphics[width=60mm]{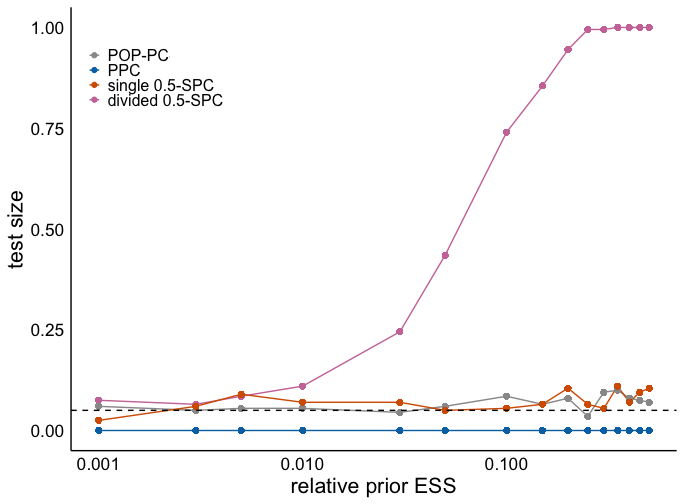}}
\subfloat[2nd moment,$\param_{\star} = 25$]{\includegraphics[width=60mm]{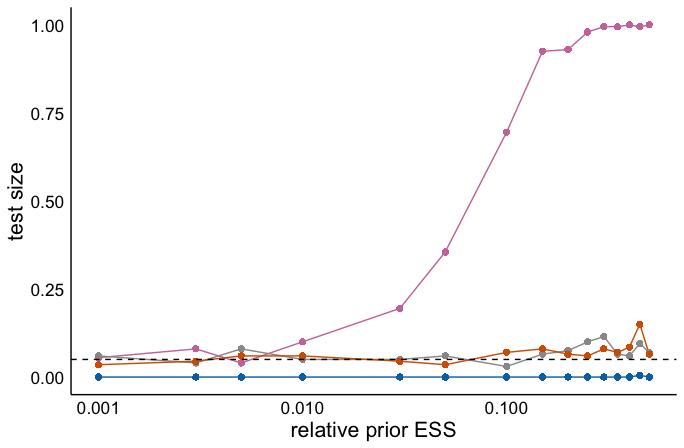}}	
\caption{Test size against relative prior ESS for POP-PC, PPC, single $0.5$-SPC and divided $0.5$-SPC with $\dspcK = \numobs^{0.49}$ for mean and 2nd moment under the conjugate Poisson model and different $\param_{\star}$ values. See caption for \cref{fig:poisson_ess} for further explanation. }
\label{fig:poisson_ess_appx}
\end{figure}

\end{document}